\documentclass[12pt]{article}
\newlength{\abstractwidth}
\usepackage{epsf}
\usepackage{color}
\usepackage{graphicx}
\usepackage{hyperref}
\hypersetup{colorlinks=true, citecolor=blue, linktoc=page}
\usepackage{dsfont}

\usepackage{booktabs}

\usepackage{amsmath, nccmath}
\usepackage{amsmath, amsthm}
\usepackage{amssymb}
\usepackage{amsfonts}
\usepackage{latexsym}
\usepackage{mathtools}

\renewcommand{\thefootnote}{\fnsymbol{footnote}}
\renewcommand{\thanks}[1]{\footnote{#1}}
\newcommand{\starttext}{
\setcounter{footnote}{0}
\renewcommand{\thefootnote}{\arabic{footnote}}}

\numberwithin{equation}{section}
\newcommand{\bea}{\begin{eqnarray}}
\newcommand{\eea}{\end{eqnarray}}
\newcommand{\be}{\begin{eqnarray}}
\newcommand{\ee}{\end{eqnarray}}
\def\beq{\begin{equation}}
\def\eeq{\end{equation}}

\newcommand{\bma}{\begin{matrix}}
\newcommand{\ema}{\end{matrix}}

\def\cA{{\cal A}}
\def\cB{{\cal B}}
\def\cC{{\cal C}}

\def\cG{{\cal G}}

\def\cI{{\cal I}}
\def\cJ{{\cal J}}
\def\cK{{\cal K}}
\def\cL{{\cal L}}
\def\cM{{\cal M}}
\def\cN{{\cal N}}

\def\cP{{\cal P}}

\def\cS{{\cal S}}
\def\cT{{\cal T}}

\def\mM{\mathfrak{M}}
\def\mN{\mathfrak{N}}

\def\mb{\mathfrak{b}}

\def\ml{\mathfrak{l}}

\def\mo{\mathfrak{o}}

\def\ms{\mathfrak{s}}

\def\CC{{\mathbb C}}

\def\RR{{\mathbb R}}

\def\tr{{\rm tr}}

\def\det{{\rm det \,}}

\def\half{{1\over 2}}
\def\thalf{\tfrac{1}{2}}
\def\p{\partial}

\def\a{\alpha}
\def\b{\beta}
\def\g{\gamma}

\def\f{\varphi}

\def\ep{\varepsilon}
\def\om{\omega}
\def\G{\Gamma}

\def\tz{{\tilde z}}
\def\tw{{\tilde w}}

\def\no{\nonumber}
\def\sm{\smallskip}

\usepackage{fix-cm}
\newcommand\semiLarge{\fontsize{16.0}{16.0}\selectfont}

\begin{document}
\starttext
\setcounter{footnote}{0}

\vspace*{0.2in}

\begin{center}

{\Large \bf Exact solutions to complex Type IIB supergravity}

\vskip 0.1in

{\semiLarge \bf for complex superalgebra $F(4)$ and its real forms}

\vskip 0.4in

{\large Eric D'Hoker${}^{(a)}$, Michael Gutperle${}^{(a)}$ and Christoph F.~Uhlemann${}^{(b)}$} 

\vskip 0.2in

{ \sl ${}^{(a)}$Mani L. Bhaumik Institute for Theoretical Physics}\\
{\sl  Department of Physics and Astronomy}\\
{\sl University of California, Los Angeles, CA 90095, USA}

 {\tt \small dhoker@physics.ucla.edu, gutperle@physics.ucla.edu}

\vskip 0.15in

{ \sl ${}^{(b)}$Theoretische Natuurkunde, Vrije Universiteit Brussel and}\\
{\sl  The International Solvay Institutes, Pleinlaan 2, B-1050 Brussels, Belgium}

{\tt \small christoph.uhlemann@vub.be}

\vskip 0.1in

\vskip 0.2in

\begin{abstract}
\vskip 0.1in

We construct the general local solutions to complexified Type IIB supergravity which are invariant under the complexified Lie superalgebra $F(4)$. The geometry is a product of complexified maximally symmetric spaces $\cM_{6  \CC}$ and $\cM_{2  \CC}$ warped over a complexified surface $\Sigma_\CC$. We classify the reality conditions that may be imposed consistently to obtain real form solutions within real forms of complex Type IIB supergravity. The latter comprise standard Type IIB, Type IIB$^\star$ and IIB$^\prime$, as well as theories  with $3$, $5$, $7$ and $9$ time-like directions. Our classification of real solutions is consistent with and exhausts the real forms of $F(4)$, whose classification we confirm by elementary methods. The geometry of each real form solution is a product of real maximally symmetric spaces $\cM_6$ and $\cM_2$ warped over a Riemann surface $\Sigma$, with various signatures.  The real solutions include, among others, known $AdS_6 \times S^2 \times \Sigma$ and $AdS_2 \times S^6 \times \Sigma$ solutions to standard Type IIB as well as new solutions of the form $dS_{1,5}\times S^2 \times \Sigma$ in Type IIB$^\star$. There are no real forms of the complex solutions with $\ms \mo (7;\RR) \oplus \ms \mo (3;\RR)$ symmetry. We discuss the relevance of the complex solutions, and of analytic continuations from $dS_{1,5}\times S^2\times\Sigma$ to  $S^6\times S^2\times\Sigma$ within complex Type IIB, in connection with holography for  the polarized IKKT model.

\end{abstract}
\end{center}

\newpage

\setcounter{tocdepth}{2} 
\tableofcontents

\baselineskip=15pt
\setcounter{equation}{0}
\setcounter{footnote}{0}

\newpage

%%%%%%%%%%%%%%%%%%%%%%%%%%%%%%%%%%%%%%%%%%%
%%%%%%%%%%%%%%%%%%%%%%%%%%%%%%%%%%%%%%%%%%%
\section{Introduction}
\setcounter{equation}{0}
\label{sec:1}
%%%%%%%%%%%%%%%%%%%%%%%%%%%%%%%%%%%%%%%%%%%
%%%%%%%%%%%%%%%%%%%%%%%%%%%%%%%%%%%%%%%%%%%

One motivation for the study of classical solutions to analytically continued or complexified quantum field theories and string theories is the use of saddle points in the semi-classical approximation to functional integrals. These saddle points are often given by complex solutions, as is familiar from the study of quantum tunneling via instanton and bounce solutions. Another motivation comes from supersymmetric localization, which can be used to derive exact results for certain Euclidean supersymmetric quantum field theories \cite{Pestun:2007rz,Freedman:2013oja,Bobev:2013cja} (see \cite{Pestun:2016zxk} for a review). Continuing from Lorentzian to Euclidean signature generally requires complexification.
Complex solutions also play a role in the Euclidean path integral formulation of quantum gravity \cite{Gibbons:1978ac, Halliwell:1989dy,Louko:1995jw,Witten:2021nzp}, in the context of time-like T-duality \cite{Hull:1998vg,Hull:1998ym}, and as holographic duals for supergroup gauge theories \cite{Dijkgraaf:2016lym}. 

\sm

Over the past year, a class of complex solutions to Type IIB supergravity was considered as holographic duals to the Ishibashi-Kawai-Kitazawa-Tsuchiya (IKKT) matrix model \cite{Ishibashi:1996xs} or, more precisely, to its ``polarized" version \cite{Bonelli:2002mb}. The polarized IKKT model is a deformation of the Euclidean IKKT model which is invariant under an $SO(7) \times SO(3)$ subgroup of $SO(10)$ and 16 supersymmetries \cite{Hartnoll:2024csr,Hartnoll:2025ecj}. These properties have led to the proposal that its holographic dual belongs to a class of complex solutions to Type IIB supergravity, with a spacetime of the form $S^6 \times S^2$ warped over a Riemann surface $\Sigma$~\cite{Komatsu:2024bop,Komatsu:2024ydh}. 

\sm

Solutions to standard Lorentzian Type IIB supergravity that are invariant under a real form of  $F(4)$ were obtained for spacetimes of the form $AdS_6 \times S^2 $ warped over a Riemann surface $\Sigma $ in  \cite{DHoker:2016ujz}. They are  dual to 5-dimensional superconformal field theories. A corresponding analysis was carried out for $AdS_2 \times S^6$ warped over~$\Sigma$ in \cite{Corbino:2017tfl}.  
These solutions were obtained by solving the BPS equations and are supersymmetric by construction. The configurations used for IKKT holography in \cite{Komatsu:2024bop}, on the other hand, were engineered from the solutions of \cite{DHoker:2016ujz} by formal analytic continuation from $AdS_6\times S^2$ to $S^6 \times S^2$ and argued to be solutions to the Type IIB field equations. 
This left the discussion of fermionic symmetries open and the connection to the $F(4)$ superalgebra speculative.

\sm

The aforementioned developments raise a number of questions regarding the solutions employed for IKKT holography, e.g.\ whether they are real forms or genuinely complex, and whether there is a real form of $F(4)$ whose maximal bosonic subalgebra is the isometry algebra  $\ms \mo (7;\RR) \, \oplus \, \ms \mo (3;\RR)$ of $S^6 \times S^2$. We will discuss these in the main part and obtain results of practical use for IKKT holography. The scope of this paper, however, is broader.

\sm

The setting for this work is fully complexified Type IIB supergravity, whose construction we will describe. Our aim is a systematic and complete construction of genuinely complex supersymmetric solutions invariant under the complexified form $F(4,\CC)$ of the Lie superalgebra $F(4)$, before systematically restricting the solutions to real forms relevant for applications.
The real forms of the solutions are obtained in real forms of complex Type IIB supergravity which include standard Type IIB as well as Type~IIB$^\star$, Type~IIB$^\prime$ and theories with $3$, $5$, $7$ and $9$ time directions.
Concretely, we will:
\begin{itemize}
	\itemsep =-0.02in
	\item[--] systematically and rigorously construct the general local form of complex solutions that are invariant under the complexified Lie superalgebra $F(4,\CC)$ in complexified Type IIB supergravity with their full fermionic symmetries;
	
	\item[--] systematically classify all real forms of the supergravity solutions within the real forms of the complexified Type IIB theory exposed previously in \cite{Bergshoeff:2007cg};
	
	\item[--] provide an independent classification of the real forms of $F(4,\CC)$, intended to resolve conflicting statements in the literature, and establish that they accommodate all real forms of the complex Type IIB solutions;
	
	\item[--] clarify conceptual aspects of the solutions engineered previously for IKKT holography and highlight the role some of the complex solutions constructed here play.
\end{itemize}
The result of the second part will be a complete list of real forms of supergravity solutions.
These are tabulated in Tables \ref{eq:real-forms-reduced-BPS}, \ref{4.table.IIB5}, \ref{4.table.IIB79}. They are constructed as local solutions to the BPS equations with explicit Killing spinors.
The implementation of global existence and regularity conditions will be left for future work. The bosonic symmetries of the real forms of $F(4)$ are summarized in (\ref{listmbsa}), and fully accommodate the real form solutions.

\sm

One additional application emerges from the role of the $AdS_6\times S^2\times\Sigma$ solutions constructed in \cite{DHoker:2016ujz} as holographic duals of 5d SCFTs realized by $(p,q)$ 5-brane junctions. Our results may be used to describe junctions of `negative 5-branes' in string theory and the moduli space of supergroup extensions of 5d SCFTs (which often have gauge theory deformations), similar to the discussions for theories based on $\mathcal N=4$ SYM in \cite{Dijkgraaf:2016lym}.
Our general complex solutions also contain real forms with the appropriate symmetries to describe the spherical 5-branes studied in \cite{Bobev:2018ugk,Bobev:2019bvq}, namely $\ms\mo(7)\oplus\ms\mo(1,2)$.

\sm

Finally, before proceeding to a more detailed outline, we note that it might be interesting to study complex extensions of other families of half-maximally supersymmetric solutions with similar structure, e.g.\ $AdS_2\times S^2\times S^4\times\Sigma$,  $AdS_4\times S^2\times S^2\times\Sigma$ and $AdS_2\times \mathds{CP}^3\times\Sigma$ in Type IIB \cite{DHoker:2007zhm,DHoker:2007mci,Corbino:2020lzq}, and 
$AdS_3\times S^3\times S^3\times\Sigma$ in M-theory \cite{DHoker:2008lup}.

\subsection{Overview and organization}

To complexify Type IIB supergravity we start with standard Type IIB supergravity in a formulation where all boson fields are real-valued and all fermion fields are Majorana-Weyl spinors, dependent on ten real (local) coordinates of spacetime \cite{Bergshoeff:2007cg}.  One then promotes each real-valued  boson field to a complex-valued field and each Majorana-Weyl spinor field to a complex Weyl spinor field, both of which depend holomorphically on ten local complex coordinates of complexified spacetime. Complexifying spacetime is required for the consistent  transformation of the complexified supergravity fields under any complexified symmetry algebra, such as $F(4;\CC)$. The resulting complexified supergravity theory will be referred to as Type~IIB$_\CC$. 

\sm

Since the spacetime metric of Type~IIB$_\CC$ is complex-valued, it can accommodate real geometries with arbitrary spacetime signatures as special cases. The consistent real forms of Type~IIB$_\CC$, as classified in \cite{Bergshoeff:2007cg}, include standard Type~IIB supergravity, henceforth referred to as Type IIB$_\RR$, and the Type IIB$^\star$ and Type IIB$^\prime$  theories which are related to Type~IIA supergravity by a time-like T-duality \cite{Hull:1998vg}. Each of these theories has spacetime signature $(1,9)$ and is accompanied by variants with signature $(5,5)$ and $(9,1)$, which we denote by an additional subscript highlighting the number of time directions, e.g.\ Type~IIB$_5^\star$ and Type~IIB$^\star_9$.
The remaining real forms are a pair of theories with spacetime signatures $(3,7)$ and $(7,3)$ which we denote as Type~IIB$_3$ and Type~IIB$_7$, respectively. These results are reviewed in section \ref{sec:2} while a derivation of the complexification procedure leading to the BPS equations of Type~IIB$_\CC$ is reviewed in appendix \ref{sec:B}.

\sm

The remainder of the paper is dedicated to obtaining the general local solutions to Type IIB$_\CC$ supergravity  that are invariant under $F(4;\CC)$ as well as the solutions to the real forms of Type IIB$_\CC$ that are invariant under the various real forms of $F(4;\CC)$. The maximal bosonic subalgebra $\ms \mo (7;\CC) \, \oplus \, \ms \mo (3;\CC)$ of $F(4;\CC)$ will require the spacetime to be of the form $\cM_{6\CC} \times \cM_{2\CC}$ warped over a surface $\Sigma_\CC$, where $\cM_{6\CC}$ and $\cM_{2\CC}$ are complexified maximally symmetric spaces of complex dimension six and two, respectively and $\Sigma _\CC$ is a complexified surface of complex dimension 2. Each real form of $\ms \mo (7;\CC) \, \oplus \, \ms \mo (3;\CC)$ will require the spacetime to be of the form $\cM_6 \times \cM_2$ warped over a Riemann surface $\Sigma$, where $\cM_6$ and $\cM_2$ are real maximally symmetric spaces of real dimension six and two. 

\sm

In section \ref{sec:3}, we construct the general local solution to Type IIB$_\CC$ invariant under $F(4;\CC)$ by adapting the strategy used in \cite{DHoker:2016ujz}. Spacetime is complexified, the corresponding spaces $\cM_{6  \CC}$, $\cM_{2 \CC}$ and $\Sigma_\CC$ have complex dimensions 6, 2, and 2, respectively, and the supergravity fields are (locally) holomorphic functions on these spaces. The general form of the fields is restricted by imposing  $\ms \mo (7;\CC) \, \oplus \, \ms \mo (3;\CC)$ symmetry and the BPS equations are required to admit 16 complex Weyl spinor supersymmetries.  In particular, the 5-form field strength  must vanish on all these solutions. Utilizing a generalized conformal gauge on $\Sigma_\CC$ the reduced BPS equations are successively decoupled from one another. The BPS equations are solved in terms of two pairs of functions $\cA_{1,2}$, $\tilde \cA_{1,2}$ on $\Sigma_\CC$. The first pair depends holomorphically on only one coordinate $w$ while the second pair depends holomorphically on the other coordiante $\tilde w$ on $\Sigma_\CC$.
Technical aspects of these calculations are relegated to appendices \ref{sec:C}, \ref{sec:D} and \ref{sec:E}. We close section \ref{sec:3} by identifying an $SL(2,\CC)$ global symmetry of Type~IIB$_\CC$ and deriving its action on the supergravity fields. 

\sm
 
In section \ref{sec:5}, we analyze the consistency of the reality conditions for the various real forms of Type~IIB$_\CC$ supergravity with the Killing spinor chirality and projection conditions. 
Upon restriction to real form solutions, $\Sigma_\CC$ will be restricted to a standard Riemann surface~$\Sigma$, and only two of the four arbitrary locally holomorphic functions will remain mutually independent while the other two will be related to the former by complex conjugation. In this way, the real form solutions to Type~IIB$_\RR$, constructed previously in  \cite{DHoker:2016ujz,Corbino:2017tfl}, will be recovered as expected. Upon restricting to a real form of Type~IIB, the $SL(2,\CC)$ global symmetry of Type~IIB$_\CC$ will restrict to a real form $SL(2,\RR)$. 

\sm

The classification of the real forms of the solutions yields a rich set of combinations of symmetric spaces, signatures, and Type IIB real forms, listed in Tables \ref{eq:real-forms-reduced-BPS}, \ref{4.table.IIB5}, \ref{4.table.IIB79}. 
These tables are consistent with the real forms of $F(4;\CC)$, whose bosonic subalgebras are,
\bea\label{listmbsa}
\ms \mo (3;\RR) ~ & \oplus & \ms \mo (1,6;\RR)
\no \\
\ms \mo (3;\RR) ~ &  \oplus & \ms \mo (2,5;\RR)
\no \\
\ms \mo (1,2;\RR) &  \oplus & ~ \ms \mo (7;\RR)
\no \\
 \ms \mo (1,2;\RR) &  \oplus & \ms \mo (3,4;\RR)
\eea
We note that, according to the classification given in \cite{Nahm:1977tg,Parker:1980af} these are the only real forms of $F(4;\CC)$. To clarify definitively some statements in the literature regarding these real forms\footnote{
In \cite{Kac} the existence of four real forms of $F(4)$ whose maximal bosonic subalgebras correspond to the left  terms in the direct sums of (\ref{listmbsa}) is proven but the right terms in (\ref{listmbsa}) are not provided. In \cite{Sorba}, the right terms of the second and third lines are listed incorrectly as $\ms \ml (2;\RR) \sim \ms \mo (1,2)$. The correct list of real forms is given in full in \cite{Nahm:1977tg,Parker:1980af}. \label{foot:2}}
we derive the classification by elementary methods in appendix \ref{sec:F}.  Two of the cases in Tables \ref{eq:real-forms-reduced-BPS}, \ref{4.table.IIB5}, \ref{4.table.IIB79}, corresponding to $AdS_6\times S^2$ and $S^6\times AdS_2$ warped over a Riemann surface, were solved previously in \cite{DHoker:2016ujz,Corbino:2017tfl}; all others are new. Notably missing from the real forms is  $S^6\times S^2\times\Sigma$ with bosonic symmetry $ \ms\mo (7;\RR) \oplus \ms\mo (3;\RR) $; this is in line with expectations and will be discussed in connection with IKKT holography in section \ref{sec:7}. 

\sm

Restricting the complex solutions to the aforementioned real forms imposes relations between the pairs of holomorphic functions and complex variables, and produces the general local solutions for the entire list of real forms in Tables \ref{eq:real-forms-reduced-BPS}, \ref{4.table.IIB5}, \ref{4.table.IIB79}.  The realization of the real forms in Table~\ref{eq:real-forms-reduced-BPS} in terms of the holomorphic data is summarized in Table~\ref{eq:real-forms-spinor-rel}.

\subsection*{Acknowledgments}
CFU thanks Nikolay Bobev, Sean Hartnoll and Shota Komatsu for related discussions. The research of ED and MG is supported in part by NSF grant PHY-22-09700.  MG~is grateful to the Centro de Ciencias de Benasque Pedro Pascual and the organizers of the ``Gravity - New quantum and string perspectives" workshop for their hospitality while this paper was being finalized.

\clearpage

%%%%%%%%%%%%%%%%%%%%%%%%%%%%%%%%%%%%%%%%%%%
%%%%%%%%%%%%%%%%%%%%%%%%%%%%%%%%%%%%%%%%%%%
\section{Complexified  Type IIB and its real forms}
\setcounter{equation}{0}
\label{sec:2}
%%%%%%%%%%%%%%%%%%%%%%%%%%%%%%%%%%%%%%%%%%%
%%%%%%%%%%%%%%%%%%%%%%%%%%%%%%%%%%%%%%%%%%%

 In this section we review the BPS equations for Type IIB$_\CC$ as well as the complex conjugation involutions that lead to the various real forms  which include standard Type IIB, the real forms Type IIB$^*$, Type IIB$^\prime$, and Type IIB$_3$ classified in \cite{Bergshoeff:2007cg}.

\subsection{Complexified Type IIB}

As previewed in the Introduction, the starting point for the complexification is a formulation of standard Type~IIB supergravity in terms of real bosonic fields and Majorana-Weyl spinor fermions on a spacetime with real local coordinates. The bosonic fields of standard Type~IIB consist of the metric $g$ (or equivalently the frame $e$), the dilaton~$\phi$, the axion~$\chi$, the two-forms $B_{(2)}$ and $C_{(2)}$ and the four-form $C_{(4)}$. In string theory, the fields $\chi, C_{(2)}$ and $C_{(4)}$ arise from the R-R sector, while all others arise from the NS-NS sector. The fermion fields comprise the gravitino $\psi$ and the dilatino $\lambda$, each of which consists of a pair of Majorana-Weyl spinors. Similarly, the local supersymmetry transformation parameter~$\ep$ of supergravity also consists of a pair of Majorana-Weyl spinors.
  
\sm

The complexification procedure consists in promoting each real bosonic field to a complex-valued field and each Majorana-Weyl spinor field to a (complex) Weyl spinor field, dependent holomorphically on the local coordinates of complexified spacetime.\footnote{When no confusion is expected to arise, and in order to avoid a proliferation of notations, we shall use the same symbol for the complexified fields as was used for the real fields.} For example, the real-valued dilaton $\phi$ of standard Type IIB supergravity will be promoted to a complex-valued field $\phi$ and the pair of dilatino Majorana-Weyl spinors $(\lambda_+, \lambda_-)$ will be promoted to a pair of (complex) Weyl spinors $(\lambda_+, \lambda_-)$ in Type IIB$_\CC$.  By construction in this process, the complex conjugates of the complexified fields never enter the field equations, the action, or the BPS equations. For example, the complex conjugate $\phi^*$ of the complexified dilaton field never enters. Thus, the dependence on the fields is often referred to as being ``holomorphic".

\sm

For a complex-valued metric there is no meaningful notion of signature, and with a  complex frame the signature of the frame metric can be chosen at will. Here, it will be convenient to retain the signature $(- + \cdots +)$ of standard Type IIB supergravity so that we can use the basis of Dirac matrices adopted in \cite{DHoker:2016ujz,Corbino:2017tfl}. 
Throughout, we shall assume that the complexified metric is non-degenerate, though we shall not spell out the precise conditions for this property (see for example \cite{Witten:2021nzp} for additional discussion of this issue).

\sm

Since the goal of this paper is to construct classical solutions invariant under 16 supersymmetries, in the presence of vanishing fermion fields, we may limit attention to the BPS equations obtained by setting the  supersymmetry variations of the dilatino and gravitino fields to zero at vanishing fermion fields. The resulting complexified BPS equations are given as follows, 
\bea
\label{1.BPS.1}
0 & = &
\nabla  _M  \ep 
- { 1 \over 8} e^{ \phi}  (\Gamma \cdot \p \chi) \Gamma _M s^2  \ep 
- {1 \over 480} e^{ \phi}  (\G \cdot  F_{(5)})  \Gamma_M s^2 \ep
\no \\ &&
-{1 \over 8} H_{(3) MNP} \Gamma^{NP} s^3 \ep
-{1 \over 48}  e^{ \phi} ( \G \cdot \tilde F_{(3)}  ) \G_M  s^1  \ep
\no \\
0 & = & (\G \cdot \p \phi) \ep 
+ e^{ \phi } (\G \cdot \p \chi) s^2 \ep 
- {1\over 12 }  ( \G \cdot H_{(3)}) s^3 \ep 
+ {1 \over 12} e^{ \phi} (\G \cdot \tilde F_{(3)} ) s^1 \ep 
\eea
Here, $M,N,P$ are Einstein indices; $\ep$ has definite chirality and satisfies $\Gamma ^{11} \ep = \ep$; and the dot product of an $n$-form $S$ with $\Gamma$ is defined by $(\Gamma \cdot S) = \Gamma ^{M_1 \cdots M_n} S_{M_1 \cdots M_n}$. The field strengths are given in terms of the potentials by solving the Bianchi identities,
\begin{align}
\label{2.doublets}
H_{(3)} &= d B_{(2)} & \tilde F_{(3)} & = F_{(3)} - \chi H_{(3)}
\no \\
F_{(3)} & = d C_{(2)} & F_{(5)} & = dC_{(4)} +{1 \over 8} \Big ( B_{(2)} \wedge d C_{(2)} - C_{(2)} \wedge dB_{(2)} \Big )
\end{align}
and the parameter $\ep$ stands for a pair of complex Weyl spinors in the following basis,\footnote{Our notations are related to those of \cite{Bergshoeff:2007cg} by setting $\cP = - I_{32} \otimes s^3$; $\cP_{1/2}=\cP_{5/2}=I_{32} \otimes s^2 $ and $\cP_{3/2} = I_{32} \otimes s^1$ where $I_{32}$ is the identity matrix in the Dirac spinor space.}
\bea
\ep = \left ( \bma \ep_+ \cr \ep_- \ema \right )
\hskip 0.5in 
s ^1 = \left ( \bma 0 & 1 \cr 1 & 0 \ema \right )
\hskip 0.5in
s ^2 = \left ( \bma 0 & 1 \cr -1 & 0 \ema \right )
\hskip 0.5in
s ^3 = \left ( \bma 1 & 0 \cr 0 & -1 \ema \right )
\eea
The derivation of the BPS equations for Type~IIB$_\CC$ by complexification of the BPS equations of Type~IIB$_\RR$ is reviewed in appendix~\ref{sec:B}.

\subsection{The real forms of Type IIB$_\CC$}
\label{sec:2.2}

In the process of constructing a real form of Type IIB$_\CC$, we restrict the coordinates of complexified spacetime to real coordinates.    We shall require that any real form of Type~IIB$_\CC$ has real-valued metric and dilaton fields, 
\bea
\label{2.gphi}
g_{MN} ^* = g_{MN} 
\hskip 1in 
\phi^* = \phi
\eea
Recall that the BPS equations of (\ref{1.BPS.1}) for Type~IIB$_\CC$ and any one of its real forms  are formulated in terms of a fixed basis of Dirac matrices with signature $(1,9)$.\footnote{For a detailed discussion of the transition to an equivalent formulation of the real forms of Type~II$_\CC$ using Dirac matrices with signature matching that of spacetime we refer to appendix B and C of \cite{Bergshoeff:2007cg}.}  Their explicit form is spelled out in appendix \ref{sec:A}, including  their complex conjugation property,
\bea
\label{2.Gamma}
(\Gamma _A)^* = \cB \, \Gamma _A \, \cB^{-1} 
\hskip 1in 
\cB =   \cC \Gamma ^0
\eea
where $A$ are frame indices, $\cB$ is the complex conjugation matrix, and $\cC$ the charge conjugation matrix. A metric $g_{MN}$ of arbitrary signature with $t$ time-like directions may be obtained by allowing the frame one-form $e^A$ to be either real-valued or purely imaginary, 
\bea
\label{2.frame}
(e^A)^* = \eta_{AA} \, e{}^A
\eea
where $\eta$ is a flat metric with $t$ time-like directions, whose sole non-vanishing entries are the diagonal ones with $\eta_{AA}= \pm 1$. The complex conjugate of the associated product $e \cdot \Gamma = e^A \Gamma_A$ is then given by,
\bea
\label{2.B.1}
(e^A \Gamma _A)^* = \cB_\eta \, ( e^A \Gamma _A) \, \cB_\eta^{-1}
\eea
where $\cB_\eta$ is the product of the charge conjugation matrix $\cC$ with the product of all the Dirac matrices that correspond to time-like directions. 
For a metric with $t$ time-like directions, which we may collect in a set $\cT$, the corresponding $\cB_\eta$ is given by,
\bea
\label{2.B.2}
\cB_\eta = u \, \cC \, \prod_{i\in\cT} \Gamma ^{i}
\hskip 1in
\cB_\eta ^* \, \cB_\eta = (-)^{\half t(t-1)} I_{32}
\eea
where $u$ is an arbitrary phase factor, namely satisfying $|u|=1$, that may be chosen for convenience.  The real forms of Type~IIB$_\CC$ with frame metric $\eta$ are obtained by imposing the following reality conditions on the spinors $\ep, \lambda, \psi$, 
\bea
\label{S-define}
\begin{array}{c} 
	\cB^{-1}_\eta \ep ^* = S \, \ep \cr  
	\cB^{-1}_\eta  \lambda ^* = S \,  \lambda \cr 
	\cB^{-1}_\eta \psi ^* = S \, \psi \end{array}
\hskip 0.9in
S =\left\{\begin{array}{rl}
	I_2 ~~ \hbox{ for }  & \text{Type~IIB$_\RR$} \\ 
	s^1 ~~ \hbox{ for }  & \text{Type~IIB$'$} \\ 
	s^2 ~~ \hbox{ for }  & \text{Type~IIB$_3$} \\ 
	s^3 ~~ \hbox{ for }  & \text{Type~IIB$^\star$}
\end{array}\right.
\eea
Compatibility of these reality conditions with the chirality condition eliminates metrics $\eta$ with  even values of $t$, while the requirement that complex conjugation be a proper involution relates the value of $t$ to the allowed matrices $S$,
\begin{align}
\label{2.parity}
t & \equiv 1 \, ({\rm mod} \, 4) & S & =I_2, s^1, s^3
\no \\
t & \equiv 3 \, ({\rm mod} \, 4) & S & =s^2 
\end{align}
The corresponding complex conjugation properties of the boson fields, other than the metric and the dilaton which were already given in (\ref{2.gphi}),  are obtained as follows. The complex conjugate equations of (\ref{1.BPS.1}) are multiplied to the left by $\cB^{-1}_\eta$ and the complex conjugates $(\Gamma_M)^*$ are recast in terms of $\Gamma_M$ and $\cB_\eta$ using (\ref{2.Gamma}) so that every term acts on the combination $\cB^{-1}_\eta  \ep^*$. A complex conjugation involution exists provided these equations are equivalent to the original BPS equations (\ref{1.BPS.1}). This comparison uniquely determines the complex conjugation properties of the bosonic fields, either as expressed with respect to Einstein indices, or expressed as forms.\footnote{When the fields are expressed with respect to frame indices additional sign changes result from the complex conjugation relations of the frame in (\ref{2.frame}).}

\sm
 
The allowed real forms of Type~IIB$_\CC$ supergravity are then accompanied by the following complex conjugation relations on the bosonic fields:
\begin{description}
\item {\bf\boldmath Type IIB$_\RR$, IIB$_5$, IIB$_9$:} In this family of theories the spinors satisfy the reality conditions of standard Type IIB supergravity with $S=I_2$, and the bosonic fields $\chi$, $B_{(2)}$, $C_{(2)}$, $C_{(4)}$ are real-valued. In addition to Type IIB$_\RR$ with $t=1$ it contains theories with 5 and 9 timelike directions which we denote as IIB$_5$ and IIB$_9$, respectively.

\item {\bf\boldmath Type IIB$^\prime$, IIB$^\prime_5$, IIB$^\prime_9$:} In this family of theories the spinor reality conditions are twisted by $S=s^1$.
The field $C_{(2)}$ is real-valued while $\chi, B_{(2)}, C_{(4)}$ are purely imaginary.
We denote the $t=1$ variant by IIB$^\prime$ and variants with $t=5,9$ by subscripts.

\item {\bf\boldmath Type IIB$^\star$, IIB$^\star_5$, IIB$^\star_9$:} Here the spinor reality conditions are twisted by $S=s^3$.
The field $B_{(2)}$ is real-valued while $\chi, C_{(2)}, C_{(4)}$ are purely imaginary.
We denote the $t=1$ variant by IIB$^\star$ and distinguish variants with $t=5,9$ by subscripts.

\item {\bf\boldmath Type IIB$_3$, IIB$_7$:} This family corresponds to $S=s^2$.
The fields $\chi$, $C_{(4)}$ are real-valued while $B_{(2)}$, $C_{(2)}$ are purely imaginary. There are two variants with $t=3,7$, which are again denoted by subscripts.
\end{description}
These are precisely the theories given in \cite[Table 1]{Bergshoeff:2007cg}.
We note that Type IIB$^\prime$ may be related to Type IIB$^\star$ by combining the reality conditions for Type IIB$^\star$ with an $SL(2)$ transformation within Type IIB$_\CC$ which swaps $B_{(2)}$ and $C_{(2)}$. The usual action of $SL(2,\RR)$ in Type IIB, extended to Type IIB$_\CC$, is not compatible with the reality conditions in Type IIB$^\star$; hence this yields a different theory.
$SL(2,\RR)$ transformations within Type IIB* were discussed e.g.\ in \cite{Hartnoll:2024csr}. For further discussion of the action of $SL(2)$, see section \ref{sec:3.9}.

\newpage

%%%%%%%%%%%%%%%%%%%%%%%%%%%%%%%%%%%%%%%%%%%
%%%%%%%%%%%%%%%%%%%%%%%%%%%%%%%%%%%%%%%%%%%
\section{Solutions to Type IIB$_\CC$ with $F(4;\CC)$ symmetry}
\setcounter{equation}{0}
\label{sec:3}
%%%%%%%%%%%%%%%%%%%%%%%%%%%%%%%%%%%%%%%%%%%
%%%%%%%%%%%%%%%%%%%%%%%%%%%%%%%%%%%%%%%%%%%

To construct the general bosonic solutions to the complexified Type IIB$_\CC$ supergravity theory which are invariant under the complexified Lie superalgebra $F(4;\CC)$, we shall solve the BPS equations for Type IIB$_\CC$ for vanishing fermion fields, given in (\ref{1.BPS.1}). 

\sm

To do so, we begin by implementing the restrictions on the bosonic fields and the supersymmetry transformation spinor imposed by invariance under the maximal bosonic subalgebra $\ms \mo (7;\CC) \oplus \ms \mo (3;\CC)$ of $F(4;\CC)$. As a result, the spacetime geometry must be of the form $\cM_{6  \CC}  \times \cM_{2  \CC}$ warped over a complexified two-dimensional  surface $\Sigma_\CC$ where $\cM_{6  \CC}$ and $\cM_{2  \CC}$ are spaces of complex dimensions 6 and 2   with complex-valued  metrics that are invariant under  $\ms \mo (7;\CC) $ and $ \ms \mo (3;\CC)$, respectively.

\subsection{Ansatz for the bosonic fields}

The complex-valued metric of Type~IIB$_\CC$ takes the following form, 
\bea
\label{3.a.1}
ds^2 = f_6^2 \, ds^2_{\cM_{6  \CC}} + f_2^2 \, ds^2_{\cM_{2  \CC}} + ds^2_{\Sigma_\CC}
\eea
where the frames and associated reduced metrics are related as follows,
\begin{align}
\label{3.a.2}
e^m&=f_6 \, \hat e^m & ds^2_{\cM_{6  \CC}} &= \eta _{mn} \, \hat e_m \hat e_n & m,n&=0,..,5
	\nonumber\\
e^i&=f_2\, \hat e^i & ds^2_{\cM_{2  \CC}} &= \delta_{ij} \, \hat e^i \hat e^j  &  i,j&=6,7
	\nonumber\\
e^a& & ds^2 _{\Sigma_\CC} &= \delta_{ab} \, e^a e^b  & a,b&=8,9
\end{align}
Here,  $\eta$ is the frame metric with signature $(-+++++)$ while  $A=(m,i,a)$ stand for frame indices. The reduced 1-form frames  $\hat e^m$ and $\hat e^i$ are complex-valued and independent of $\Sigma_\CC$;  they will be given below.  The flux fields $ F_{(5)}, \tilde F_{(3)}$ and $H_{(3)}$ are given as follows,
\bea
\label{3.a.3}
F_{(5)} =0 
\hskip 0.9in 
\tilde F_{(3)} = g_a \, e^a \wedge e^{67} 
\hskip 0.9in
H_{(3)} = h_a e^a \wedge e^{67} 
\eea
where $e^{67} = e^6 \wedge e^7$. These fields may also be seen as resulting from the Ansatz for the flux field potentials $B_{(2)} = b_2 \, \hat e^{67} $ and $C_{(2)} = c_2 \, \hat e^{67} $, where $b_2$ and $c_2$ are scalar functions. 
It will often be convenient to use the following notation for the differentials of $\phi$ and $\chi$, 
\bea
\label{3.a.4}
d \phi = \f_a \, e^a \hskip 1in d\chi = \chi_a \, e^a
\eea
The functions $f_2, f_6, g_a, h_a, b_2, c_2, \f_a, \chi_a$ and the axion $\chi$ and dilaton $\phi$ fields are all holomorphic functions on $\Sigma_\CC$ so that the geometry is warped over $\Sigma_\CC$. 

\sm

We close this subsection by giving a construction of the reduced complex frames $\hat e^m$ and $\hat e^i$ for 
$\cM_{6 \CC}$ and $\cM_{2 \CC}$, respectively, as well as of the canonical connections $\hat \om^m{}_n$ and $\hat \om^i{}_j$ that will be of use later. Consider the complex Lie group $SO(n+1;\CC)$, its subgroup $SO(n;\CC)$ embedded in the defining representation, and the quotient $\cM_{n  \CC} = SO(n+1;\CC)/SO(n;\CC)$. The space $\cM_{n \CC}$ is the complex sphere which may be embedded into $\CC^{n+1}$ as follows,
\bea
\label{eq:quad}
\cM_{n \CC}  = \big \{ (z_1, \cdots, z_{n+1}) \in \CC^{n+1} \hbox{ such that } z_1^2+ \cdots + z_{n+1}^2=1 \big \}
\eea 
with metric $ds^2_{\mathcal M_{n\CC}} = k^{2}(dz_1^2 + \cdots + dz_{n+1}^2)$ where $k$ determines the curvature. In the sequel, it will be convenient to introduce a diagonal metric $\eta$ with non-zero entries $\pm 1$ and define $\cM_{n\CC}$ as the quadric $z^p\eta_{pq}z^q=1$ in $\CC^{n+1}$ with metric $ds^2_{\mathcal M_{n\CC}}=k^{2}dz^p\eta_{pq}dz^q$. This will come in handy when restricting $\cM_{n \CC}$  to its real forms  $\cM_n$ in sections \ref{sec:5} and \ref{sec:6}.

\sm

The Lie algebra decomposes as follows,
\bea
\ms \mo (n+1;\CC) = \ms \mo (n;\CC) \oplus \mM_{n\CC}
\eea
Adopting a flat metric $\eta$ of arbitrary signature,  the generators of $\ms \mo (n+1;\CC)$ consist of the generators $t_{AB} =- t_{BA}  $ of $\ms \mo (n;\CC)$  and $u_A $ of $\mM_{n\CC}$ for $A,B=1, \cdots, n$ and obey the following structure relations,
\bea
{} [ t_{AB}, t_{CD} ] & = & \eta_{AD} \, t_{BC} + \eta _{BC} \, t_{AD} - \eta_{AC} \, t_{BD} - \eta _{BD} \, t_{AC}
\no \\
{} [ t_{AB} , u_C] & = & \eta_{BC} \, u_A - \eta _{AC} \, u_B
\no \\
{} [ u_A, u_B] & = & -k^{-2} t_{AB}
\eea  
The Maurer-Cartan form for $SO(n+1;\CC)$,
\bea
\Omega=  \thalf \, \hat \omega ^{AB} \, t_{AB} +   \hat e ^A \, u_A
\eea 
decomposes into the canonical connection $\hat \om^{AB}$ and the canonical frame $\hat e^A$ of $\cM_{n\CC}$,  while the Maurer-Cartan equation $d  \Omega + \Omega \wedge \Omega=0$ decomposes into the relations for vanishing torsion and constant curvature of $\hat e^A$ and $\hat \om^{AB}$, respectively given by, 
\bea
\label{3.torcur}
d \hat e^A + \eta_{BC} \, \hat \om^{AB} \wedge \hat e^C & = & 0
\no \\
d \hat \om^{AB} + \eta_{CD} \, \hat \om^{AC} \wedge \hat \om^{DB} & = &  k^{-2}\hat e^A \wedge \hat e^B
\eea
Specializing to the cases $n=6$ and $n=2$, and setting $A=m$ and $A=i$, provides the canonical frames $\hat e^m$ and $\hat e^i$ for $\cM_{6\CC}$ and $\cM_{2\CC}$, respectively, and also gives the canonical connections $\hat \om$ that will be used in the next subsection.

\subsection{Killing spinors}
\label{sec:Killingk1k2}

Flatness of the $\ms \mo (n+1;\CC)$ Maurer-Cartan form $\Omega$ implies that the equation $(d + \Omega ) \chi=0$ is integrable. Considering the connection $\Omega$  in a spinor representation of $\ms \mo (n+1;\CC)$  gives the Killing spinor equation on the space $\cM_{n\CC}$. The Killing spinor equation may be integrated locally by solving the Maurer-Cartan equation locally in terms of a gauge transformation $U: \cM_{n\CC} \to SO (n+1;\CC)$ such that $\Omega = U^{-1} dU$ and $\chi = U^{-1} \chi_0$ where $\chi_0$ is a spinor that is constant on $\cM_{n\CC}$. This explicit solution will not  be needed here.

\sm

The Killing spinor equations on the spaces $\cM_{6\CC}$ and $\cM_{2\CC}$ may be expressed as follows,
\bea
\left ( d + \frac{1}{4} \hat \om^{mn} \, \gamma _{mn} 
+ {k_1 \over 2} \, \hat e^m \, \gamma _m \, \gamma _{(1)} \right ) \chi_6 & = & 0
\no \\
\left ( d + \frac{1}{4} \hat \om^{ij} \, \gamma _{ij} 
+ {k_2 \over 2} \, \hat e^i \gamma _i \, \gamma _{(2)} \right ) \chi_2 & = & 0
\eea
where $\gamma_{mn} = \half [\gamma_m, \gamma_n]$ and $\gamma _{ij} = \half [\gamma _i, \gamma_j]$ in the basis of Dirac matrices spelled out in appendix \ref{sec:A} where the definitions of the chirality matrices $\gamma_{(1)}$ and $\gamma _{(2)}$ are also given. The spaces  $\cM_{6\CC}$ and $\cM_{2\CC}$ have constant curvature  $-k_1^2$ and $-k^2_2$, respectively. 

\sm

The dimensions of the spaces of Killing spinors are 8 for $\chi_6$ and 2 for $\chi_2$.  To construct a basis for the Killing spinors of $\cM_{6\CC} \times \cM_{2\CC}$, we take the tensor product of the two spinors and label the basis spinors by the eigenvalues $\eta_1= \pm 1$ of $\gamma_{(1)}$ and $\eta_2= \pm 1$ of $\gamma_{(2)}$, leaving a four-fold degeneracy for $\chi_6$ (see \cite{DHoker:2016ujz} for a more detailed discussion). Finally, it will be useful to express the Killing spinor equations in terms of covariant derivatives, in frame basis where $d  + \frac{1}{4} \hat \om^{mn} \, \gamma _{mn} = \hat e^m \, \hat \nabla _m$ and $d + \frac{1}{4} \hat \om^{ij} \, \gamma _{ij} = \hat e^i \, \hat \nabla _i$, 
\begin{align}
\label{3.b.1}
\left ( \hat \nabla _m - \frac{k_1}{2} \eta _1 \, \gamma _m \otimes I_2 \right ) \chi ^{\eta _1, \eta _2} & =  0
	\nonumber \\
\left ( \hat \nabla _i - \frac{k_2}{2} \eta _2 \, I_8 \otimes \gamma _i \right ) \chi ^{\eta _1, \eta _2}  & =  0	
\end{align}
Decomposing the doublet of supersymmetry transformation spinors $\ep$  in the above basis, 
\begin{align}
\label{3.b.2}
\ep & = \sum_{\eta_1,\eta_2=\pm}\chi^{\eta_1,\eta_2} \otimes \zeta_{\eta_1,\eta_2}
&
\ep & = \left ( \bma \ep_+ \cr \ep_- \ema \right )
&
\zeta_{\eta_1,\eta_2} & = \left ( \bma \zeta_{+, \eta_1,\eta_2} \cr  \zeta_{-, \eta_1,\eta_2} \ema \right )
\end{align}
the coefficients $\zeta_{+, \eta_1,\eta_2}$ and $ \zeta_{-, \eta_1,\eta_2}$ are themselves 2-component spinors. The doublets $\zeta_{\eta_1,\eta_2}$ are acted upon by the matrices $s^1, s^2, s^3$ of (\ref{2.doublets}). 
When explicitly referring to components we use $\zeta_{s,\eta_1,\eta_2,\sigma}$, where $s=\pm$ refers to the doublet components, $\eta_1,\eta_2=\pm 1$ to the basis on $\cM_{6\CC}$, $\cM_{2\CC}$, and $\sigma$ to the 2-spinor components.
With the $\ms \mo (7;\CC) \oplus \ms \mo(3;\CC)$ invariant Ansatz for the bosonic fields and  the spinor doublet $\ep$, we can now reduce the BPS equations (\ref{1.BPS.1}).

\subsection{The reduced BPS equations}
\label{sec:3.3}

In this subsection, we present the BPS equations of (\ref{1.BPS.1}) reduced to the $\ms \mo (7;\CC) \oplus \ms \mo(3;\CC)$ invariant Ans\"atze for the boson fields in (\ref{3.a.1}) to (\ref{3.a.4}) and  the  spinor doublet $\ep$ in (\ref{3.b.2}). We relegate the details of their derivation to appendix \ref{sec:C}. To streamline the presentation,  we shall use the $\tau$-matrix notation introduced in \cite{Gomis:2006cu} and used subsequently in \cite{DHoker:2016ujz}  to express operations on the $\eta_1,\eta_2$ indices. Defining $\tau^{(ij)} = \tau^i \otimes \tau^j$ where $\tau^0$ stands for the identity matrix and $\tau^i=\sigma^i$ for the Pauli matrices, the action on $\zeta$ is as follows,
\begin{align}
\label{3.d.1}
(\tau^{(ij)} \zeta)_{\eta_1,\eta_2} = 
\sum_{\eta_1',\eta_2' = \pm } (\tau^i)_{\eta_1 \eta_1'} \, (\tau^j)_{\eta_2 \eta_2'} \, \zeta_{\eta_1' \eta_2'}
\end{align}
The chirality condition $\big ( \Gamma_{11} \otimes I_2 \big ) \ep =- \ep$ with $\Gamma ^{11}$ given in (\ref{eq:Gamma-star}), reduces to,  
\bea
\label{3.chiral}
\big ( I_2 \otimes \tau^{(11)} \otimes \gamma _{(3)} \big ) \zeta = \zeta 
\hskip 1in
\big ( I_2 \otimes  \gamma_{(3)}  \big )  \zeta_{-\eta_1,-\eta_2} =  \zeta_{\eta_1,\eta_2}
\eea
where the first factor $I_2$  is the identity in doublet space.

\sm

For the gravitino BPS equation, we shall also use the  relations between the covariant derivatives $\nabla$ and $\hat \nabla$ obtained in appendix C of  \cite{DHoker:2016ujz}, 
\bea
\label{3.d.2}
\nabla_m \, \ep = \hat \nabla_m \, \ep + \frac{D_af_6}{2f_6} \, \Gamma_m\Gamma^a \ep
\hskip 1in
\nabla_i \, \ep = \hat \nabla_i \, \ep + \frac{D_af_2}{2f_2} \, \Gamma_i\Gamma^a \ep
\eea
Eliminating the covariant derivatives $\nabla_m \ep $ and $\nabla _i \ep$ in (\ref{1.BPS.1}) in favor of $\hat \nabla_m \ep $ and $\hat \nabla _i \ep $, respectively,  using (\ref{3.d.2}) and then eliminating the latter with the help of the Killing spinor equations of (\ref{3.b.1}), we obtain the following reduced BPS equations, 
\begin{align}
\label{3.c.1}
(m)&& 
	0&=- \frac{i k_1}{2f_6}\tau^{(20)}\zeta
	+\left(\frac{D_af_6}{2f_6}
	-\frac{1}{8}e^\phi\chi_a s^2 \right) \tau^{(01)} \gamma^a \zeta
	-\frac{i}{8}e^\phi g_a  s^1  \gamma^a \zeta
\no \\
(i)&&
	0&=
	\frac{k_2}{2f_2}\tau^{(03)}\zeta
	+\left(\frac{D_a f_2}{2f_2}
	-\frac{1}{8}e^\phi\chi_a s^2 \right) \tau^{(01)} \gamma^a \zeta
	+\left(- \frac{i}{4}h_a s^3 +\frac{i}{8}e^\phi g_a  s^1 \right)  \gamma^a \zeta
\no \\
(a)&&
	0&=\left(D_a+\frac{i}{2}\hat \omega_a \gamma _{(3)} \right)\zeta
	+\frac{1}{8}e^\phi\chi_b s^2 \gamma^b\gamma_a  \zeta
	+\left(- \frac{i}{4}h_a s^3
	+\frac{i}{8}e^\phi g_b \, s^1 \gamma^b\gamma_a  \right)\tau^{(01)}\zeta	
\no \\
(\lambda) &&	0&=\big ( \varphi_a-e^\phi\chi_a s^2 \big ) \tau^{(01)} \gamma^a \zeta 
	- \frac{i}{2} \big ( h_a s^3 + e^\phi g_a s^1 \big )  \gamma^a\zeta
\end{align}
where factors of $\tau^{(00)}$ have been suppressed as they equal the identity.

\subsection{Discrete symmetries and projection}

All the discrete symmetries of the reduced BPS equations (\ref{3.c.1}) are obtained as follows. 
\begin{itemize}
\itemsep =0in 
\item 
Invariance of the $(a)$ and $(\lambda)$ equations is achieved as follows. When $[\tau^{(01)}, T]=0$ the transformations must be of the form  $I_2 \otimes T \otimes I_2$ or $I_2 \otimes T \otimes \gamma _{(3)}$, while when $\{\tau^{(01)}, T\}=0$ the transformations must be of the form  $s^2 \otimes T \otimes I_2$ or $s^2 \otimes T \otimes \gamma _{(3)}$.
\item 
Combining the above result with invariance of the $(m)$ and $(i)$ equations imposes further conditions on the allowed matrices $T$ depending on whether $T$ commutes or anti-commutes with $\tau^{(20)}$ and $\tau^{(03)}$. 
\end{itemize}
The result is a discrete symmetry group with 8 real elements, which is generated by the identity along with the following three elements, 
\bea
\label{3.group1}
&& I_2 \otimes \tau^{(11)} \otimes \gamma _{(3)} \hskip 0.6in \gamma _{(3)} = \sigma ^3
\no \\
&& i \, s^2 \otimes \tau^{(32)} \otimes I_2
\no \\
&& i \, I_2 \otimes \tau^{(20)} \otimes I_2
\eea
Clearly, the last element anti-commutes with the first two, so that the group is non-Abelian. Its maximal Abelian subgroup has 4 elements and is isomorphic to the group consisting of the identity and the following three elements,
\bea
\label{3.group2}
\cI & = & I_2 \otimes \tau^{(11)} \otimes \gamma _{(3)}
\no \\
\cJ & = & i \, s^2 \otimes \tau^{(32)} \otimes I_2
\no \\
\cI \cJ & = & i \, s^2 \otimes \tau^{(23)} \otimes\gamma_{(3)}
\eea
We recognize $\cI$ as the chirality matrix acting on the spinor doublet $\zeta$ given in (\ref{3.chiral}), and the second as the projector by the same name defined in \cite{DHoker:2016ujz}.

\sm

As was shown in \cite{DHoker:2016ujz}, restrictions on $\zeta$ by linearly acting and mutually commuting discrete symmetries solves the BPS equations in different sectors. Thus we impose the chirality constraint and diagonalize the generator $\cJ$ by imposing the conditions,
\bea
\label{3.IJ}
\big ( I_2 \otimes \tau^{(11)} \otimes \gamma _{(3)} \big ) \zeta & = & \zeta
\no \\
\big ( i \, s^2 \otimes \tau^{(32)} \otimes I_2 \big ) \zeta & = & \nu \zeta
\eea
where $\nu$ can take the values $\nu = \pm 1$. In components, these conditions become,
\bea
\big ( I_2 \otimes \tau^{(11)} \otimes \gamma_{(3)} \big ) \zeta _{s, \eta_1, \eta_2, \sigma} & = & 
\sigma \, \zeta_{s , - \eta_1, - \eta_2, \sigma} = \zeta _{s , \eta_1, \eta_2, \sigma}
\no \\
\big ( i\, s^2 \otimes \tau^{(32)} \otimes I_2 \big ) \zeta _{s, \eta_1, \eta_2, \sigma} & = & 
s \, \eta_1 \, \eta_2 \, \zeta _{- s, \eta_1, - \eta_2, \sigma} = \nu \, \zeta_{s, \eta_1, \eta_2, \sigma}
\eea
We parametrize the independent components by complex $\alpha$, $\beta$, $\gamma$, $\delta$ as follows,
\begin{align}
\label{eq:zeta-comp}
\alpha &=\zeta_{++++}=+\zeta_{+--+}=+\nu\zeta_{-+-+}=+\nu\zeta_{--++}
	\notag\\
\beta &=\zeta_{+---}=-\zeta_{+++-}=-\nu\zeta_{-+--}=+\nu\zeta_{--+-}
	\notag\\
\gamma &=\zeta_{-+++}=+\zeta_{---+}=-\nu \zeta_{++-+}=-\nu\zeta_{+-++}
	\notag\\
\delta &=\zeta_{----}=-\zeta_{-++-}=+\nu\zeta_{++--}=-\nu\zeta_{+-+-}
\end{align}
Finally, we shall express the reduced BPS equations in a convenient  basis for the frame components  $e^a$ with $a=8,9$, given as follows,
\begin{align}
\label{eq:e-z-def}
	e^z&=\tfrac{1}{2}\left(e^8+ie^9\right)
	&
	e^{\tilde z}&=\tfrac{1}{2}\left(e^8-ie^9\right)
	&
	\delta_{z\tilde z}&=2  & \delta^{z\tilde z}&=\tfrac{1}{2}
\end{align}
where we stress  that $e^z$ and $e^{\tilde z}$ are independent and not related to each other by complex conjugation.
The associated Dirac matrices may be chosen as follows,
\bea
\gamma ^z = \left ( \bma 0 & 1 \cr 0 & 0 \ema \right ) = \thalf \gamma _\tz 
\hskip 1in 
\gamma ^\tz = \left ( \bma 0 & 0 \cr 1 & 0 \ema \right ) = \thalf \gamma _z
\eea

\subsection{Reduced BPS equations on the restricted spinor space}

The dilatino BPS equation $(\lambda)$ in (\ref{3.c.1}) yields the following component relations,
\begin{align}
\label{2.BPS.dil}
	0 &= \varphi_z \beta  -  e^\phi\chi_z \delta +\frac{i\nu}{2}\left( h_z \delta - e^\phi g_z \beta \right)
\no \\	
	0 &=\varphi_z \delta + e^\phi \chi_z \beta  +\frac{i\nu}{2}\left( h_{z} \beta + e^\phi g_z \delta \right)	
\no \\
	0&= \varphi_\tz \alpha-  e^\phi\chi_\tz \gamma
	+\frac{i\nu}{2} \left( h_\tz \gamma - e^\phi g_\tz \alpha \right)
\no \\
	0&= \varphi_\tz \gamma+e^\phi\chi_\tz \alpha +\frac{i\nu}{2}\left( h_\tz \alpha + e^\phi g_\tz \gamma \right)
\end{align}
The $(m)$ equations  in (\ref{3.c.1}) give the following relations,
\begin{align}
\label{2.BPS.m}
	0&=+\frac{k_1}{2f_6} \alpha + \frac{D_z f_6}{2 f_6} \beta 
	-\frac{1}{8}  e^\phi \chi_z \delta -\frac{i\nu}{8}  e^\phi g_z \beta 
\no \\
	0&=+\frac{k_1}{2f_6}\gamma + \frac{D_z f_6}{2 f_6} \delta
	+\frac{1}{8}  e^\phi \chi_z \beta + \frac{i\nu}{8} e^\phi g_z  \delta 
\no \\
	0&=-\frac{k_1}{2f_6} \beta+ \frac{D_\tz f_6}{2 f_6} \alpha
	-\frac{1}{8} e^\phi\chi_\tz \gamma -\frac{i\nu}{8}   e^\phi g_\tz \alpha 
\no \\
	0&=-\frac{k_1}{2f_6} \delta+ \frac{D_\tz f_6}{2 f_6} \gamma 
	+\frac{1}{8}  e^\phi \chi_\tz \alpha + \frac{i\nu}{8}  e^\phi g_\tz \gamma 
\end{align}
while the $(i)$ components in (\ref{3.c.1}) are
\begin{align}
\label{2.BPS.i}
	0&=-\frac{k_2 \nu }{2 f_2}\gamma + \frac{D_zf_2}{2 f_2} \beta
	-\frac{1}{8}  e^{\phi } \chi_z \delta  + \frac{i\nu}{8} e^{\phi } g_z \beta + \frac{i\nu}{4}   h_z \delta
\no \\
	0&=+\frac{k_2 \nu }{2 f_2}\alpha + \frac{D_zf_2}{2 f_2} \delta 
	+\frac{1}{8}   e^{\phi } \chi_z \beta - \frac{i\nu}{8} e^{\phi } g_z \delta  +\frac{i\nu}{4}   h_z 	\beta
\no \\
	0&=-\frac{k_2 \nu }{2 f_2} \delta +\frac{D_\tz f_2}{2 f_2} \alpha 
	-\frac{1}{8}   e^{\phi } \chi_\tz \gamma + \frac{i\nu}{8} e^{\phi } g_\tz \alpha	+\frac{i\nu}{4}  h_\tz \gamma
\no \\
	0&=+\frac{k_2 \nu }{2 f_2}\beta + \frac{D_\tz f_2}{2 f_2} \gamma +\frac{1}{8}  e^{\phi } \chi_\tz \alpha  
	-\frac{i\nu}{8} e^{\phi } g_\tz \gamma	 + \frac{i\nu}{4}  h_\tz \alpha 
\end{align}
The differential equation $(a)$ in  (\ref{3.c.1}) yields, for the $z$ and $\tz$ components, respectively,
\begin{align}
\label{2.BPS.a}
	0&=D_\tz \alpha+\frac{i}{2} \hat \omega_\tz \alpha + \frac{i\nu}{4} h_\tz \gamma   
\no \\
	0&=D_\tz \gamma+\frac{i}{2} \hat \omega_\tz \gamma+\frac{i\nu}{4} 	 h_\tz  \alpha
\no \\	
	0&=D_z \beta-\frac{i}{2} \hat \omega_z \beta+\frac{i\nu}{4}  h_z \delta
\no \\
	0&=D_z\delta-\frac{i}{2} \hat \omega_z\delta+\frac{i\nu}{4} h_z \beta
\end{align}
as well as
\begin{align}
\label{2.BPS.b}
	0&=D_z\alpha+\frac{i}{2} \hat \omega_z \alpha + \frac{1}{4} e^{\phi } \chi _z \gamma 
	+\frac{i\nu}{4}  e^{\phi} g_z \alpha + \frac{i\nu}{4}  h_z \gamma 
\no \\
	0&=D_z\gamma+\frac{i}{2} \hat \omega_z\gamma - \frac{1}{4}  e^{\phi } \chi _z \alpha  
	-\frac{i\nu}{4}  e^{\phi}g_z \gamma + \frac{i\nu}{4} h_z \alpha
\no \\	
	0&=D_\tz \beta-\frac{i }{2} \hat \omega_\tz \beta + \frac{1}{4}  e^{\phi } \chi _\tz \delta  
	+\frac{i\nu}{4}  	e^{\phi } g_\tz \beta  + \frac{i\nu}{4}  h_{\tilde z} \delta 
\no \\
	0&=D_\tz \delta-\frac{i}{2} \hat \omega_\tz \delta - \frac{1}{4}  e^{\phi } \chi _\tz \beta 
	-\frac{i\nu}{4} e^{\phi } g_\tz  \delta  + \frac{i\nu}{4} h_\tz \beta 
\end{align}

\subsection{Solving for $f_6$ and $f_2$}

We shall now show that the system of equations in (\ref{2.BPS.m}) and (\ref{2.BPS.i}) is equivalent to the following system of equations,
\bea
\label{3.f6}
f_6 & = & c_6 (\a \delta - \b \g)
\no \\
f_2 & = & c_2 (\a \b +  \g \delta )
\eea
where $c_6$ and $c_2$ are arbitrary complex constants, together with the system,
\bea
\label{3.BPS.m}
0 & = & { k_1 \over 2 c_6} - { 1 \over 8} e^\phi \chi_z (\b^2 + \delta^2) -{i \nu \over 4} e^\phi g_z \b \delta
\no \\
0 & = & { k_1 \over 2 c_6} - { 1 \over 8} e^\phi \chi_\tz (\a^2 + \g^2) -{i \nu \over 4} e^\phi g_\tz \a \g
\eea
and
\bea
\label{3.BPS.i}
0 & = & { k_2 \nu \over 2 c_2} + { 1 \over 8} e^\phi \chi_z (\b^2 + \delta^2) -{i \nu \over 4} e^\phi g_z \b \delta
+ { i \nu \over 4} h_z (\b^2 - \delta^2)
\no \\
0 & = & { k_2 \nu \over 2 c_2} + { 1 \over 8} e^\phi \chi_\tz (\a^2 + \g^2) -{i \nu \over 4} e^\phi g_\tz \a \g
+ { i \nu \over 4} h_\tz (\a^2 - \g^2)
\eea
To prove the first equation in (\ref{3.f6}), we take a linear combination of the first pair of equations in (\ref{2.BPS.m}) by multiplying the first by $- \gamma $ and the second by $\alpha$, and proceed similarly with the second set of equations in (\ref{2.BPS.m}), to obtain,
\bea
\label{3.mf6}
0 & = & { D_z f_6 \over 2 f_6} (\a \delta - \b \g) + {1 \over 8} e^\phi \chi_z (\a \b + \g \delta) 
+ { i \nu \over 8} e^\phi g_z (\a \delta + \b \g) 
\no \\
0 & = & { D_\tz f_6 \over 2 f_6} (\a \delta - \b \g) - {1 \over 8} e^\phi \chi_tz (\a \b + \g \delta) 
- { i \nu \over 8} e^\phi g_\tz (\a \delta + \b \g) 
\eea
Corresponding combinations may be obtained by taking linear combinations of the differential equations in (\ref{2.BPS.a}) and (\ref{2.BPS.b}) which give,
\bea
\label{3.Df6}
D_z (\a \delta - \b \g) & = & - {1 \over 4} e^\phi \chi_z (\a \b + \g \delta) 
- { i \nu \over 4} e^\phi g_z (\a \delta + \b \g) 
\no \\
D_\tz (\a \delta - \b \g) & = &  {1 \over 4} e^\phi \chi_\tz (\a \b + \g \delta) 
+ { i \nu \over 4} e^\phi g_\tz (\a \delta + \b \g) 
\eea
Eliminating the fields $\chi$ and $g$ between (\ref{3.mf6}) and (\ref{3.Df6}) yields,
\bea
{ D_z f_6 \over  f_6} (\a \delta - \b \g) -  D_z (\a \delta - \b \g)  & = & 0
\no \\
{ D_\tz f_6 \over  f_6} (\a \delta - \b \g) -  D_\tz (\a \delta - \b \g)  & = & 0
\eea
which are readily integrated to give the first line of (\ref{3.f6}). The second line of (\ref{3.f6}) may be shown analogously. The equations in (\ref{3.f6}) solve one pair of linear combinations of the four equations in (\ref{2.BPS.m}) and one pair of linear combinations of the four equations in (\ref{2.BPS.i}). The complementary linear combinations in  (\ref{3.BPS.m}) are obtained by multiplying the first equation of (\ref{2.BPS.m}) by $\delta$ and the second by $- \beta$, and similarly for the second pair of equations in (\ref{2.BPS.m}). Proceeding analogously for (\ref{2.BPS.i}) gives (\ref{3.BPS.i}).

\subsubsection{Implications of the dilatino equations}

The remaining reduced BPS equations may be simplified further.  
Eliminating  $\f_z$ and $\f_\tz$ from the dilatino equations (\ref{2.BPS.dil}) gives the following relations, 
\bea
\label{3.dil.1}
0 & = & e^\phi \chi_z (\b^2 + \delta ^2) +{i \nu \over 2} h_z (\b^2 - \delta^2) + i \nu e^\phi g_z \b \delta
\no \\
0 & = & e^\phi \chi_\tz (\a^2 + \g ^2) +{i \nu \over 2} h_\tz (\a^2 - \g^2) + i \nu e^\phi g_\tz \a \g
\eea
Eliminating the combinations $\b^2 - \delta ^2$ and $\a^2-\g^2$ from these equations using (\ref{3.BPS.i}) gives,
\bea
\label{3.BPS.new}
0 & = & { k_2 \nu \over 2 c_2} - { 3 \over 8} e^\phi \chi_z (\b^2 + \delta^2) -{3 i \nu \over 4} e^\phi g_z \b \delta
\no \\
0 & = & { k_2 \nu \over 2 c_2} - { 3 \over 8} e^\phi \chi_\tz (\a^2 + \g^2) -{3 i \nu \over 4} e^\phi g_\tz \a \g
\eea
Comparing the equations of (\ref{3.BPS.new}) with those of (\ref{3.BPS.m}) imposes the relation,
\bea
\label{3.k1k2}
{ k_2 \nu \over  c_2} = { 3 k_1 \over  c_6}
\eea
Henceforth, we may omit equations (\ref{3.BPS.i}) and (\ref{3.BPS.new}) while keeping all of the dilatino equations.  Using (\ref{3.dil.1}), we further simplify the relations (\ref{3.BPS.m}) as follows,
\bea
\label{3.BPS.i1}
0 & = & { k_1 \over  c_6} + { 1 \over 4} e^\phi \chi_z (\b^2 + \delta^2) + { i \nu \over 4} h_z (\b^2 - \delta^2)
\no \\
0 & = & { k_1 \over  c_6} + { 1 \over 4} e^\phi \chi_\tz (\a^2 + \g^2) + { i \nu \over 4} h_\tz (\a^2 - \g^2)
\eea

\subsubsection{Summary}

In summary, the combined system of equations (\ref{2.BPS.m}) and (\ref{2.BPS.i}) is equivalent to the combined system of (\ref{3.f6}),  (\ref{3.k1k2}) and (\ref{3.BPS.i1}). In the process of showing this equivalence we have used the other BPS equations

\subsection{Solving the differential equations of (\ref{2.BPS.a})}

To solve the differential equations of (\ref{2.BPS.a}) we need expressions for the metric and connection on $\Sigma_\CC$. Since the metric on $\Sigma_\CC$ is complex-valued, the choice of conformally flat gauge has to be modified from the  case where the metric is Riemannian. A careful derivation  is given in Appendix \ref{sec:D}. The result is that one may choose two complex frame directions $z , \tz$ and a complex-valued function $\rho$ on $\Sigma_\CC$ such that, 
\begin{align}
\label{3.conf}
e^z&=\rho dw &
D_z&=\rho^{-1}\partial_w & \hat \omega_z&= + i\rho^{-2} \, \partial_w\rho &
\no \\
e^{\tilde z}&=\rho d\tilde w &
D_{\tilde z}&=\rho^{-1}\partial_{\tilde w} & \hat \omega_{\tilde z}&=-i\rho^{-2} \, \partial_{\tilde w}\rho
\end{align}
where $w$ and $\tilde w$ are mutually independent local complex coordinates on $\Sigma_\CC$. In particular, $w$ and $\tilde w$ are not complex conjugates of one another.\footnote{Note that $w, \tilde w$ are not arbitrary local complex coordinates on $\Sigma_\CC$:  they are subject to the requirement that the complex metric on $\Sigma _\CC$ be \textit{diagonal} in these coordinates, analogous to the choice of complex coordinates on an ordinary Riemann surface that puts the Riemannian metric in conformal gauge.}

\sm

To solve the differential equations of (\ref{2.BPS.a}) we form linear combinations  to obtain the following equations, 
\begin{align}
0&=D_{\tilde z}(\alpha^2-\gamma^2)+i \hat \omega_{\tilde z} (\alpha^2-\gamma^2)
\no \\
0 &=D_z(\beta^2-\delta^2)-i \hat \omega_z (\beta^2-\delta^2)
\end{align}
We now use the conformally flat metric (\ref{3.conf}) to recast the equations as follows, 
\bea
\p_\tw \big ( \rho (\a^2 - \g^2 ) \big ) & = & 0
\no \\
\p_w \big ( \rho ( \beta^2 - \delta ^2) \big ) & = & 0
\eea
These equations may be solved as follows,  
\bea
\label{3.bd}
 \rho (\a^2 - \g^2) & = & 2 \kappa_1
 \no \\
  \rho ( \beta^2 - \delta ^2) & = & 2 \tilde \kappa _1 
 \eea
where  $ \kappa_1$ is independent of $\tw$ and $ \tilde \kappa_1$ is independent of  $w$.\footnote{Note that this restriction on the dependence of $\kappa_1$ and $\tilde \kappa_1$ on $w$ and $\tilde w$ is reminiscent of holomorphicity conditions if the tilde corresponded to complex conjugation. For the solutions considered here, however, no  complex conjugates enter as all supergravity fields are holomorphic on $\Sigma _\CC$. Instead, the restriction is tied  to the condition that the metric be \textit{diagonal} in the coordinates $w, \tilde w$.}

\sm

Forming linearly independent combinations of the two pairs of equations in (\ref{2.BPS.b})  and using the dilatino  equations to eliminate $h_z$  in favor of $\f_z$ and $\chi_z$, and $h_\tz$ in favor of $\f_\tz$ and $\chi_\tz$ along with the function $\phi$ and the spinors $\a, \b\, \g, \delta$, we obtain, 
\begin{align}
	0&=D_z(\beta\delta)	-i  \hat \omega_z \beta  \delta -\varphi_z \, \beta  \delta
	-\tfrac{1}{2} e^\phi\chi_z (\b^2 - \delta^2) 
	\notag\\
	0&=D_\tz (\alpha\gamma)+i \hat \omega_\tz \alpha  \gamma - \varphi_\tz \alpha  \gamma 
	-\tfrac{1}{2} e^\phi\chi_{\tilde z} (\a^2 -\g^2) 
\end{align}
Using the expressions for the dilaton and axion derivatives in conformal gauge,
\begin{align}
\f_z & = \rho^{-1} \p_w \phi & \chi_z = \rho^{-1} \p_w \chi
\no \\
\f_\tz & = \rho^{-1} \p_\tw \phi & \chi_\tz = \rho^{-1} \p_\tw \chi
\end{align}
and the relations (\ref{3.bd}), the differential equations may be recast in the following form, after multiplying through by $\rho^2 \, e^{- \phi}$, 
\bea
\p_w \big ( e^{ - \phi} \rho \b \delta + \chi \tilde \kappa_1 \big ) & = & 0
\no \\
\p_\tw \big ( e^{- \phi} \rho \a \g + \chi \kappa_1 \big ) & = & 0
\eea
where we have used the fact that $\p_w \tilde \kappa_1= \p_\tw \kappa_1=0$. In terms of two further functions, $\tilde \kappa_2 ( \tw)$ and $\kappa_2(w)$, we may now integrate these equations in turn as follows,
\begin{align}
\label{3.bd2}
\rho \beta \delta&=e^{\phi}\left( \tilde \kappa_1\chi - \tilde \kappa_2 \right)
\no \\
\rho \alpha \gamma&=e^{\phi}\left(\kappa_1 \chi - \kappa_2 \right)
\end{align}
This completely solves the equations  (\ref{2.BPS.a}).

\subsubsection{Solutions for $f_6, f_2, g_{z,\tz}, h_{z,\tz}$ in terms of $\rho$ and $\tau_\pm$}
\label{sec:ffgghh}

Introducing the following linear combinations of the axion-dilaton fields,
\bea
\label{4.sub.1}
\tau_\pm = \chi \pm i e^{-\phi} 
\hskip 1in 
\xi = \frac{\kappa_2}{\kappa_1}
\hskip 1in 
 \tilde \xi = \frac{\tilde \kappa_2}{\tilde \kappa_1}  
\eea
where $\xi$ and $\tilde \xi$ are scalars independent of $\tw$ and $w$, respectively, 
the solutions  (\ref{3.bd}) and (\ref{3.bd2}) may be combined as follows,
\bea
\label{4.sub.2}
\rho \big (\a \pm i \g \big )^2 & = & \mp \, 4 \, \kappa_1 \,  \frac{\tau_\mp - \xi}{\tau_+ - \tau_-}
\no \\
\rho \big (\b \pm i \delta \big )^2 & = & \mp \, 4  \, \tilde \kappa_1 \, \frac{\tau _\mp - \tilde  \xi }{\tau_+ - \tau_-}
\eea
The relations of (\ref{4.sub.2}) give the spinors $\a, \b, \g, \delta$, as well as all other fields,  in terms of  $\rho$ and $\tau_\pm$.  Using (\ref{3.f6}) and (\ref{4.sub.2}), the radii $f_6$ and $f_2$ are given as follows,
\bea
\label{3.radii}
\rho f_6 & = & - 2 \, c_6  \, { (\kappa_1 \tilde \kappa_1)^\half \over \tau_+ - \tau_-}  
\Big \{ (\tau_- - \xi )^\half (\tau_+ - \tilde \xi)^\half -
 (\tau_+ - \xi )^\half (\tau_- - \tilde \xi)^\half \Big \}
 \no \\
\rho f_2 & = & 2i \, c_2\,{ (\kappa_1 \tilde \kappa_1)^\half \over \tau_+ - \tau_-}  
\Big \{ (\tau_- - \xi )^\half (\tau_+ - \tilde \xi)^\half +
 (\tau_+ - \xi )^\half (\tau_- - \tilde \xi)^\half \Big \}
 \eea
 Similarly, the flux fields may be expressed solely in terms of these fields,
 \bea
\label{4.sub.3}
\rho \big ( h_z \pm i e^\phi g_z) & = & 2 i \nu e^{\phi } \, {\b \pm i \delta \over \b \mp i \delta} \, \p_w \tau _\pm
= \pm 4i \nu \left ( { \tau_\mp - \tilde \xi \over \tau_\pm - \tilde \xi} \right )^\half { \p_w \tau_\pm \over \tau_+ - \tau_-} 
\no \\
\rho \big ( h_\tz \pm i e^\phi g_\tz) & = & 2 i \nu e^{\phi } \, {\a \pm i \g \over \a \mp i \g} \, \p_\tw \tau _\pm
= \pm 4i \nu  \left ( { \tau_\mp -  \xi \over \tau_\pm -  \xi} \right )^\half { \p_\tw \tau_\pm \over \tau_+ - \tau_-} 
\eea
by solving the dilatino equations for the flux fields in terms of $\phi, \chi$ and $\rho$ and $\tau_\pm$.

\subsection{Solving the differential equations for $\rho$ and $\tau_\pm$}
\label{sec:3-solving}

Having expressed the spinors $\a , \b, \g, \delta$ and the flux fields in terms of $\rho$ and $\tau_\pm$ using (\ref{3.radii}) and (\ref{4.sub.3}), respectively, the remaining equations are  (\ref{2.BPS.b}) and (\ref{3.BPS.i1}), giving us six differential equations for three functions in two variables. These equations  may be expressed conveniently in terms of $\a \pm i \gamma$ and $\b \pm i \delta$ and, in turn,  these combinations may be eliminated in terms of 
$\rho$ and $\tau_\pm$ using (\ref{4.sub.2}). 

We proceed in two steps. First, we express the equations in (\ref{2.BPS.b}) in a suitable way by taking linear combinations and using the relations in (\ref{3.bd}) and (\ref{3.bd2}), leading to
\begin{align}\label{3.8.1}
	0&=\frac{\left(D_z+i\hat\omega_z\right)\left(\alpha^2-\gamma^2\right)}{\alpha^2-\gamma^2}+\frac{\chi-\xi}{2}e^{2 \phi }\chi_z
	+\frac{1}{X}\big(\varphi_z-e^{2 \phi } (\chi-\tilde\xi)\chi_z\big)
	\no\\
	0&=\frac{\left(D_{\tilde z}-i\hat\omega_{\tilde z}\right)\left(\beta^2-\delta^2\right)}{\beta^2-\delta^2}+\frac{\chi-\tilde\xi}{2} e^{2 \phi } \chi_{\tilde z}+X \left(\varphi_{\tilde z}-e^{2 \phi } (\chi-\xi)\chi_{\tilde z}\right)
	\no\\
	0&=\frac{\left(D_z+i\hat\omega_z\right)\left(\alpha\gamma\right)}{\alpha^2-\gamma^2}-\frac{1}{4} e^{\phi } \chi_z
	-\frac{1}{2 X}e^{\phi } ((\chi-\tilde\xi)\varphi_z+\chi_z)
	\no\\
	0&=\frac{\left(D_{\tilde z}-i\hat\omega_{\tilde z}\right)\left(\beta\delta\right)}{\beta^2-\delta^2}
	-\frac{1}{4} e^{\phi }\chi_{\tilde z}-\frac{1}{2} X e^{\phi } ((\chi-\xi)\varphi_{\tilde z}+\chi_{\tilde z})
\end{align}
The remaining Killing spinor components aside from $X$ can be expressed directly using (\ref{3.bd}) and (\ref{3.bd2}), while expressing $X$, defined as
\begin{align}
	X&= \frac{\beta^2+\delta^2}{\alpha^2+\gamma^2}\frac{\kappa_1}{\tilde \kappa_1}
\end{align}
through (\ref{4.sub.2}) entails taking a square root.
The equations in (\ref{3.BPS.i1}) can be expressed as
\begin{align}\label{3.8.3}
	0&=-\frac{k_1}{c_6}\frac{\beta ^2+\delta ^2}{\tilde\kappa_1^2}e^{-\phi }\rho ^2+\big(1-e^{2 \phi }
	(\chi-\tilde\xi)^2\big)\chi_z
	+2(\chi-\tilde\xi)\varphi_z
	\no\\
	0&=-\frac{k_1}{c_6}
	\frac{\alpha ^2+\gamma ^2}{\kappa_1^2}e^{-\phi }\rho ^2 +\left(1-e^{2 \phi }
	(\chi-\xi)^2\right)\chi_{\tilde z}+2 (\chi-\xi)\varphi_{\tilde z}
\end{align}
They also involve the combinations $\alpha^2+\gamma^2$ and $\beta^2+\delta^2$, which we express using (\ref{4.sub.2}) as
\begin{align}\label{eq:Lambda-def}
	\alpha^2+\gamma^2&=\frac{4\epsilon_\kappa\kappa_1}{\rho(\tau_+-\tau_-)}\sqrt{-\big(\tau_+-\xi\big)\big(\tau_--\xi\big)} & \epsilon_\kappa&=\frac{\sqrt{\kappa_1^2}}{\kappa_1}
	\no\\
	\beta^2+\delta^2&=\frac{4\tilde\epsilon_\kappa\tilde\kappa_1}{\rho(\tau_+-\tau_-)}\sqrt{-\big(\tau_+-\tilde\xi\big)\big(\tau_--\tilde\xi\big)} & \tilde\epsilon_\kappa&=\frac{\sqrt{\tilde\kappa_1^2}}{\tilde\kappa_1}
\end{align}
These expressions also fix $X$. To accurately track the behavior under sign reversals in $\kappa_{1,2}$ and $\tilde\kappa_{1/2}$, we introduced $\epsilon_\kappa$ and $\tilde\epsilon_\kappa$. This will be useful for the discussion of real forms.

\sm

The second step is to express the equations concisely. The equations in (\ref{3.8.1}) become
\bea
\label{4.sub.8}
(\tau _\mp - \xi)  \p_w \ln \hat \rho^2  + \p_w \xi  \mp { \p_w \tau_\mp  \over \tau_+ - \tau_-}
\left \{ \tau _\pm - \xi -{2 \over X}  ( \tau _\pm - \tilde  \xi )    \right \}  & = & 0
\no \\
(\tau _\mp - \tilde  \xi )  \p_\tw \ln \hat \rho^2  + \p_\tw \tilde  \xi \mp {\p_\tw \tau_\mp  \over \tau_+ - \tau_-}
\left \{ \tau _\pm - \tilde  \xi  -2 X  ( \tau _\pm -\xi )    \right \}  & = & 0
\eea
where $X$ and an additional shorthand $\hat\rho$ are given by
\begin{align}
	\label{3.Xrho}
	X &=\frac{\tilde\epsilon_\kappa}{\epsilon_\kappa}\sqrt{\frac{\tau_+-\tilde\xi}{\tau_+-\xi}\cdot\frac{\tau_--\tilde\xi}{\tau_--\xi}} 
	&
	\rho^2 &= { \kappa_1 \tilde \kappa _1 \, \hat \rho^2 \over (\tau_+- \tau_-)^\half}   
\end{align}
while the equations in (\ref{3.8.3}) become
\bea
\label{4.sub.9}
{  k_1 \over  c_6} \tilde\epsilon_\kappa\kappa_1 \, \hat \rho^2  
\left ( { 1 \over \tau_- - \tilde  \xi } - {1 \over \tau_+ - \tilde  \xi } \right )^{3 \over 2} 
+ \p_w \left ( { 1 \over \tau_- - \tilde  \xi } + {1 \over \tau_+ - \tilde  \xi } \right ) & = & 0
\no \\
{  k_1 \over c_6} \epsilon_\kappa\tilde \kappa_1 \, \hat \rho^2  
\left ( { 1 \over \tau_- - \xi} - {1 \over \tau_+ - \xi } \right )^{3 \over 2} 
+ \p_\tw \left ( { 1 \over \tau_- - \xi} + {1 \over \tau_+ - \xi} \right ) & = & 0
\eea
The equations in (\ref{4.sub.8}), (\ref{4.sub.9}) and in (\ref{3.8.1}), (\ref{3.8.3}) transform consistently under the transformations $(\beta,\delta)\rightarrow (-\delta,\beta)$ and $(\alpha,\gamma)\rightarrow (-\gamma,\alpha)$, which induce $\tilde\kappa_{1,2}\rightarrow -\tilde\kappa_{1,2}$ and $\kappa_{1,2}\rightarrow -\kappa_{1,2}$, respectively, via (\ref{3.bd}) and (\ref{3.bd2}): (\ref{3.8.3}) and (\ref{4.sub.9}) are invariant while the only effect in (\ref{3.8.1}) and (\ref{4.sub.8}) is a sign reversal in $X$.

\subsubsection{Decoupling}
\label{sec:3-decoupling}

Decoupling these equations involves a further change of variables from $\tau_\pm$ to $Z_\pm$ and from $\hat \rho^2$ to $R^2$, given as follows,
\bea
\label{4.Ztau1}
Z_\pm^2 = { \tau _\pm - \tilde  \xi  \over \tau_\pm - \xi}
\hskip 1in 
{ 1 \over R^2} = - {  k_1 \over c_6} \, 
 { (Z_-+\varepsilon_\kappa Z_+)^{3 \over 2} \, \hat \rho^2 \over X^\half   (Z_--\varepsilon_\kappa Z_+)^{1 \over 2} ( \xi  - \tilde \xi)^\half  } 
 \hskip 0.5in
 \varepsilon_\kappa=\frac{\tilde\epsilon_\kappa}{\epsilon_\kappa} \hskip 0.2in 
\eea
In appendix \ref{sec:E}, we show that, in terms of the function $R^2$ and the combinations,
\bea
\label{3.XY}
X=\varepsilon_\kappa Z_+ Z_- \hskip 1in Y=-\frac{(Z_- +\varepsilon_\kappa  Z_+)^2}{X}
\eea
the equations may be organized in a manner in which the decoupling will be readily seen. A first pair of equations involves only $R$ and $X$ and is given by,
\bea
\label{4.dec}
 \p_w \Big ( 2 R^2 X^\half (\tilde  \xi  - \xi) \Big )  & = &  -\epsilon_\kappa(\kappa _2 - \kappa _1 \tilde  \xi)  
  \no \\
\p_\tw \Big ( 2R^2 X^{-\half} (  \xi  - \tilde \xi) \Big ) & = & + \epsilon_\kappa(\tilde \kappa _2 - \tilde \kappa _1  \xi)   
\eea
A second pair of equations also involves only $R$ and $X$ and reads, 
\bea
\label{4.sum}
 \p_w \ln \big ( R^2 X^{- \half} \big )   - X \, \p_w \ln (\tilde  \xi -\xi) & = & 0
\no \\
\p_\tw \ln \big ( R^2 X^\half \big )  - X^{-1}  \p_\tw \ln (\tilde  \xi -\xi) & = & 0
\eea
and the third pair consists of inhomogeneous linear differentials  in the function $Y$,
\bea
\label{4.linear}
\p_w Y  +2(Y+3)X  \p_w \ln (\tilde  \xi  - \xi) 
+ (Y+6)  \p_w \ln \Big ( X (\tilde \xi - \xi) \Big ) 
& = &  0
\no \\
\p_\tw Y + 2(Y+3) /X  \p_\tw \ln (\tilde  \xi  - \xi)  +   (Y+6) \p_\tw \ln \Big ( X^{-1} (\tilde \xi - \xi) \Big ) 
& = &  0
\eea

\subsubsection{Integrating}
\label{sec:3-integrating}

The set of equations in (\ref{4.dec}) may be integrated by quadrature since the functions on the right side of each equation are considered to be given.  To integrate them, we parametrize the one-forms as local derivatives of scalar functions, as follows,
\begin{align} 
\label{3.Akappa}
-\epsilon_\kappa\kappa _1 & = \p_w \cA_1 & \epsilon_\kappa\tilde \kappa _1 & = \p_\tw \tilde \cA_1  
\no \\
-\epsilon_\kappa\kappa _2 & = \p_w \cA_2  & \epsilon_\kappa\tilde \kappa _2 & = \p_\tw \tilde \cA_2  
\no \\
\xi & = \frac{\kappa_2}{\kappa_1} & \tilde  \xi & = \frac{\tilde \kappa_2}{\tilde \kappa_1}
\end{align}
where the functions $\cA_1, \cA_2$  are independent of $\tw$ while $\tilde \cA_1, \tilde \cA_2$ are independent of $w$. Given the differentials $\kappa_1, \kappa_2, \tilde \kappa_1, \tilde \kappa_2$, these four functions are unique up to additive constants.   In terms of this parametrization, the general solution to  the equations (\ref{4.dec}) is as follows,
\bea
\label{4.dec1}
2R^2 X^{+\half} (\tilde  \xi  - \xi) & = & \tilde \cA_0  + \cA_2 - \tilde  \xi  \cA_1 
\no \\
2R^2 X^{-\half} ( \xi  - \tilde \xi) & = &  \cA_0 + \tilde \cA_2 - \xi \tilde \cA_1 
\eea
where $\cA_0$ and $\tilde \cA_0$ are as of yet undetermined functions independent of $\tw$ and $w$, respectively. Eliminating the combinations $R^2 X^{\pm \half}$ from (\ref{4.sum}) using the results of (\ref{4.dec1}), which is carried out in detail in appendix \ref{sec:E},  determines the functions $\cA_0$ and $\tilde \cA_0$ as follows, 
\bea
\label{3.AA0}
\cA_0 & = & -\xi \cA_1 + \cA_2
\no \\
\tilde \cA_0 & = & -\tilde  \xi  \tilde \cA_1 + \tilde \cA_2
\eea
In summary, the equations of (\ref{4.dec1}) may be expressed solely in terms of $\tilde \cA_1, \cA_1, \tilde \cA_2,\cA_2$ and their derivatives using the following convenient combinations, 
\begin{align}
\label{4.dec2}
2 R^2 X^{+\half} (\tilde  \xi  - \xi) & = \tilde \cL & \tilde \cL &=  -\tilde  \xi  ( \tilde \cA_1 + \cA_1) + \tilde \cA_2  + \cA_2
\no \\
2R^2 X^{-\half} (  \xi  - \tilde \xi) & =  \cL  & \cL & = - \xi ( \cA_1 + \tilde \cA_1) +  \cA_2 + \tilde \cA_2 
\end{align}
while $R^2$ and $X$ themselves are given by,
\bea
\label{4.RX}
 R^4 = - { \cL \tilde \cL  \over 4 (\xi - \tilde \xi)^2}
\hskip 1in 
X = - { \tilde \cL \over \cL}
\eea
To solve the equations (\ref{4.linear}) we recast the second term in each equation of (\ref{4.linear}) with the help of (\ref{4.sum}) to obtain the following equations, 
\bea
\p_w \Big ( Y (\tilde  \xi  - \xi) R^4 \Big ) & = & 
 { 3 \over 2} \Big ( -\cA_2 \p_w \cA_1 + \cA_1 \p_w \cA_2  + \tilde \cA_1 \p_w \cA_2 - \tilde \cA_2 \p_w \cA_1  \Big )
\no \\
\p_\tw \Big ( Y (\xi  - \tilde   \xi) R^4 \Big ) & = & 
 { 3 \over 2}  \Big ( -\tilde \cA_2 \p_\tw \tilde \cA_1  + \tilde \cA_1 \p_\tw \tilde \cA_2 
 + \cA_1 \p_\tw \tilde \cA_2  - \cA_2 \p_\tw \tilde \cA_1  \Big )
\eea
Defining $\cB$ and $\tilde \cB$ by,
\bea
\label{4.BB}
\p_w \cB & = & \cA_2 \p_w \cA_1 - \cA_1 \p_w \cA_2
\no \\
\p_\tw \tilde \cB & = &  \tilde \cA_2 \p_\tw \tilde \cA_1 - \tilde \cA_1 \p_\tw \tilde \cA_2 
\eea
we have,
\bea
\label{4.ZpZm}
Y (\tilde  \xi  - \xi) R^4 
= {3 \over 2} \Big (  -\cB + \tilde \cB + \tilde \cA_1 \cA_2 - \cA_1 \tilde \cA_2  \Big )
\eea
or, explicitly in terms of the functions $\cA$ and $\cB$,
\bea
\label{4.Y}
Y = - 6 \, { \tilde \xi - \xi \over \cL \tilde \cL} \, \Big ( - \cB + \tilde \cB + \tilde \cA_1 \cA_2 - \cA_1 \tilde \cA_2  \Big )
\eea
This completes the solution to the BPS equations of complexified Type IIB.

\subsection{Summary of the solution}\label{sec:sol-summary}

In this section we complete the supergravity solution by expressing $\rho$ and $\tau_\pm$ in terms of the functions $\cA_1, \cA_2, \tilde \cA_1, \tilde \cA_2$ and their composites $\cB, \tilde \cB$ and $\cL, \tilde \cL$, defined in (\ref{4.BB}) and (\ref{4.dec2}), respectively. We then similarly discuss the remaining supergravity fields. In the process, we will carefully track branch choices of square roots arising in the expressions.

\sm

To begin with, the functions $\tau_\pm$ are determined in terms of $Z_\pm$ by equation (\ref{EE.2}) which we repeat here for convenience, 
\bea
\label{EE.2a}
\tau_\pm - \xi = { \xi  - \tilde \xi \over Z_\pm^2-1}
\hskip 1in
\tau_\pm - \tilde  \xi  = { Z_\pm ^2 ( \xi  - \tilde \xi ) \over Z_\pm^2-1}
\eea 
Having expressed $X $ and $R^4$ in terms of $\xi, \tilde \xi, \cL,\tilde \cL$ via the equations of (\ref{4.RX}), and $Y$ via (\ref{4.Y}), it suffices to express $\rho$ and $Z_\pm^2$ in terms of $X,Y,R$. To this end we solve (\ref{3.XY}) by
\begin{align}
	\label{3.Zpm}
	Z^2_\pm&=-\frac{1}{2}X(Y+2)\pm \frac{1}{2}XYT
\end{align}
where $T$ is defined up to a sign by
\begin{align}
	T^2&=\frac{Y+4}{Y}
\end{align}
We solve for $\rho^2 $ using the second equation of (\ref{3.Xrho}) and the second equation of (\ref{4.Ztau1}), which leads to
\bea
\rho^2 = { c_6 \over k_1} { \kappa_1 \tilde \kappa_1 \over R^2 \, Y} 
\left ( X+{ 1 \over X} +Y+2 \right )^\half
\eea
The branch choice associated with the square root can be absorbed into a redefinition of $c_6$ (which will only appear through its square in the other metric functions).

The solution to the BPS equations determines certain combinations of the remaining metric functions without branch or sign ambiguities. Two convenient combinations determined directly by (\ref{3.f6}) and (\ref{3.bd}), (\ref{3.bd2}) in the form of (\ref{4.sub.2}) are
\begin{align}\label{f6pmf2}
	\left(\frac{f_6}{c_6}\pm i\frac{f_2}{c_2}\right)^2&=
	-(\alpha\pm i\gamma)^2(\beta\mp i\delta)^2
	=
	\frac{16\kappa_1\tilde\kappa_1Z_\pm^2}{\rho^2 XY(Y+4)}
	\left(X+\frac{1}{X}+Y+2\right)
\end{align}
A particularly simple combination determined directly by (\ref{3.f6}), (\ref{3.bd}), (\ref{3.bd2}) is
\begin{align}
	\rho^2f_2f_6&=2c_2c_6e^\phi\left(\tilde\kappa_1\kappa_2-\kappa_1\tilde\kappa_2\right)
\end{align}
The combinations in (\ref{f6pmf2}) unambiguously determine, using (\ref{3.Zpm}),
\begin{align}\label{f6f2sq-diff}
	\frac{f_6^2}{c_6^{2}}-\frac{f_2^2}{c_2^{2}}&=-\frac{8\kappa_1\tilde\kappa_1}{\rho^2 Y}\frac{Y+2}{Y+4}\left(X+\frac{1}{X}+Y+2\right)
\end{align}
The remaining linear combination can be evaluated by taking the product of the expressions in (\ref{f6pmf2}) and then the square root. This leads to
\begin{align}\label{f6f2sq-sum}
	\frac{f_6^2}{c_6^{2}}+\frac{f_2^2}{c_2^{2}}
	&=\frac{16\eta\kappa_1\tilde\kappa_1}{\rho^2 Y(Y+4)}\left(X+\frac{1}{X}+Y+2\right)
\end{align}
where we introduced $\eta\in\{\pm 1\}$ to track the choice of branch for the square root.\footnote{This branch choice arises at the level of the BPS equations and we keep it as parameter in the following. Adding the Bianchi identities for the two-form fields ultimately forces $\eta=-1$ \cite{DGU-to-appear}.\label{foot:eta}}
The same combination can be determined from (\ref{3.f6}) through $\alpha^2+\gamma^2$ and $\beta^2+\delta^2$ in (\ref{eq:Lambda-def}), which entails a corresponding choice of square root branch.
From (\ref{f6f2sq-diff}), (\ref{f6f2sq-sum}) we obtain
\begin{align}
	f_6^2&=- 4c_6^2\frac{\kappa_1\tilde\kappa_1}{\rho^2 Y}\,\frac{Y+2-2\eta}{Y+4}\left(X+\frac{1}{X}+Y+2\right)
	\no\\
	f_2^2&=+4c_2^2\frac{\kappa_1\tilde\kappa_1}{\rho^2Y}\,\frac{Y+2+2\eta}{Y+4}\left(X+\frac{1}{X}+Y+2\right)
\end{align}
The metric functions can alternatively be obtained from (\ref{3.radii}), where the square-root branches have to be chosen so as to be compatible with the above combinations;
the discussion here shows that these choices reduce to a single additional sign variable $\eta$.

\sm

The string-frame metric functions given above can be converted to Einstein frame using the relation $g^\text{E} _{MN} = e^{-\phi/2} g^\text{S}_{MN}$. This results in
\begin{align}
	\label{3.Eins}
	(\rho^\text{E})^2 & = +\frac{c_6}{k_1}\kappa_1 \tilde \kappa_1
	\left(-\frac{i(\tilde\xi-\xi)}{2R^4Y} T\right)^{\frac{1}{2}}
	\no\\
	(f_6^E)^2 & =  +2c_6^2\frac{i\kappa_1\tilde\kappa_1(\tilde\xi-\xi)}{(\rho^\text{E})^2 }T^{-\eta}
	\no \\
	(f_2^E)^2 & = -2c_2^2\frac{i\kappa_1\tilde\kappa_1(\tilde\xi-\xi)}{ (\rho^\text{E})^2}
	T^\eta
\end{align}
The expressions for the metric in Einstein frame will be useful for investigating the $SL(2,\CC)$ duality symmetry of Type IIB$_\CC$.

\sm

To further streamline the expressions, the composite quantities $X$, $Y$ and $R$ can be expressed concisely by introducing  $\cG$ and $\kappa^2$ defined by
\begin{align}\label{kappa2-G-def}
	\cG&=i(-\cB+\tilde\cB+\tilde\cA_1\cA_2-\cA_1\tilde\cA_2)
\no\\
	\kappa^2&=i(\partial_w\cA_1\partial_{\tilde w}\tilde\cA_2-\partial_{\tilde w}\tilde\cA_1\partial_{w}\cA_2)=-\partial_w\partial_{\tilde w}\cG
\end{align}
Then we have $\partial_w\cG=-i\partial_w\cA_1\cL$ and $\partial_{\tilde w}\cG=+i\partial_{\tilde w}\tilde\cA_1\tilde\cL$. From (\ref{4.RX}) and (\ref{4.ZpZm}),
\begin{align}
	Y&=\frac{6\kappa^2\cG}{\partial_w\cG\partial_{\tilde w}\cG}
	&
	R^4Y&=-\frac{3i\cG}{2(\tilde\xi-\xi)}
	&
	X&=\frac{\partial_w\cA_1\partial_{\tilde w}\cG}{\partial_{\tilde w}\tilde\cA_1\partial_w\cG}
\end{align}
Using also $\kappa_1\tilde\kappa_1(\tilde\xi-\xi)=i\kappa^2$, the combination $T^2$ becomes
\begin{align}\label{eq:T-def}
	T^2&=1+\frac{2}{3}\frac{\partial_w\cG\partial_{\tilde w}\cG}{\kappa^2\cG}
\end{align}
The Einstein-frame metric functions then take the form
\begin{align}
	\label{3.Eins-2}
	(\rho^\text{E})^2 & = -\frac{c_6}{k_1}\kappa^2\sqrt{-\frac{T}{3\cG}}
	&
	(f_6^E)^2 & =  -\frac{2c_6^2\kappa^2}{(\rho^\text{E})^2}\frac{1}{T^{\eta}}
	&(f_2^E)^2 & =  +\frac{2c_2^2\kappa^2}{(\rho^\text{E})^2}T^{\eta}
\end{align}
The axion-dilaton combinations $\tau_\pm$ become
\begin{align}\label{eq:tau-pm-sum}
	\tau_\pm&=\frac{\partial_{\tilde w}\tilde\cA_2(T\pm 1)\partial_w\cG+\partial_w\cA_2(T\mp 1)\partial_{\tilde w}\cG}{\partial_{\tilde w}\tilde\cA_1(T\pm 1)\partial_w\cG+\partial_w\cA_1(T\mp 1)\partial_{\tilde w}\cG}
\end{align}

\subsection{$SL(2,\CC)$ symmetry}
\label{sec:3.9}

In this last subsection, we shall show that the following $SL(2,\CC)$ transformation of the functions $\cA_1, \cA_2, \tilde \cA_1, \tilde \cA_2$,
\bea
\label{3.AA}
\left ( \bma \cA_2' \cr \cA_1' \ema \right ) 
= M \left ( \bma \cA_2 \cr \cA_1 \ema \right )
\hskip 0.8in 
\left ( \bma \tilde \cA_2' \cr \tilde \cA_1' \ema \right ) 
= M \left ( \bma \tilde \cA_2 \cr \tilde \cA_1 \ema \right )
\hskip 0.8in 
M= \left ( \bma a & b \cr c & d \ema \right )
\eea
with $a,b,c,d \in \CC$ and $\det M=1$, induces a complexified version of the $SL(2,\RR)$ symmetry of standard Type IIB supergravity. It follows from (\ref{4.BB}) and (\ref{3.AA})  that $\p_w \cB$ and $\p_\tw \tilde \cB$ are invariant. We shall choose suitable integration constants such that $\cB - \tilde \cB$ is invariant.  The combinations $\kappa^2$ and $\cG$ defined in (\ref{kappa2-G-def}) are then also invariant, and so is $T^2$ defined in (\ref{eq:T-def}). This implies that the Einstein-frame metric functions in (\ref{3.Eins-2}) are invariant.
The transformation of $\tau_\pm$ can be determined from (\ref{eq:tau-pm-sum}), which leads to
\bea
\label{3.tautr}
\tau' _\pm = { a \tau_\pm + b \over c \tau_\pm +d } 
\eea
The individual combinations $\kappa_1, \tilde \kappa_1, \xi, \tilde \xi, \cL$ and $\tilde \cL$, the latter defined in (\ref{4.dec2}), transform as follows,
\begin{align}
	\label{3.Mobius}
	\kappa_1 ' & = (c \, \xi +d) \kappa_1 &  
	\xi ' & = { a \, \xi + b \over c \, \xi +d} & 
	\cL' & = { \cL \over c \, \xi +d}  
	\no \\
	\tilde \kappa_1 ' & = (c \, \tilde \xi +d) \tilde \kappa_1 &
	\tilde \xi ' & = { a \, \tilde \xi + b \over c \, \tilde \xi +d} 
	& \tilde \cL' & = { \tilde \cL \over c \, \tilde \xi +d}
\end{align}
The transformation of the flux fields then follows from (\ref{4.sub.3}),
\bea
\rho ' \big ( h_z ' \pm i e^{\phi'} g_z' \big ) & = & 
(  c \tau_\mp +d )^\half ( c \tau_\pm +d) ^{- \half}  \rho  \big ( h_z  \pm i e^{\phi} g_z \big ) 
\no \\
\rho ' \big ( h_\tz ' \pm i e^{\phi'} g_\tz' \big ) & = & 
(   c \tau_\pm +d )^\half ( c \tau_\mp +d)^{- \half}  \rho  \big ( h_\tz  \pm i e^{\phi} g_\tz \big ) 
\eea

We note that this $SL(2,\CC)$, which arises at the level of Type IIB$_\CC$ supergravity, reduces to a real form in the various real forms of Type IIB$_\CC$ supergravity, as will be discussed in section \ref{sec:SL2R}, and further to a discrete subgroup in the corresponding string theories.

\newpage

%%%%%%%%%%%%%%%%%%%%%%%%%%%%%%%%%%%%%%%%%%%
%%%%%%%%%%%%%%%%%%%%%%%%%%%%%%%%%%%%%%%%%%%
\section{Real forms of $F(4)$ solutions: classification}
\setcounter{equation}{0}
\label{sec:5}
%%%%%%%%%%%%%%%%%%%%%%%%%%%%%%%%%%%%%%%%%%%
%%%%%%%%%%%%%%%%%%%%%%%%%%%%%%%%%%%%%%%%%%%

Complexified Type~IIB supergravity, as defined in section \ref{sec:2},  involves complex-valued fields with holomorphic dependence on the coordinates of complexified spacetime. Since these fields enter in the absence of their  complex conjugate fields, the theory cannot be invariant under complex conjugation. To obtain a complex conjugation symmetry, we must first restrict to one of the real forms of Type~IIB$_\CC$ supergravity discussed  in section \ref{sec:2}. Within a given real form of Type~IIB$_\CC$ we can then restrict to real solutions. 
In this section we classify the real forms of the solutions. We will not invoke any restriction based on the allowed real forms of the Lie superalgebra $F(4;\CC)$. 
Instead, we will show that our classification of real solutions is compatible with and exhausts the real forms of $F(4;\CC)$.

\subsection{Conditions for real solutions}

We begin by spelling out, for a given real form of Type~IIB$_\CC$ supergravity,  all ingredients and conditions involved in specifying a real solution whose spacetime manifold consists of $\cM_6$ and $\cM_2$ warped over a two-dimensional surface $\Sigma$.  They are obtained by specifying,
\begin{enumerate}
\itemsep =-0.03in
\item one of the real forms of Type~IIB$_\CC$ theory; 
\item the signatures of $\cM_6$, $\cM_2$ and $\Sigma$;
\item the signs of the curvatures of $\cM_6$ and $\cM_2$ given by $-k_1^2$ and $-k_2^2$, respectively.
\end{enumerate}
In the remainder of this section, we spell out these conditions in more detail, and enforce compatibility between them and with the chirality and projection conditions of (\ref{3.IJ}). This will lead to a classification of real forms of the solutions in section \ref{sec:5-summary}.

\subsection{Specifying the real form of Type~IIB$_\CC$}

We recall that the reality condition of (\ref{S-define}) for a real form of Type~IIB$_\CC$ on the supersymmetry transformation parameter $\ep$,  given by,
\bea
\label{4.B.ep}
\cB^{-1} _\eta \ep^* = S \ep \hskip 1in S=I_2, s^1,s^2,s^3
\eea
involves the complex conjugation matrix $\cB_\eta$ defined in (\ref{2.B.2}) and the matrix $S$ given in (\ref{2.parity}), which both depend on $t$.  The matrix $\cB_\eta$ acts only on the spacetime spinor indices, and does not act on the doublet indices. Therefore $\cB_\eta$ commutes with $S$. 

\sm

For the discussions to come it will be convenient to introduce some notation and express the condition (\ref{4.B.ep}) in terms of the reduced spinors $\zeta$.
The signatures of the metrics on $\cM_6$ and $\cM_2$ enter into the complex conjugation formula for the Killing spinors 
$\chi^{\eta_1, \eta_2}$ onto which the supersymmetry transformation parameter $\ep$ was decomposed in (\ref{3.b.2}), 
\bea
\label{4.ep}
\ep = \sum _{\eta_1, \eta_2= \pm} \chi^{\eta_1, \eta_2} \otimes \zeta _{\eta_1 , \eta_2} 
\eea
and will result in a complex conjugation formula for $\ep$ given by,
\bea
\label{4.epstar}
\cB_\eta ^{-1} \ep ^* = \sum _{\eta_1, \eta_2= \pm} 
\chi^{\eta_1, \eta_2} \otimes \big (  \mb_\eta^{-1} \zeta^* \big ) _{\eta_1 , \eta_2}  
\eea
The reduced complex conjugation matrix $\mb_\eta^{-1}$ on the reduced spinor $\zeta$ will depend on the signatures of  $\cM_6$, $\cM_2$ and $\Sigma$ and on the parameters $k_1^2$ and $k_2^2$ that specify the curvatures of $\cM_6$ and $\cM_2$. 
Combining (\ref{4.B.ep}) with the decompositions (\ref{4.ep}) and (\ref{4.epstar}), leads to the reality condition at the level of the reduced spinors
\begin{align}
\label{4.4.a}
\cK \zeta ^* &= \zeta 
&
\cK &= S^{-1}  \otimes \mb_\eta^{-1}  
\end{align}
where $S$ is given in (\ref{S-define}) and acts on the doublet of reduced spinors $\zeta$ while $\mb^{-1} _\eta$, acting on the components, is left to be determined.

\sm

In the next two subsections we give two different derivations for $\mb^{-1} _\eta$ which emphasize different aspects: the first is direct and connects to the reduced BPS equations, the second connects to the 10d discussion.  
Agreement between them will provide a consistency check.

\subsection{Determining $\mb_\eta$ from the reduced BPS equations}

The first derivation of $\mb_\eta$ is by inspecting the conjugation symmetries of the reduced BPS equations (\ref{3.c.1}) at the level of the reduced spinors.
That is, we determine $\mb_\eta^{-1}$ such that
\begin{align}\label{eq:cK-rep}
	\zeta&\rightarrow \cK \zeta^\star
	&
	\cK &= S^{-1}  \otimes \mb_\eta^{-1}  
\end{align}
combined with complex conjugation of the equations (\ref{3.c.1}) is a symmetry.
Since the reality conditions for the bosonic fields in the real forms of IIB$_\CC$ can be derived from the BPS conditions, this is expected to produce the same $\mb_\eta$ as the reality conditions for $\epsilon$.

\sm

The ingredients determining the behavior of (\ref{3.c.1}) under complex conjugation can be built into $\mb_\eta^{-1}$ successively. They are
\begin{itemize}
	\itemsep =-0.03in
	\item[--] the constants $k_1$, $k_2$ encoding the curvature of $\cM_6$ and $\cM_2$
	\item[--] the reality properties of $f_2$ and $f_6$
	\item[--] the reality properties of the component fields $\chi_a$, $g_a$ and $h_a$
\end{itemize}
The last point deserves discussion.
While the  real form of IIB$_\CC$ dictates the reality properties of the fields $\chi$, $H_{(3)}$ and $\tilde F_{(3)}$, the reality properties of $\chi_a$, $h_a$ and $g_a$ depend in addition on the signatures of $\cM_2$ and $\Sigma$ through the expansions (\ref{3.a.3}), (\ref{3.a.4}),
\begin{align}
\tilde F_{(3)} &= g_a \, e^a \wedge e^{67} 
&
H_{(3)} &= h_a e^a \wedge e^{67} 
&	
d \phi &= \f_a \, e^a 
&
d\chi = \chi_a \, e^a
\end{align}
We recall that the signature of the frame metric is kept $(1,9)$, while the spacetime signature of the real form of Type IIB$_\CC$ is realized by imaginary frame components.
This is encoded in the reality condition (\ref{2.frame}), which we repeat here for convenience,
\begin{align}\label{eq:frame-conj-rep}
	(e^A)^* &= \eta_{AA} \, e{}^A
\end{align}
where $\eta_{AA}=\pm 1$ encode the signature of the spacetime metric.
If $\Sigma$ has signature $(1,1)$ or $(2,0)$, one or both components of the $\Sigma$-frame $e^a$ are imaginary;
if the signature of $\cM_2$ is $(1,1)$, the volume form $e^{67}$ in (\ref{3.a.3}) is imaginary, otherwise it is real.
This determines the reality properties of $\chi_a$, $h_a$, $g_a$ in terms of the reality properties of $\chi$, $H_{(3)}$, $\tilde F_{(3)}$.

\sm

With these preparations one can directly determine  $\mb_\eta^{-1}$ in a factorized form which systematically takes into account the signatures of $\cM_6$, $\cM_2$, $\Sigma$. Namely,
\begin{align}\label{eq:mb-Ki}
\mb_\eta ^{-1} &= K_1 \, K_2 \, K_3 \, \tau ^{(02)} \otimes \sigma ^2 
\end{align}
In (\ref{3.c.1}), the reality properties of $k_1$, $k_2$ and $f_6$, $f_2$ only enter in the combinations $k_1/f_6$ and $k_2/f_2$. They are accounted for by,
\begin{align}\label{eq:K1-def}
	K_1&=\begin{cases}
		\tau^{(00)} \otimes I_2 ~~ \hbox{ for } & (\mathds{k}_1,\mathds{k}_2)=(1,i) \\
		\tau^{(01)} \otimes I_2 ~~ \hbox{ for } & (\mathds{k}_1,\mathds{k}_2)=(1,1) \\
		\tau^{(10)}\otimes I_2  ~~ \hbox{ for } & (\mathds{k}_1,\mathds{k}_2)=(i,i) \\
		\tau^{(11)} \otimes I_2 ~~ \hbox{ for } & (\mathds{k}_1,\mathds{k}_2)=(i,1)
	\end{cases}
\end{align}
where we define $\mathds{k}_{1}$ and $\mathds{k}_{2}$, which determine the actual curvature of the six- and two-dimensional spaces in the 10d geometry, by,
\begin{align}\label{eq:ds-k12-def}
	\mathds{k}_1&=k_1\frac{|f_6|}{f_6}
	&
	\mathds{k}_2&=k_2\frac{|f_2|}{f_2}
\end{align}
$K_2$ and $K_3$ account for the signature of $\cM_2$ and $\Sigma$, respectively, and are given by,
\begin{align}\label{eq:K23-def}
	K_{2} &=	\left\{
	\begin{array}{rl}
		\tau^{(00)} \otimes I_2 ~~ \hbox{ for } &  \cM_{0,2}, \cM_{2,0}\\ 
		-i\tau^{(32)} \otimes I_2 ~~ \hbox{ for } &  \cM_{1,1}
	\end{array}\right.
	&
	K_{3}&=	\left\{
	\begin{array}{rl} \tau^{(00)} \otimes I_2 ~~ \hbox{ for } &   \Sigma_{0,2}\\ 
		i \tau^{(20)} \otimes \sigma ^1  ~~ \hbox{ for } &   \Sigma _{1,1}\\
		\tau^{(00)} \otimes \sigma ^3 ~~ \hbox{ for } &  \Sigma_{2,0}
	\end{array}\right.	
\end{align}
Verifying that (\ref{eq:cK-rep}) with this $\mb_\eta ^{-1}$ is a conjugation symmetry of (\ref{3.c.1}) is straightforward.

\sm
Finally, we note for later use that each one of the factors $K_1$, $K_2$, $K_3$ and $\tau^{(02)} \otimes \sigma ^2$ is  real and commutes with the chirality operator $\tau^{(11)} \otimes \sigma ^3$ and, therefore, so does $\mb_\eta^{-1}$.

\clearpage

\subsection{Determining $\mb_\eta$ from 10d conjugation relations}

The second derivation of $\mb_\eta$ proceeds from the conjugation relations of the 10d spinors, through a discussion of the charge conjugation matrices on $\cM_6$ and $\cM_2$ for all possible signatures, to the conjugation relation for the reduced spinors.

\subsubsection{Charge conjugation matrices for arbitrary signatures}

 We shall make the following choice for the complex conjugation matrices,\footnote{Since the dimensions of $\cM_6$, $\cM_2$ and $\Sigma$ are even, both $B_{\eta(1)} $ and $B_{\eta (1)}  \gamma_{(1)}$ act as charge conjugation matrices, but the signs in the equations (\ref{4.Bstar}) are reversed between the two choices, and similarly for $B_{\eta(2)}$ and $\Sigma$. The choice made here reduces to the one made in \cite{DHoker:2016ujz} for the signatures $(1,5)$, $(0,2)$ and $(0,2)$ for $\cM_6$, $\cM_2$ and $\Sigma$, respectively.}
\bea
\label{4.Bstar}
\big ( \hat e^m \, \gamma _m \big )^* & = & B_{\eta (1)} \, \big ( \hat e^m \, \gamma _m \big ) \,  B_{\eta (1)}  ^{-1}
\no \\
\big ( \hat e^i \, \gamma _i \big )^* & = & - B_{\eta (2)} \, \big ( \hat e^i \, \gamma _i \big ) \,  B_{\eta (2)}  ^{-1}
\no \\
\big ( e^a \, \gamma _a \big )^* & = & - B_{\eta (3)} \, \big ( e^a \, \gamma _a \big ) \,  B_{\eta (3)}  ^{-1}
\eea
As a function of the signature of $\cM_6$  the matrices $B_{\eta (1)}^{-1}$ are given as follows in terms of the complex conjugation matrix $B_{(1)}^{-1}$ for signature $(1,5)$ given in appendix \ref{sec:A} by,
\begin{align}
\label{4.MB.6}
& \cM_{0,6} & B_{\eta(1)}^{-1}  &= \gamma _{(1)} \gamma ^0 B_{(1)}^{-1}  & \a_1 & = 1
\no \\
& \cM_{1,5} & B_{\eta(1)}^{-1}  &= B_{(1)}^{-1} & \a_1 & = 0
\no \\
& \cM_{2,4} & B_{\eta(1)}^{-1}  &= \gamma _{(1)} \gamma ^1 B_{(1)}^{-1} & \a_1 & = 1
\no \\
& \cM_{3,3} & B_{\eta(1)}^{-1}  &= \gamma ^1 \gamma^2  B_{(1)}^{-1} & \a_1 & = 0
\no\\
& \cM_{4,2} & B_{\eta(1)}^{-1}  &= \gamma _{(1)} \gamma ^1 \gamma ^2 \gamma ^3 B_{(1)}^{-1}  & \a_1 & = 1
\no \\
& \cM_{5,1} & B_{\eta(1)}^{-1}  &= \g^1 \g^2 \g^3 \g^4 B_{(1)}^{-1} & \a_1 & = 0
\end{align}
where $\a_1$ is defined in (\ref{4.alpha}) below. The matrices $B_{\eta (2)}^{-1} $ are given as follows in terms of the complex conjugation matrix $B_{(2)}^{-1}$ for signature $(0,2)$ given in appendix \ref{sec:A}, 
\begin{align}
\label{4.MB.2}
& \cM_{0,2} & B_{\eta (2)}^{-1}  &= B_{(2)}^{-1} = \sigma ^2  & \a_2 & = 1
\no \\
& \cM_{1,1} & B_{\eta(2)}^{-1}  &= \gamma_{(2)} \gamma ^7 B_{(2)}^{-1} = \sigma ^3   & \a_2 & = 0
\no \\
& \cM_{2,0} & B_{\eta(2)}^{-1}  &= \gamma ^6 \gamma ^7 B_{(2)}^{-1} = \sigma ^1  & \a_2 & = 1
\end{align}
where $\a_2$ is defined in (\ref{4.alpha}), and $B_{\eta(3)}^{-1}$ are given by, 
\begin{align}
\label{4.MB.3}
& \Sigma_{0,2} & B_{\eta (3)}^{-1}  &= B_{(3)}^{-1} = \sigma ^2   & \a_3 & = 1
\no \\
& \Sigma_{1,1} & B_{\eta(3)}^{-1}  &= \gamma_{(3)} \gamma ^9 B_{(3)}^{-1} = \sigma ^3  & \a_3 & = 0
\no \\
& \Sigma_{2,0} & B_{\eta(3)}^{-1}  &= i \gamma ^8 \gamma ^9 B_{(3)}^{-1} = i \sigma ^1  & \a_3 & = 1
\end{align}
where $\a_1,\a_2$ and $\a_3$ can take the values $0$ or $1$ determined as follows,
\bea
\label{4.alpha}
\gamma _{(1)} B_{\eta(1)} & = & (-)^{\a_1} B_{\eta(1)} \g_{(1)} 
\no \\
\gamma _{(2)} B_{\eta(2)} & = & (-)^{\a_2} B_{\eta(2)} \g_{(2)} 
\no \\
\gamma _{(3)} B_{\eta(3)} & = & (-)^{\a_3} B_{\eta(3)} \g_{(3)} 
\eea
The conjugation matrices $B_{\eta(1)}, B_{\eta(2)}$ and $B_{\eta(3)}$ depend on the signatures of $\cM_6$, $\cM_2$ and $\Sigma$, as is indicated by their subscripts $\eta$, but do not depend on the values of $k_1^2$ and $ k_2^2$.  

\sm

The complex conjugation matrix $\cB_\eta$, which was defined (up to an arbitrary phase factor) by (\ref{2.B.1}) for a frame  metric $\eta$ of arbitrary signature, and prominently figures in the reality condition (\ref{4.B.ep}), may be constructed from the matrices $B_{\eta(1)}, B_{\eta(2)}$ and $B_{\eta(3)}$. The matrix $\cB_\eta$ is given by the tensor product, 
\bea
\cB_\eta = \tilde B_{\eta(1)} \otimes \tilde B_{\eta(2)} \otimes \tilde B_{\eta(3)}
\eea
where $\tilde B_{\eta (k)}$ may be either $B_{\eta (k)}$ itself or the product thereof by the corresponding chirality matrix $B_{\eta (k)} \gamma_{(k)}$ for $k=1,2,3$. To determine which option is realized, we enforce the definition (\ref{2.B.1}) for each group of indices $A=m,i,a$. For $A=m$, we have $\Gamma _m = \gamma _m \otimes I_2 \otimes I_2$, so that $\tilde B_{\eta (1)} = B_{\eta (1)}$.  For $A=i$, we have $\Gamma_i = \gamma_{(1)} \otimes \gamma _i \otimes I_2$ which requires, 
\bea
\cB_\eta (\hat e^i \Gamma _i) \cB_\eta ^{-1} = B_{\eta(1)} \gamma _{(1)} B_{\eta (1)}^{-1} \otimes
\tilde B_{\eta(2)} (\hat e^i \gamma _i)  \tilde B_{\eta (2)}^{-1} \otimes I_2= \gamma_{(1)} \otimes (\hat e^i \gamma_i)^* \otimes I_2
\eea
Using the first line of (\ref{4.alpha}) and extracting the second factor in the tensor product, 
\bea
\tilde B_{\eta(2)} (\hat e^i \gamma _i)  \tilde B_{\eta (2)}^{-1} = (-)^{\a_1} (\hat e^i \gamma_i)^*
= (-)^{1-\a_1} B_{\eta(2)} (\hat e^i \gamma _i)  B_{\eta (2)}^{-1}
\eea
we see that this condition is solved by $\tilde B_{\eta (2)} = B_{\eta(2)} \gamma _{(2)}^{\a_1+1}$. Proceeding analogously for $A=a$ using  the second line in (\ref{4.alpha}), we obtain $\tilde B_{\eta (3)} = B_{\eta(3)} \gamma _{(3)}^{\a_1+\a_2+1}$ and the final result,
\bea
\cB_\eta = 
B_{\eta(1)} \otimes \left ( B_{\eta(2)} \, \gamma _{(2)}^{\a_1+1} \right )  \otimes 
\left ( B_{\eta(3)} \, \gamma _{(3)}^{\a_1+\a_2+1} \right )
\eea
Swapping the orders of the matrices $B_{\eta (k)}$ and $\gamma _{(k)}$ is immaterial as its effect is to multiply  $\cB_\eta$ by a factor of $\pm 1$. Compatibility of the reality condition with the chiral projection of any one of the real forms of Type~IIB$_\CC$ requires $[\cB_\eta, \Gamma_{11} ]=0$ which in turn implies,
\bea
\a_1+\a_2+\a_3 \equiv 0 ~ ({\rm mod} \, 2)
\eea
It will be convenient to use $\a_2$ and $\a_3$ as  independent variables, and to eliminate $\gamma_{(3)}$ using the chirality condition to obtain an equivalent expression for $\cB_\eta$, 
\bea
\cB_\eta =  \left ( B_{\eta (1)} \, \gamma_{(1)}^{1+\a_3} \right ) \otimes \left ( B_{\eta(2)} \, \gamma ^{\a_2} _{(2)} \right ) 
\otimes B_{\eta (3)} 
\eea

\subsubsection{Evaluation of the  matrices $\mb_\eta$}

The complex conjugation properties of the Killing spinors $\chi^{\eta_1, \eta_2}$ may be read off from their definition in (\ref{3.b.1}), by contracting the equations with $\hat e^m$ and $\hat e^i$, respectively, and then using the conjugation properties of (\ref{4.Bstar}).\footnote{The vanishing torsion conditions on the first line of (\ref{3.torcur})  imply the complex conjugation property $( \hat \om^{mn} \gamma_{mn})^* = B_\eta ( \hat \om^{mn} \gamma_{mn}) B_\eta^{-1} $. }
Inspection of the Killing spinor equations shows that $(\cB_{\eta (1)}^{-1} \otimes \cB_{\eta(2)}^{-1}) (\chi^{\eta_1 , \eta_2})^*$ satisfies the Killing spinors equations for $\chi^{k_1^2 \eta_1, - k_2^2 \eta_2}$, so that these spinors may be chosen to be proportional to one another, as shown in \cite{DHoker:2016ujz}. Normalizing the proportionality by the conditions,

\bea
\left ( \cB_{\eta (1)}^{-1} \otimes \cB_{\eta(2)}^{-1} \right ) (\chi^{+,+})^* = \chi^{k_1^2, - k_2^2}
\eea
and using the action of the chirality matrices, 
\bea
\big ( \gamma _{(1)} \otimes I_2 \big ) \chi^{\eta_1, \eta_2} & = & \chi^{- \eta_1, \eta_2}
\no \\
\big ( I_8 \otimes \gamma _{(2)} \big ) \chi^{\eta_1, \eta_2} & = & \chi ^{ \eta_1, - \eta_2}
\eea
we find,
\bea
\left ( \cB_{\eta (1)}^{-1} \otimes \cB_{\eta(2)}^{-1} \right ) (\chi^{\eta_1, \eta_2})^* 
= \eta _1 ^{\a_1} \, \eta _2 ^{\a_2} \, \chi^{k_1^2 \eta_1 , - k_2^2 \eta_2}
\eea
where $\a_1, \a_2$ were defined in (\ref{4.alpha}). Using these relations, we obtain,
\bea
\cB_\eta ^{-1} \ep ^* = 
\sum_{\eta_1, \eta_2} \eta_1^{\a_2+\a_3} \, \eta _2 ^{\a_2} \, \gamma _{(1)}^{1+\a_3} \otimes \gamma _{(2)} ^{\a_2} 
\, \chi ^{k_1^2 \eta _1, - k_2^2 \eta_2} \otimes B_{\eta (3)}^{-1} \, \zeta^*_{\eta_1, \eta_2}
\eea
Treating the cases for the four different values of $(\a_2,\a_3)$ individually, and making a suitable change of variables $(\pm k_1^2 \eta_1, \pm k_2^2 \eta_2) \to (\eta_1,  \eta_2)$, we obtain the following expressions, 
\begin{align}
(\a_2, \a_3) & = (0,0) &
\cB_\eta ^{-1} \ep ^* & = 
\sum_{\eta_1, \eta_2}  \chi ^{\eta _1, \eta_2} \otimes B_{\eta (3)}^{-1} \, \zeta^*_{-k_1^2 \eta_1, - k_2^2 \eta_2}
\no \\
(\a_2, \a_3) & = (0,1) &
\cB_\eta ^{-1} \ep ^* & = 
\sum_{\eta_1, \eta_2}  \chi ^{\eta _1, \eta_2} \otimes (k_1^2 \eta_1) \,  B_{\eta (3)}^{-1} \, \zeta^*_{k_1^2 \eta_1, - k_2^2 \eta_2}
\no \\
(\a_2, \a_3) & = (1,0) &
\cB_\eta ^{-1} \ep ^* & = 
\sum_{\eta_1, \eta_2}  
\, \chi ^{\eta _1,  \eta_2} \otimes (- k_1 ^2 \, \eta_1 \, k_2^2 \,  \eta _2 ) \, B_{\eta (3)}^{-1} \, \zeta^*_{-k_1^2 \eta_1, k_2^2 \eta_2}
\no \\
(\a_2, \a_3) & = (1,1) &
\cB_\eta ^{-1} \ep ^* & = 
\sum_{\eta_1, \eta_2}  
\, \chi ^{\eta _1,  \eta_2} \otimes (k_2^2 \, \eta _2 ) \, B_{\eta (3)}^{-1} \, \zeta^*_{k_1^2 \eta_1, k_2^2 \eta_2}
\end{align}
We may represent the matrix elements by the $\tau$ matrices introduced in (\ref{3.d.1}). In particular, the dependence on the parameters $k_1^2, k_2^2$ may be represented by the action to the left of the following matrix assignment, 
\begin{align}
\label{eq:K1-hat-def}
	\hat K_1&=\begin{cases}
		\tau^{(00)} \otimes I_2 ~~ \hbox{ for } & (k_1,k_2)=(1,i) \\
		\tau^{(01)} \otimes I_2 ~~ \hbox{ for } & (k_1,k_2)=(1,1) \\
		\tau^{(10)}\otimes I_2  ~~ \hbox{ for } & (k_1,k_2)=(i,i) \\
		\tau^{(11)} \otimes I_2 ~~ \hbox{ for } & (k_1,k_2)=(i,1)
	\end{cases}
\end{align}
Using the $\tau$-matrix and $K_1$-matrix notation, we obtain the following expressions for $\mb_\eta^{-1} $,
\begin{align}\label{eq:mb-alpha}
(\a_2, \a_3) & = (0,0) & \mb_\eta^{-1} & =  \hat K_1 \, \tau^{(10)} \otimes \sigma ^3
\no \\
(\a_2, \a_3) & = (0,1) & \mb_\eta^{-1} & =  - i \hat K_1 \, \tau^{(30)} \otimes B_{\eta (3)}^{-1} 
\no \\
(\a_2, \a_3) & = (1,0) & \mb_\eta^{-1} & = - \hat K_1 \, \tau^{(22)} \otimes \sigma^3 
\no \\
(\a_2, \a_3) & = (1,1) & \mb_\eta^{-1} & =  \hat K_1 \, \tau^{(02)}  \otimes B_{\eta (3)}^{-1} 
\end{align}
Recall from (\ref{4.MB.2}) and (\ref{4.MB.3}) that $\cM_{0,2}$ and $\cM_{2,0}$ correspond to $\a_2=1$ while $\cM_{1,1}$ has $\a_2=0$. Similarly, $\Sigma_{0,2}$ and $\Sigma _{2,0}$ correspond to $\a_3=1$ while $\Sigma_{1,1}$ has $\a_3=0$.
Then these results are consistent with the factorized form of $\mb_\eta^{-1} $ in (\ref{eq:mb-Ki}) with (\ref{eq:K1-def}), (\ref{eq:K23-def}).

\subsection{Consistency conditions}
\label{sec:4.5}

The reality condition (\ref{4.B.ep}) on the  spinor $\ep$ for each real form of Type~IIB$_\CC$ supergravity translates into a reduced reality condition on the reduced spinor $\zeta$, given in (\ref{4.4.a}),
\bea
\label{4.5.a}
\cK \zeta ^* = \zeta 
\hskip 0.7in 
\cK = S^{-1}  \otimes \mb_\eta^{-1}  
\hskip 0.7in 
\cK^* = \cK
\eea
In the previous section, we had already established that $\mb_\eta^{-1}$ is real and commutes with the chirality operator $\cI$. Since $S$ is also real and manifestly commutes with $\cI$, it follows that $[\cI, \cK]=0$. The remaining conditions are the projection relation $\cJ \zeta = \nu \zeta$, where $\cJ = i \, s^2  \otimes \tau^{(32)} \otimes I_2$, and the self-consistency of the reality condition (\ref{4.5.a}). They  impose the following conditions (using the fact that $\cJ^2=I_{16}$), 
\bea
\cJ \cK \cJ \zeta =   \cK \zeta   \hskip 1in   
\cK^* \cK\zeta=\cK^2 \zeta & = & \zeta
\eea
We now analyze these conditions in turn, using the expressions (\ref{eq:mb-Ki}), (\ref{eq:K1-def}), (\ref{eq:K23-def}), and then take their intersection.

\subsubsection{Consistency with the $\cJ$-projection}

From the expressions (\ref{eq:K1-def}) and (\ref{eq:K23-def}) we readily evaluate the following relations, 
\begin{align}
\cJ S \cJ &=S \cdot \begin{cases} +1 & \text{IIB$_\RR$ \hbox{ or } IIB$_3$} \\ -1 & \text{IIB$^\star$ \hbox{ or } IIB$^\prime$}\end{cases}
\no \\
\cJ K_1 \cJ & =K_1 \cdot \begin{cases} +1 & (\mathds{k}_1,\mathds{k}_2)\in \{(1,i),(i,1)\} \\ -1 & (\mathds{k}_1,\mathds{k}_2)\in\{(1,1),(i,i)\}\end{cases}
\no \\
\cJ K_2 \cJ &=K_2 
\no \\
\cJ K_3 \cJ &=K_3 \cdot
\left\{\begin{array}{rl} +1 & \Sigma_{2,0} \hbox{ or } \Sigma_{0,2} \\ -1 & \Sigma_{1,1}\end{array}\right.
\end{align} 
Combining these observations, we arrive at the following options for $\cK$ to be compatible with $\cJ$,
\begin{align}\label{eq:cJ-cK-comm-res}
	\text{IIB$_\RR$, IIB$_3$}:
	&& (\mathds{k}_1,\mathds{k}_2)&\in \{(1,i),(i,1)\} &&\{\Sigma_{2,0} \hbox{ or } \Sigma_{0,2}\}\times\{\cM_{0,2} \hbox{ or } \cM_{1,1}\}
	\nonumber\\
	&& & &&\Sigma_{1,1}\times\cM_{2,0}
	\nonumber\\
	&&(\mathds{k}_1,\mathds{k}_2)&\in\{(1,1),(i,i)\} &&\Sigma_{1,1}\times\{\cM_{0,2} \hbox{ or } \cM_{1,1}\}
	\nonumber\\
	&& & &&\{\Sigma_{2,0} \hbox{ or } \Sigma_{0,2}\}\times \cM_{2,0}
	\nonumber\\[1mm]
	\text{IIB$^\star$, IIB$^\prime$}:&&
	 (\mathds{k}_1,\mathds{k}_2)&\in\{(1,1),(i,i)\} &&\{\Sigma_{2,0} \hbox{ or } \Sigma_{0,2}\}\times\{\cM_{0,2} \hbox{ or } \cM_{1,1}\}
	\nonumber\\
	&& & &&\Sigma_{1,1}\times \cM_{2,0}
	\nonumber\\
	&& (\mathds{k}_1,\mathds{k}_2)&\in \{(1,i),(i,1)\} &&\Sigma_{1,1}\times\{\cM_{0,2} \hbox{ or } \cM_{1,1}\}
	\nonumber\\
	&& & &&\{\Sigma_{2,0} \hbox{ or } \Sigma_{0,2}\}\times \cM_{2,0}
\end{align}

\subsubsection{Self-consistency of complex conjugation}

Since $\cK$ is real, the self-consistency condition of the complex conjugation condition (\ref{4.5.a}) reduces to $\cK^2 \zeta=\zeta$, which may be computed  from $\cK^2 = S^{-2} \otimes \mb_\eta^{-2}$ with,
\bea
S^{-2} = \left\{\begin{array}{rl} +I_2 ~~ \hbox{ for } &  \text{IIB}_\RR, \text{IIB}^\star, \text{IIB}^\prime
				\\ -I_2 ~~ \hbox{ for }  & \text{IIB}_3\end{array}\right\}
\eea
The remaining $ \mb_\eta^{-2}$ can be determined straightforwardly from the explicit expressions given above in (\ref{eq:mb-Ki}), (\ref{eq:K1-def}), (\ref{eq:K23-def}). For $\Sigma_{0,2}$ or $\Sigma _{2,0}$ this leads to
\begin{align}
\cK^2 \zeta&= \zeta
	\left\{\begin{array}{rl} \mathds{k}_2^2~, & \mathcal M_{0,2}, \mathcal M_{2,0} \\ \mathds{k}_1^2~,  & \mathcal M_{1,1}
	\end{array}\right\}
	\left\{\begin{array}{rl} 
		-1~, & \Sigma_{0,2}
		\\
		1~, & \Sigma_{2,0}
			\end{array}\right\}
	\left\{\begin{array}{rl} +1~,  & \text{IIB}_\RR, \text{IIB}^\star, \text{IIB}^\prime
				\\ -1~, & \text{IIB}_3\end{array}\right\}			
\end{align}
and for $\Sigma _{1,1}$, 
\begin{align}
	\cK^2 \zeta&= \zeta 
	\left\{\begin{array}{rl} 
	- \mathds{k}_1^2 \mathds{k}_2^2~, & \{\cM_{0,2}\text{\ or\ }\cM_{2,0}\} \times \Sigma_{1,1} \\ 
	1~,  & \cM_{1,1} \times \Sigma_{1,1}\end{array}\right\}
	\left\{\begin{array}{rl} +1~,  & \text{IIB}_\RR, \text{IIB}^\star, \text{IIB}^\prime
				\\ -1~, & \text{IIB}_3\end{array}\right\}
\end{align}

\subsection{Real forms of $F(4)$ solutions}\label{sec:5-summary}

In this section we summarize all real forms of the solutions which are compatible with the consistency conditions in the real forms of Type IIB$_\CC$. These have signatures $(1,9)$, $(3,7)$, $(5,5)$, $(7,3)$, $(9,1)$. The geometries of the real forms with signature $(t,10-t)$ are related to the geometries of real forms with signature $(10-t,t)$ by an overall signature reversal. This leads to physically distinct solutions with different causal structures.\footnote{A change from mostly-plus to mostly-minus signature convention for the metric would entail a corresponding change of signs in the definition of spacelike and timelike vectors. Here these definitions remain unchanged  and are identical for solutions with signatures $(t,10-t)$ and $(10-t,t)$.} 

\sm

The data determining viability of a real form are $\mathds{k}_{1/2}$ and the signatures of $\cM_2$ and $\Sigma$. 
We also give explicit geometric realizations in which $\cM_6$ and $\cM_2$ are substituted by concrete symmetric spaces.
To set this up, we first discuss the symmetric spaces and then give the classification of real forms of the $F(4)$ solutions in the real forms of Type IIB$_\CC$.

\subsubsection{Symmetric spaces}

The symmetric spaces we will use are $S^n$, $H_n$ and $AdS_{r,s}$, $dS_{r,s}$. In the following we denote $r+s=n$. These spaces can be realized as embeddings into $\RR^{n+1}$ with coordinates $x^i$ for $i=1,\cdots, n+1$, and flat metric $\mu$,
\bea
\label{eq:embeddings}
S^n & = & \{ \mu_{ij}  x^i x^j =+1;~ ds^2 = \mu_{ij} dx^i dx^j; ~ \mu= (+_{n+1} )  \} 
\no \\[0.3mm]
H_n & = & \{\mu_{ij}  x^i x^j =-1;~ ds^2 = \mu_{ij} dx^i dx^j; ~ \mu = (-_1, \,  +_n)  \} 
\no \\[0.3mm]
AdS_{r,s} & = & \{ \mu_{ij}  x^i x^j =- 1;~ ds^2 = \mu_{ij} dx^i dx^j; ~ \mu = (-_{r+1}, \,  +_s )  \} 
\no \\[0.3mm]
dS_{r,s} & = & \{ \mu_{ij}  x^i x^j =+1;~ ds^2 = \mu_{ij} dx^i dx^j; ~ \mu = (-_{r}, \,  +_{s+1})  \}  
\eea
where $+_r$ and $-_r$ denotes $r$ diagonal entries with values $+1$ and $-1$, respectively. 
They naturally arise as real slices of the quadric (\ref{eq:quad}) in $\CC^{n+1}$.

\sm

Clearly, $dS_{0,n}=S^n$ and $AdS_{0,n}=H^n$. 
We further note that the subspaces in $\RR^{n+1}$ defined by the constraints are identical as point sets for $AdS_{p,q}$ and $dS_{q,p}$.
The ambient metrics on $\RR^n$, however, differ by an overall minus sign. As a result, the induced metrics on the hypersurfaces are identical up to a sign. We denote this geometric relation by,
\begin{align}\label{eq:ds2-sym-rel}
	-dS_{p,q}&=AdS_{q,p}
	&
	-AdS_{p,q}&=dS_{q,p}
\end{align}
Similar relations were discussed e.g.\ in \cite{2016JGP...104..163P,2020arXiv200200810B}.

\subsubsection{Real forms in IIB$_\RR$, IIB$^\star$/IIB$^\prime$ and IIB$_3$}

We start the discussion of real forms of the solutions with the Type IIB$_\CC$ real forms IIB$_\RR$, IIB$^\star$/IIB$^\prime$ and IIB$_3$, corresponding, respectively, to $t=1$ and $t=3$ timelike directions.
In addition to the consistency constraints discussed in the previous subsections, which constrain the signatures of $\cM_2$ and $\Sigma$, the number of time-like directions in $\mathcal M_2\times\Sigma$ should not exceed one for IIB$_\RR$, IIB$^\star$, IIB$^\prime$ and three for IIB$_3$.

\sm
The solutions to the consistency conditions we then find, based on the classification of real forms of the reduced BPS equations, are given in Table \ref{eq:real-forms-reduced-BPS}.
\begin{table}[h]
	\centering
$
	\begin{array}{cc|c|c|c|c|c}
		\toprule
\mathds{k}_1 & \mathds{k}_2 & \, f_2^2\mathcal M_2 \, & \quad \Sigma \quad & \text{real form} & \text{geometry} & \text{bosonic subalgebra} \\
	\midrule\midrule
1 & i & (0,2) & {(0,2)} & \text{IIB}_\RR & AdS_{1,5}\times S^2\times\Sigma_{0,2} & 
	\ms \mo (2,5)   \oplus   \ms \mo (3) 
\\
i & 1 & (1,1) & (0,2) & \text{IIB}_\RR & S^6\times AdS_{1,1}\times\Sigma_{0,2} &
	\ms \mo (7)  \oplus \ms \mo (1,2) \\
\midrule
i & i & (0,2) & (0,2) & \text{IIB}^\star\text{, IIB}^\prime & dS_{1,5}\times S^2\times\Sigma_{0,2} &
	\ms \mo (1,6)  \oplus  \ms \mo (3)  
\\
1 & i & (0,2) & (1,1) & \text{IIB}^\star\text{, IIB}^\prime & H^6\times S^2\times\Sigma_{1,1} &
	\ms \mo (1,6)  \oplus  \ms \mo (3) 
\\
i & 1 & (0,2) & (1,1) & \text{IIB}^\star\text{, IIB}^\prime & S^6\times H^2\times\Sigma_{1,1} &
	\ms \mo (7)   \oplus  \ms \mo (1,2) 
\\
i & i & (1,1) & (0,2) & \text{IIB}^\star\text{, IIB}^\prime & S^6\times dS_{1,1}\times\Sigma_{0,2} & 
	\ms \mo (7) \oplus \ms \mo (1,2) \\
\midrule
i & 1 & (0,2) & (0,2) & \text{IIB}_3 & dS_{3,3}\times H^2\times \Sigma_{0,2} &
	\ms \mo (3,4)  \oplus \ms \mo (1,2)  
\\
1 & 1 & (0,2) & (1,1) & \text{IIB}_3  & AdS_{2,4}\times H^2\times\Sigma_{1,1} &
	\ms \mo (3,4)  \oplus \ms \mo (1,2) 
\\
	i & i & (0,2) & (1,1) & \text{IIB}_3 & dS_{2,4}\times S^2\times\Sigma_{1,1} &
	\ms \mo (2,5)   \oplus  \ms \mo (3) 
\\
1 & i & (0,2) & (2,0) & \text{IIB}_3 & AdS_{1,5}\times S^2\times \Sigma_{2,0}& 
	\ms \mo (2,5)   \oplus  \ms \mo (3)  
\\
1 & i & (1,1) & (0,2) & \text{IIB}_3 &AdS_{2,4}\times dS_{1,1}\times\Sigma_{0,2} &
	\ms \mo (3,4)  \oplus \ms \mo (1,2) 
\\
i & 1 & (1,1) & (2,0) & \text{IIB}_3 & S^6\times AdS_{1,1}\times\Sigma_{2,0} &
	\ms \mo (7) \oplus \ms \mo (1,2)
\\
i & 1 & (2,0) & (0,2) & \text{IIB}_3 & dS_{1,5}\times (-S^2)\times\Sigma_{0,2} &
	\ms \mo (1,6)  \oplus  \ms \mo (3)
\\
1 & 1 & (2,0) & (1,1) & \text{IIB}_3 & H^6\times (-S^2)\times\Sigma_{1,1} &
	\ms \mo (1,6)  \oplus  \ms \mo (3) 
\\
i & i & (2,0) & (1,1) & \text{IIB}_3 & S^6\times (-H^2)\times \Sigma_{1,1} &
	\ms \mo (7) \oplus \ms \mo (1,2)
\\
		\bottomrule
\end{array} 
$
\caption{Real forms of $F(4)$ solutions in Type~IIB$_\RR$, IIB$^\star$/IIB$^\prime$ and IIB$_3$. The first two columns encode the Killing spinor parameters $k_{1/2}$ and reality properties of $f_{6/2}$ through $\mathds{k}_{1/2}$ defined in (\ref{eq:ds-k12-def}). Columns 3 and 4 give the signatures of $f_2^2ds^2_{\cM_2}$, denoted as $f_2^2\mathcal M_2$, and of $\Sigma$. In the last two columns we give a geometric realization and the corresponding maximal bosonic subgroup of the Lie superalgebra $F(4;\CC)$, where here and in the subsequent tables  $\ms \mo (p,q)= \ms\mo(p,q; \RR)$ is implied.
	\label{eq:real-forms-reduced-BPS}}
\end{table}

\sm

It is reassuring that the symmetries of the BPS solutions  in Type~IIB$_\RR$, IIB$^\star$, IIB$^\prime$ and IIB$_3$ agree with the bosonic subalgebras of the real forms of  $F(4)$ in (\ref{listmbsa}).
The only real forms in standard Type IIB$_\RR$ are the $AdS_6\times S^2$ and $AdS_2\times S^6$ cases discussed in \cite{DHoker:2016ujz,Corbino:2017tfl} before, for which globally regular solutions were constructed subsequently in \cite{DHoker:2017mds,DHoker:2017zwj,Corbino:2018fwb}. 
We emphasize that Table \ref{eq:real-forms-reduced-BPS} gives consistent real forms of the reduced BPS equations, for which we provided the general local solution; which of them produce globally viable solutions is a separate question which we plan to come back to in future work.

\sm

The last 3 rows in Table \ref{eq:real-forms-reduced-BPS} involve timelike $f_2^2ds^2_{\cM_2}$ of signature $(2,0)$. The geometric realization for these cases warrants some discussion. 
For the realizations given in the table we chose $f_2$ imaginary with real reduced frame $\hat e^i$ in (\ref{3.a.2}).
This means the signature of the $(6,7)$ components of the 10d metric in (\ref{3.a.1}), i.e.\ of $f_2^2ds^2_{\cM_2}$, is opposite to the signature of $ds^2_{\cM_2}$.
This is indicated by the minus signs in the third column: if e.g.\ $\cM_2=S^2$ with $f_2^2<0$, we denote the space as $(-S^2)$.
It also means that $k_2^2$, which encodes the curvature of $\cM_2$, is given by $k_1^2=-\mathds{k}_2^2$;
this dictates the choices of $\cM_2$ in these three cases.

\sm

An equivalent realization can be obtained by taking $f_2$ real with imaginary frame components $\hat e^i$.
This leads to $k_1=+\mathds{k}_2$,\footnote{The change in $k_2$ can also be understood in the reduced Killing spinor equation (\ref{3.b.1}): the reduced frame enters in $\hat\nabla_i=\hat e_i^M \nabla_M$, and changing from real to imaginary $\hat e^i$ amounts to a redefinition of $k_2$.} which results in the replacements
\begin{align}\label{eq:H2S2rep}
	\cM_2=S^2&\rightarrow \cM_2=AdS_{2,0}
	&
	\cM_2=H^2&\rightarrow \cM_2=dS_{2,0}
\end{align}
in the last 3 rows of Table \ref{eq:real-forms-reduced-BPS}.
These replacements can be understood in terms of (\ref{eq:ds2-sym-rel}): they only change $ds^2_{\cM_2}$ by an overall sign.
This sign is compensated in the 10d metric (\ref{3.a.1}), which contains the combination $f_2^2ds^2_{\cM_2}$, by the fact that the replacements in (\ref{eq:H2S2rep}) are accompanied by a sign reversal in $f_2^2$ compared to the realization with imaginary $f_2$.

In summary, one may use the replacements (\ref{eq:ds2-sym-rel}) in the geometric realizations in Table \ref{eq:real-forms-reduced-BPS}, which yields an alternative notation for the same 10d metric.

\subsubsection{Real forms with $t=5,7,9$}

Extending the classification to theories with $t=5$ follows the same strategy. 
One difference lies in the constraints on the total number of timelike directions in $\cM_2\times\Sigma$, which are less stringent and allow new options with more time directions.

The results of analyzing the consistency conditions for the $t=5$ cases IIB$_5$ and  IIB$_5^\star$/IIB$_5^\prime$ are given in Table \ref{4.table.IIB5}, with
$\mathds{k}_{1/2}$ again defined in (\ref{eq:ds-k12-def}). The discussion of the cases with timelike $f_2^2\cM_{2,0}$ carries over unchanged from the $t=1,3$ real forms.

\begin{table}[h]
	\centering
$
	\begin{array}{cc|c|c|c|c|c}
		\toprule
\mathds{k}_1 & \mathds{k}_2 & \, f_2^2 \cM_2 \, & \quad \Sigma \quad & \text{real form} & \text{geometry} & \text{bosonic subalgebra} \\
		\midrule\midrule
1 & i & (0,2) & (0,2) & \text{IIB}_5 & AdS_{5,1}\times S^2\times\Sigma_{0,2} & 
	\ms \mo (1,6)  \oplus  \ms \mo (3)
	\\
i & 1 & (0,2) & (2,0) & \text{IIB}_5 & dS_{3,3 }\times H_2 \times\Sigma_{2,0} & 
\ms \mo (3,4)   \oplus   \ms \mo (1,2) 
\\
i & 1 & (1,1) & (0,2) & \text{IIB}_5 & dS_{4,2}\times AdS_{1,1}\times\Sigma_{0,2} &
	\ms \mo (3,4)  \oplus \ms \mo (1,2) 
\\
1 & i & (1,1) & (2,0) & \text{IIB}_5 & AdS_{2,4}\times dS_{1,1}\times\Sigma_{2,0} &
	\ms \mo (3,4)  \oplus \ms \mo (1,2) 
\\
	1 & 1 & (1,1) & (1,1) & \text{IIB}_5 & AdS_{3,3}\times AdS_{1,1} \times\Sigma_{1,1} & 
	\ms \mo (3,4)   \oplus   \ms \mo (1,2) 
	\\
	i & i & (1,1) & (1,1) & \text{IIB}_5 & dS_{3,3}\times dS_{1,1} \times\Sigma_{1,1} &
	\ms \mo (3,4)  \oplus \ms \mo (1,2) 
\\
1 & i & (2,0) & (0,2) & \text{IIB}_5 & AdS_{3,3}\times (-H_2) \times\Sigma_{0,2} & 
	\ms \mo (3,4)   \oplus   \ms \mo (1,2) 
\\
i & 1 & (2,0) & (2,0) & \text{IIB}_5 & dS_{1,5}\times (-S^2) \times\Sigma_{2,0} &
	 \ms \mo (1,6)  \oplus  \ms \mo (3)
\\		
\midrule
i & i & (0,2) & (0,2) & \text{IIB}^\star_5 \text{, IIB}^\prime_5 & dS_{5,1}\times S^2\times\Sigma_{0,2} &
	\ms \mo (2,5)  \oplus  \ms \mo (3)
\\
1 & i & (0,2) & (1,1) & \text{IIB}^\star _5 \text{, IIB}^\prime_5 & AdS_{4,2} \times S^2\times\Sigma_{1,1} &
	\ms \mo (2,5)  \oplus  \ms \mo (3) 
\\
i & 1 & (0,2) & (1,1) & \text{IIB}^\star _5 \text{, IIB}^\prime_5 & dS_{4,2}\times H^2\times\Sigma_{1,1} &
	\ms \mo (3,4)   \oplus  \ms \mo (1,2) 
\\
	1 & 1 & (0,2) & (2,0) & \text{IIB}^\star_5 \text{, IIB}^\prime_5 & AdS_{3,3}\times H^2\times\Sigma _{2,0} & \ms \mo (3,4)   \oplus \ms\mo(1,2)
\\
i & i & (1,1) & (0,2) & \text{IIB}^\star_5 \text{, IIB}^\prime_5 & dS_{4,2}\times dS_{1,1}\times\Sigma_{0,2} & 
	\ms \mo (3,4) \oplus \ms \mo (1,2) \\
	1 & 1 & (1,1) & (2,0) & \text{IIB}^\star_5 \text{, IIB}^\prime_5 & AdS_{2,4}\times AdS_{1,1}\times \Sigma _{2,0} & \ms \mo (3,4)   \oplus \ms\mo(1,2)\\
	1 & i & (1,1) & (1,1) & \text{IIB}^\star_5 \text{, IIB}^\prime_5 & AdS_{3,3}\times dS_{1,1}\times \Sigma _{1,1} & \ms \mo (3,4)   \oplus \ms\mo(1,2)\\
	i & 1 & (1,1) & (1,1) & \text{IIB}^\star_5 \text{, IIB}^\prime_5 & dS_{3,3}\times AdS_{1,1}\times \Sigma _{1,1} & \ms \mo (3,4)   \oplus \ms\mo(1,2)\\
	1 & i & (2,0) & (1,1) & \text{IIB}^\star_5 \text{, IIB}^\prime_5 & AdS_{2,4}\times (-H^2)\times \Sigma _{1,1} & \ms \mo (3,4)   \oplus \ms\mo(1,2)\\
	1 & 1 & (2,0) & (2,0) & \text{IIB}^\star_5 \text{, IIB}^\prime_5 & AdS_{1,5}\times (-S^2)\times\Sigma _{2,0} & \ms \mo (2,5)   \oplus \ms\mo(3)\\
	i & i & (2,0) & (0,2) & \text{IIB}^\star_5 \text{, IIB}^\prime_5 & dS_{3,3}\times (-H^2)\times\Sigma _{0,2} & \ms \mo (3,4)   \oplus \ms\mo(1,2)\\
	i & 1 & (2,0) & (1,1) & \text{IIB}^\star_5 \text{, IIB}^\prime_5 & dS_{2,4}\times (-S^2)\times \Sigma _{1,1} & \ms \mo (2,5)   \oplus \ms\mo(3)\\
		\bottomrule
\end{array} 
$

\caption{
Real forms of $F(4)$ solutions with 5 time-like directions, $t=5$, in Type~IIB$_5$, IIB$^\star_5$ and IIB$^\prime_5$.
\label{4.table.IIB5}}
\end{table}

\sm

The set of real forms of the solutions for each Type IIB real form with $t=5$ has a symmetry, which results from the fact that for $t=5$ we have $(t,10-5)=(10-t,t)$. For a given real form, an overall signature reversal in the metric yields another real form. As a result, the entries in Table \ref{4.table.IIB5} are related in pairs.
The overall signature reversal can be combined with (\ref{eq:ds2-sym-rel}), which leads to
\begin{align}
	\left(AdS_{p,q},dS_{p,q},\Sigma_{r,2-r}\right)&\longrightarrow 
	\left(-AdS_{p,q},-dS_{p,q},-\Sigma_{r,2-r}\right)
	=
	\left(dS_{q,p},AdS_{q,p},\Sigma_{2-r,r}\right)
\end{align}
where the first replacement reverses the signature and the equality follows from (\ref{eq:ds2-sym-rel}). The bosonic symmetry algebras are preserved along this chain.
Applying this transformation maps the Type~IIB$_5$ and IIB$^\star_5$/IIB$^\prime_5$ parts of Table \ref{4.table.IIB5} into themselves.
An example of related cases is the pair
\begin{align}
	\left(~AdS_{5,1}\times S^2 \times\Sigma _{2,0}~,~
	dS_{5,1}\times (-S^2) \times\Sigma _{0,2}\right)
\end{align}
Overall signature reversal combined with the relations in (\ref{eq:ds2-sym-rel}) maps the two geometries into each other. 
The causal structures differ, e.g.\ in that tangent vectors along $\Sigma$ are timelike in the first case and spacelike in the second.

\sm

\begin{table}
	\centering
$
	\begin{array}{cc|c|c|c|c|c}
		\toprule
		\mathds{k}_1 & \mathds{k}_2 & \, f^2\mathcal M_2 \, & \quad \Sigma \quad & \text{real form} & \text{geometry} & \text{bosonic subalgebra} \\
		\midrule
		\midrule
		1 & i & (0,2) & (2,0) & \text{IIB}_7 & AdS_{5,1}\times S^2 \times\Sigma _{2,0} & \ms \mo (1,6)  \oplus  \ms \mo (3)\\
		1 & 1 & (0,2) & (1,1) & \text{IIB}_7 &
		(-S^6)\times H^2 \times\Sigma _{1,1} & \ms \mo (7)  \oplus  \ms \mo (1,2)\\
		i & i & (0,2) & (1,1) & \text{IIB}_7 & (-H^6)\times S^2\times\Sigma _{1,1} & \ms \mo (1,6)  \oplus  \ms \mo (3)\\
		i & 1 & (1,1) & (2,0) & \text{IIB}_7 & dS_{4,2}\times AdS_{1,1}\times\Sigma _{2,0} & \ms \mo (3,4)  \oplus  \ms \mo (1,2)\\
		1 & i & (1,1) & (0,2) & \text{IIB}_7 & (-S^6)\times dS_{1,1}\times\Sigma _{0,2} & \ms \mo (7)  \oplus  \ms \mo (1,2)\\
		1 & i & (2,0) & (2,0) & \text{IIB}_7 & AdS_{3,3}\times(-H^2)\times\Sigma _{2,0} & \ms \mo (3,4)  \oplus  \ms \mo (1,2)\\
		1 & 1 & (2,0) & (1,1) & \text{IIB}_7 & AdS_{4,2}\times (-S^2)\times\Sigma _{1,1} & \ms \mo (5,2)  \oplus  \ms \mo (3)\\
		i & i & (2,0) & (1,1) & \text{IIB}_7 & dS_{4,2}\times(-H^2) \times\Sigma _{1,1} & \ms \mo (3,4)  \oplus  \ms \mo (1,2)\\
		i & 1 & (2,0) & (0,2) & \text{IIB}_7 & dS_{5,1}\times (-S^2) \times\Sigma _{0,2} &  \ms \mo (2,5)  \oplus  \ms \mo (3)
		\\
\midrule
		1 & i & (1,1) & (2,0) & \text{IIB}_9 & (-S^6)\times dS_{1,1}\times\Sigma_{2,0} &  \ms \mo (7)  \oplus  \ms \mo (1,2)\\
		i & 1 & (2,0) & (2,0) & \text{IIB}_9 & dS_{5,1}\times (-S^2)\times\Sigma_{2,0} & \ms \mo (2,5)  \oplus  \ms \mo (3)\\
\midrule
	1 & 1 & (1,1) & (2,0) & \text{IIB}^\star_9, \text{IIB}^\prime_9 & (-S^6)\times AdS_{1,1}\times \Sigma _{2,0} & \ms \mo (7)  \oplus  \ms \mo (1,2)\\
	1 & 1 & (2,0) & (1,1) & \text{IIB}^\star_9, \text{IIB}^\prime_9 & (-S^6)\times (-H^2)\times\Sigma _{1,1} & \ms \mo (7)  \oplus  \ms \mo (1,2)\\
	1 & i & (2,0) & (2,0) & \text{IIB}^\star_9, \text{IIB}^\prime_9 & AdS_{5,1}\times(-S^2)\times \Sigma _{2,0} &  \ms \mo (1,6)  \oplus  \ms \mo (3)\\
	i & i & (2,0) & (1,1) & \text{IIB}^\star_9, \text{IIB}^\prime_9 & (-H^6)\times(-S^2)\times\Sigma _{1,1} & \ms \mo (1,6)  \oplus  \ms \mo (3)\\
	\bottomrule
\end{array} 
$
\caption{Real forms of $F(4)$ solutions with $t=7,9$ in Type~IIB$_7$, IIB$_9$, IIB$^\star_9$/IIB$^\prime_9$.\label{4.table.IIB79}}
\end{table}

The remaining real forms are those with $t=7$ and $t=9$, in the Type~IIB$_7$, IIB$_9$ and IIB$_9^\star$/IIB$_9^\prime$ real forms of Type IIB$_\CC$.
These are given in Table \ref{4.table.IIB79}.
We again encounter cases with $f_2^2\cM_2$ of signature $(2,0)$, to which the discussion from the previous real forms carries over. We now also encounter cases where $f_6^2\cM_6$ has signature $(6,0)$. The discussion is fully analogous: one may use the relations (\ref{eq:ds2-sym-rel}) without changing the 10d metric.

\sm

At the level of the geometries, the real forms in Table~\ref{4.table.IIB79} are related to those for $t=1$ and $t=3$ in Table~\ref{eq:real-forms-reduced-BPS} by an overall sign reversal combined with the relations (\ref{eq:ds2-sym-rel}).
This in particular includes the partners of the known $AdS_{1,5}\times S^2\times\Sigma$ and $S^6\times AdS_{1,1}\times\Sigma$ solutions in standard Type IIB.

\clearpage

%%%%%%%%%%%%%%%%%%%%%%%%%%%%%%%%%%%%%%%%%%%
%%%%%%%%%%%%%%%%%%%%%%%%%%%%%%%%%%%%%%%%%%%
\section{Real forms of $F(4)$ solutions: examples}
\label{sec:6}
%%%%%%%%%%%%%%%%%%%%%%%%%%%%%%%%%%%%%%%%%%%
%%%%%%%%%%%%%%%%%%%%%%%%%%%%%%%%%%%%%%%%%%%

The consistent real forms of the general complex solutions were classified in section~\ref{sec:5} and tabulated in Tables \ref{eq:real-forms-reduced-BPS}, \ref{4.table.IIB5}, \ref{4.table.IIB79}.
In this section we determine the explicit conjugation relations for each case in Table \ref{eq:real-forms-reduced-BPS}. 
We list the conjugation  relations  for the two-dimensional spinor components $\alpha,\beta,\gamma, \delta$, the holomorphic $\kappa_{1,2}, \tilde \kappa_{1,2}$, and the  data $(w,\tilde w,\cA_{1,2},\tilde \cA_{1,2})$.
For a selection of cases, the supergravity fields will be given explicitly, based on the general complex solution to Type~IIB$_\CC$ supergravity obtained in section \ref{sec:3}.
The remaining cases in Table~\ref{eq:real-forms-reduced-BPS} and the remaining Tables~\ref{4.table.IIB5}, \ref{4.table.IIB79} can be discussed along the same lines.

\subsection{Realization by holomorphic data}
\label{sec:real-forms-hol-data}

Table \ref{eq:real-forms-reduced-BPS} gives the combinations of $\mathds{k}_1$, $\mathds{k}_2$,  the signatures of $\cM_6, \cM_2$ (in combination with $f_6^2$ and $f_2^2$) and $\Sigma$, and the real form of Type IIB$_\CC$ for which a complex conjugation symmetry exists. For those options, the reality condition $\cK \zeta^* =\zeta$ is consistent with the chirality condition $\cI \zeta = \zeta$ and the projection  $\cJ \zeta = \nu \zeta$. 
For these cases, the reality conditions for the 10d spinors can be solved in terms of conjugation relations between the spinor components $\alpha$, $\beta$, $\gamma$, $\delta$ defined in (\ref{eq:zeta-comp}).
These relations are given in Table \ref{eq:real-forms-spinor-rel}.
We note that, although Type~IIB$^\star$ and Type~IIB$^\prime$ appear undistinguished in Table \ref{eq:real-forms-reduced-BPS}, they have different conjugation relations and are therefore distinguished in Table \ref{eq:real-forms-spinor-rel}.

\begin{table}[ht]
	\centering
$
\begin{array}{c|c|c|c}
\toprule
	\text{geometry} & \text{real form} & \{\alpha,\beta,\gamma,\delta\}^\star & 
		\Lambda_\rho\{\tilde 	\kappa_1,\kappa_1,\tilde \kappa_2, \kappa_2\}^\star\\
\midrule\midrule
	AdS_{1,5}\times S^2\times\Sigma _{0,2} & \text{IIB}_\RR & 
		\{-\delta  \nu ,\gamma  \nu ,\beta  \nu ,-\alpha  \nu \} & 
		\{-\kappa_1,-\tilde \kappa_1,-\kappa_2,-\tilde \kappa_2\} \\
	S^6\times AdS_{1,1}\times\Sigma _{0,2} & \text{IIB}_\RR & 
		\{\beta ,\alpha ,\delta ,\gamma \} & 
		\{+\kappa_1,+\tilde \kappa_1,+\kappa_2,+\tilde \kappa_2\} 
		\\[0.5mm]
\midrule
	dS_{1,5}\times S^2\times\Sigma _{0,2} & \text{IIB}^\star & 
		\{-\beta ,-\alpha ,\delta ,\gamma \} & 
		\{+\kappa_1,+\tilde \kappa_1,-\kappa_2,-\tilde \kappa_2 \} \\
		& \text{IIB}^\prime & \{-\delta ,-\gamma ,-\beta ,-\alpha \} & 
		\{-\kappa_1,-\tilde \kappa_1,+\kappa_2,+\tilde \kappa_2 \} \\
	H^6\times S^2\times\Sigma _{1,1} & \text{IIB}^\star & 
		\{\alpha ,\beta ,-\gamma ,-\delta \} & 
		\{+\tilde \kappa_1,+\kappa_1,-\tilde \kappa_2,-\kappa_2\} \\
		 & \text{IIB}^\prime & \{\gamma ,\delta ,\alpha ,\beta \} & 
		 \{-\tilde \kappa_1,-\kappa_1,+\tilde \kappa_2,+\kappa_2\} \\
	S^6\times H^2\times\Sigma _{1,1} & \text{IIB}^\star & 
		\{\alpha ,-\beta ,-\gamma ,\delta \} & 
		\{+\tilde \kappa_1,+\kappa_1,-\tilde \kappa_2,-\kappa_2\} \\
		& \text{IIB}^\prime & \{\gamma ,-\delta ,\alpha ,-\beta \} & 
		\{-\tilde \kappa_1,-\kappa_1,+\tilde \kappa_2,+\kappa_2\} \\
	S^6\times dS_{1,1}\times \Sigma _{0,2} & \text{IIB}^\star & 
		\{-\delta  \nu ,-\gamma  \nu ,-\beta  \nu ,-\alpha  \nu \} & 
		\{-\kappa_1,-\tilde \kappa_1,+\kappa_2,+\tilde \kappa_2 \} \\
		& \text{IIB}^\prime & \{\beta  \nu ,\alpha  \nu ,-\delta  \nu ,-\gamma  \nu \} & 
		\{+\kappa_1,+\tilde \kappa_1,-\kappa_2,-\tilde \kappa_2 \} 
		\\[0.5mm]
\midrule
	dS_{3,3}\times H^2\times \Sigma _{0,2} & \text{IIB}_3 & 
		\{\beta  \nu ,\alpha  \nu ,\delta  \nu ,\gamma  \nu \} & 
		\{+\kappa_1,+\tilde \kappa_1,+\kappa_2,+\tilde \kappa_2\} \\
	AdS_{2,4}\times H^2 \times\Sigma _{1,1} & \text{IIB}_3 & 
		\{-\alpha  \nu ,-\beta  \nu ,-\gamma  \nu ,-\delta  \nu \} & 
		\{+\tilde \kappa_1,+\kappa_1,+\tilde \kappa_2,+\kappa_2\} \\
	dS_{2,4}\times S^2\times\Sigma _{1,1} & \text{IIB}_3 & 
		\{-\alpha  \nu ,\beta  \nu ,-\gamma  \nu ,\delta  \nu \} & 
		\{+\tilde \kappa_1,+\kappa_1,+\tilde \kappa_2,+\kappa_2\} \\
	AdS_{1,5}\times S^2\times \Sigma _{2,0} & \text{IIB}_3 & 
		\{\beta  \nu ,\alpha  \nu ,\delta  \nu ,\gamma  \nu \} & 
		\{+\kappa_1,+\tilde \kappa_1,+\kappa_2,+\tilde \kappa_2\} \\
	AdS_{2,4}\times dS_{1,1}\times\Sigma _{0,2} & \text{IIB}_3 & 
		\{\delta ,-\gamma ,-\beta ,\alpha \} & 
		\{-\kappa_1,-\tilde \kappa_1,-\kappa_2,-\tilde \kappa_2\} \\
	S^6\times AdS_{1,1}\times\Sigma _{2,0} & \text{IIB}_3 & 
		\{\delta ,-\gamma ,-\beta ,\alpha \} & 
		\{-\kappa_1,-\tilde \kappa_1,-\kappa_2,-\tilde \kappa_2\} \\
	dS_{1,5}\times (-S^2)\times\Sigma _{0,2} & \text{IIB}_3 & 
		\{\beta  \nu ,\alpha  \nu ,\delta  \nu ,\gamma  \nu \} & 
		\{+\kappa_1,+\tilde \kappa_1,+\kappa_2,+\tilde \kappa_2\} \\
	H^6\times(-S^2)\times\Sigma _{1,1} & \text{IIB}_3 & 
		\{-\alpha  \nu ,-\beta  \nu ,-\gamma  \nu ,-\delta  \nu \} & 
		\{+\tilde \kappa_1,+\kappa_1,+\tilde \kappa_2,+\kappa_2\} \\
	S^6\times (-H^2)\times\Sigma _{1,1} & \text{IIB}_3 & 
		\{-\alpha  \nu ,\beta  \nu ,-\gamma  \nu ,\delta  \nu \} & 
		\{+\tilde \kappa_1,+\kappa_1,+\tilde \kappa_2,+\kappa_2\} \\
	\bottomrule
	\end{array}
$

\caption{Realization of the real forms summarized in Table~\ref{eq:real-forms-reduced-BPS} in terms of conjugation relations for the Killing spinor components $\alpha$, $\beta$, $\gamma$, $\delta$ and the holomorphic data $\kappa_{1,2}$, $\tilde\kappa_{1,2}$.\label{eq:real-forms-spinor-rel}}
\end{table}

\sm

The conjugation conditions for $\alpha$, $\beta$, $\gamma$, $\delta$ can further be expressed in terms of relations among $\kappa_{1,2},\tilde\kappa_{1,2}$ defined in (\ref{3.bd}), (\ref{3.bd2}). This depends on the reality properties of $\rho$, i.e.\ whether $\rho^2$ is positive or negative, which we do not fix at this point.
We only assume
\begin{align}
	\rho^\star &=\Lambda_\rho \rho
\end{align}
with $\Lambda_\rho=+1$ for real $\rho$ with $\rho^2>0$, and $\Lambda_\rho=-1$ for imaginary $\rho$ with $\rho^2<0$.
We further assume that the dilaton is real and $\chi$ satisfies the reality condition for the appropriate real form of Type IIB$_\CC$.
The resulting relations between $\kappa_{1,2}$, $\tilde\kappa_{1,2}$ are also given in Table~\ref{eq:real-forms-spinor-rel}. 
Once the conjugation relations among $\kappa_{1,2}$, $\tilde\kappa_{1,2}$ are fixed, they can be realized through a choice of relation between $w$, $\tilde w$ and relations among $\cA_{1,2}$, $\tilde\cA_{1,2}$ in (\ref{3.Akappa}).

\sm

The reality properties of $\rho$ can be used to realize the signature on $\Sigma$:
Euclidean $\Sigma_{0,2}$ can be realized by $w^\star=\tilde w$ with $\Lambda_\rho=1$, while $\Sigma_{2,0}$ with two timelike directions can be realized by $w^\star=\tilde w$ with $\Lambda_\rho=-1$.
For $\Sigma_{1,1}$, one of $e^8$ and $e^9$ is real and the other imaginary. This means $e^z$ and $e^{\tilde z}$ in (\ref{eq:e-z-def}) are both real or both imaginary. This translates, via (\ref{3.conf}), to real and independent $w$, $\tilde w$, which in turn can be interpreted as light cone coordinates.

In Table~\ref{eq:real-forms-spinor-rel} the cases involving $\Sigma_{1,1}$ are distinguished by conjugation relations which relate each of $\kappa_{1,2}$, $\tilde\kappa_{1,2}$ to itself (up to a sign) upon complex conjugation. That is, they are individually real or imaginary.
Looking at (\ref{3.Akappa}), these relations are indeed solved by taking $w$, $\tilde w$ real and independent, with $\cA_{1,2}$ and $\tilde\cA_{1,2}$ real functions of the respective variables.

\subsection{Recovering Type IIB$_\RR$ solutions}

As consistency check we recover the supergravity fields for the two cases in standard Type IIB, discussed previously in \cite{DHoker:2016ujz,Corbino:2017tfl}: $AdS_{1,5}\times S^2\times\Sigma_{0,2}$ and $S^6\times AdS_{1,1}\times\Sigma_{0,2}$. 
To provide a blueprint for other cases we will be explicit and detailed.
To streamline the notation we drop the explicit subscripts for the signature of the individual spaces in this subsection.

Both cases have $\Sigma$ with signature $(0,2)$, which can be realized by real $\rho$ with $w^\star=\tilde w$ and  $\Lambda_\rho=1$. This determines the relation between $\tilde\kappa_1$ and $\kappa_2$.
Both cases also need real $f_6$; through the expression for $f_6$ in (\ref{3.f6}) combined with the conjugation relations among $\alpha$, $\beta$, $\gamma$, $\delta$ in Table~\ref{eq:real-forms-spinor-rel} this fixes the reality properties of $c_6$. We find
\begin{align}\label{AdS26-real-1}
	AdS_6\times S^2\times\Sigma:&&  (\alpha^\star,\beta^\star)&=\nu(-\delta,\gamma)
	&
	\kappa_{1/2}^\star&=-\tilde\kappa_{1/2}
	&
	c_6&\in\RR
	\nonumber\\
	S^6\times AdS_{2}\times\Sigma:&&  (\alpha^\star,\gamma^\star)&=(\beta,\delta)
	&
	\kappa_{1/2}^\star&=+\tilde\kappa_{1/2}
	&
	c_6&\in i\,\RR
\end{align}
The constant $c_2$ is then fixed by (\ref{3.k1k2}),
where $k_2/k_1$ is imaginary for both cases. This makes $f_2$ in (\ref{3.f6}) real for both cases. 
From the conjugation relations between $\kappa_{1,2}$, $\tilde\kappa_{1,2}$ we conclude that $\epsilon_\kappa$, $\tilde\epsilon_\kappa$ defined in (\ref{eq:Lambda-def}) satisfy $\epsilon_\kappa^\star = \mp\tilde\epsilon_\kappa$.
Being sign variables, this means 
\begin{align}\label{AdS26-eps}
	\epsilon_\kappa&=\mp\tilde\epsilon_\kappa
	&
	\varepsilon_\kappa&=\mp 1
\end{align}
where $\varepsilon_\kappa$ was defined in (\ref{4.Ztau1}) and the upper/lower signs are for $AdS_6\times S^6$ / $S^6\times AdS_2$. 
With these relations the equations for $\tau_\pm$ and $\rho$ in (\ref{4.sub.8}), (\ref{4.sub.9}) decompose into two sets of complex conjugate equations.
The conjugation relations for $\kappa_{1,2}$, $\tilde\kappa_{1,2}$ can be summarized as $\kappa_{1/2}^\star=\varepsilon_\kappa\tilde\kappa_{1/2}$ and
are solved in terms of $(\cA_{1,2},\tilde\cA_{1,2})$ introduced in (\ref{3.Akappa}) by
\begin{align}\label{AdS26-real-2}
	(\cA_{1})^\star&= -\varepsilon_\kappa\tilde\cA_{1}
	&
	(\cA_{2})^\star&= -\varepsilon_\kappa\tilde\cA_{2}
\end{align}
This makes $\kappa^2$ and $\cG$ in (\ref{kappa2-G-def}) real, leading to real $T^2$, $Y$ and $R^4$ while $X$ is a phase.

\sm

For the two-form fields, the signature of $\cM_2$ plays a crucial role, and we start the discussion at a similarly general level.
The dilatino equations in (\ref{3.c.1}) (or (\ref{4.sub.3})) determine $h_{z,\tilde z}$ and $g_{z,\tilde z}$ in terms of the remaining fields and the Killing spinor components as
\begin{align}\label{eq:g-h-sol}
	h_z&= \frac{2 i (2 \beta  \delta \phi_z+e^\phi\chi_z (\beta^2 -\delta^2 ))}{\nu  \left(\beta ^2+\delta ^2\right)}
	&
	h_{\tilde z}&=\frac{2 i (2 \alpha  \gamma  \phi_{\tilde z}+e^\phi\chi_{\tilde z} (\alpha^2 -\gamma^2 ) )}{\nu  \left(\alpha ^2+\gamma ^2\right)}
	\nonumber\\
	g_z&=\frac{2 i \left((\delta^2-\beta ^2) e^{-\phi}\phi_z+2 \beta  \delta \chi_z\right)}{\nu  \left(\beta ^2+\delta ^2\right)}
	&
	g_{\tilde z}&=\frac{2 i \left((\gamma^2-\alpha ^2)e^{-\phi}\phi_{\tilde z}+2 \alpha  \gamma  \chi_{\tilde z}\right)}{\nu  \left(\alpha ^2+\gamma ^2\right)}
\end{align}
From the conjugation relations among $\alpha$, $\beta$, $\gamma$, $\delta$ we conclude
\begin{align}
	(h_z)^\star&=\pm h_{\tilde z} & (g_z)^\star&=\pm g_{\tilde z}
\end{align}
with plus for $AdS_6\times S^2$ and minus for $S^6\times AdS_2$.
That is, $h_ae^a$ and $g_a e^a$ are real for $AdS_6\times S^2$ and imaginary for $S^6\times AdS_2$.
This leads to real $H_{(3)}$ and $\tilde F_{(3)}$ via (\ref{3.a.3}) in both cases:
For $S^6\times AdS_{2}\times\Sigma$, the time direction in the metric is in $\mathcal M_2$ and misaligned with the Clifford algebra where $\Gamma_0$ is timelike.
This mismatch has to be compensated by imaginary frame components, one in $\cM_6$ and one in $\cM_2$. This in particular means  $e^{67}$ in (\ref{3.a.3}) is imaginary, so that the combinations $H_{(3)}$ and $\tilde F_{(3)}$  are real also for $S^6\times AdS_{2}\times\Sigma$.

\subsubsection{Explicit expressions}

We further compare the explicit expressions for the supergravity fields to the $AdS_6\times S^2$ and $AdS_2\times S^6$ solutions of \cite{DHoker:2016ujz,Corbino:2017tfl}.
The Einstein-frame metric functions were given in combined form in \cite[(5.2),(5.4),(5.6)]{Corbino:2017tfl}.
Where confusion could arise we distinguish quantities defined there by a hat.
Then the results of \cite{Corbino:2017tfl} can be written as
\begin{align}\label{eq:metric-def-recall}
	\hat f_2^2&=\frac{c^2\hat\kappa^2}{9\hat\rho^2}\frac{1}{\hat T}
	&
	\hat f_6^2&=\frac{c^2\hat\kappa^2}{\hat\rho^2}\hat T
	&
	\hat \rho^2&=c\,\hat \kappa^2\sqrt{\frac{\hat T}{6\Lambda\hat\cG}}
\end{align}
where $\Lambda=+1$ for $AdS_6\times S^2$ and $\Lambda=-1$ for $AdS_2\times S^6$, and
\begin{align}\label{eq:LambdaR-def-recall}
	\hat\cG&=|\cA_+|^2-|\cA_-|^2+\hat\cB+\hat\cB^\star & \partial_w\hat\cB&=\cA_+\partial_w\cA_--\cA_-\partial_w\cA_+
	\no\\
	\hat\kappa^2&=-|\partial_w\cA_+|^2+|\partial_w\cA_-|^2&
	\hat T&=\frac{1+\Lambda \hat R}{1-\Lambda \hat R}=\sqrt{1+\frac{2|\partial_w\hat\cG|^2}{3\hat\kappa^2\hat\cG}}
\end{align}
The axio-dilaton and two-form are were given in \cite[(5.9),(5.15)]{Corbino:2017tfl} as
\begin{align}\label{eq:B-cC-recall}
	B&=\frac{1+i\hat\tau}{1-i\hat\tau}=\frac{\partial_w\cA_+\partial_{\bar w}\hat\cG-\Lambda \hat R \partial_{\bar w}\bar\cA_-\partial_w\hat\cG}{\Lambda \hat R\partial_{\bar w}\bar\cA_+\partial_w\hat\cG-\partial_w\cA_-\partial_{\bar w}\hat\cG}
	\nonumber\\
	\cC&=\frac{4ic}{9}\Lambda\left[\frac{\partial_{\bar w}\bar\cA_-\partial_w\hat\cG}{\hat\kappa^2}-2\Lambda \hat R\frac{\partial_{\bar w}\hat\cG\partial_w\cA_++\partial_w\hat\cG\partial_{\bar w}\bar\cA_-}{(1+\Lambda \hat R)^2\hat\kappa^2}-\bar\cA_--2\cA_+\right]
\end{align}
Redefining $\tilde R=\Lambda \hat R$ absorbs all factors of $\Lambda$ except in the square root in $\hat\rho^2$ and as an overall factor in $\cC$ (where it can be compensated by an $SL(2,\RR)$ transformation).

\sm

We will recover these expressions from the general complex solution given in section~\ref{sec:sol-summary} with the reality conditions in (\ref{AdS26-real-1}), (\ref{AdS26-real-2}). To this end, we identify $\cA_{1/2}$ and $\cA_\pm$ as follows,
\begin{align}
	\cA_1&=\sqrt{-\varepsilon_\kappa}\left(a_1 \cA_+ - a_1^\star\cA_-\right)
	&
	\cA_2&=\sqrt{-\varepsilon_\kappa}\left(a_2 \cA_+ - a_2^\star\cA_-\right)
	\nonumber\\
	\tilde\cA_1&=\sqrt{-\varepsilon_\kappa}\left(a_1^\star \cA_+^\star - a_1\cA_-^\star\right)
	&
	\tilde\cA_2&=\sqrt{-\varepsilon_\kappa}\left(a_2^\star \cA_+^\star - a_2\cA_-^\star\right)
\end{align}
where $a_1$, $a_2$ are complex constants.
Then the expressions for $\kappa^2$ and $\cG$ in (\ref{kappa2-G-def}) become
\begin{align}
	\kappa^2&=-\varepsilon_\kappa i(a_1^\star a_2-a_1a_2^\star)\hat\kappa^2
	&
	\cG&=-\varepsilon_\kappa i(a_1^\star a_2-a_1a_2^\star)\hat\cG
\end{align}
This matches, up to an overall factor, the previous definitions for $AdS_6\times S^2$ and $AdS_2\times S^6$ denoted here as $\hat\kappa$ and $\hat G$.
Moreover, the composite quantities $T^2$ and $\hat T^2$ match (cf.\ (\ref{eq:T-def})).

\sm

The axion-dilaton combinations $\tau_\pm$ in (\ref{eq:tau-pm-sum}) are complex conjugate to each other if $T$ is real. With the above replacements $\tau_+$ becomes
\begin{align}
	\tau_+&=\frac{\partial_{\tilde w}\cG  (T-1)\left(a_2\partial_w\cA_+-a_2^\star \partial_w\cA_-\right)+\partial_w\cG  (T+1) (a_2^\star \partial_{\tilde w}\tilde\cA_+-a_2\partial_{\tilde w}\tilde\cA_-)}{\partial_{\tilde w}\cG  (T-1)
		\left(a_1  \partial_w\cA_+-a_1^\star\partial_w\cA_-\right)+\partial_w\cG  (T+1) (a_1^\star\partial_{\tilde w}\tilde\cA_+-a_1\partial_{\tilde w}\tilde\cA_-)}
\end{align}
Matching this to $\hat\tau$ in (\ref{eq:B-cC-recall}) leads to
\begin{align}
	T&=-\hat T & a_1&\in\RR &a_2&=-ia_1
\end{align}
where we used that both $T$ and $\hat T$ are only defined up to a sign for the first identification. These relations lead to $\kappa^2=-2a_1^2\varepsilon_\kappa \hat\kappa^2$ and $\cG=-2a_1^2\varepsilon_\kappa \cG$.

\sm

We now turn to the metric.
For both $AdS_6\times S^2$ and $S^6\times AdS_2$ we have $k_1^2/k_2^2=-1$ and $c_2^2=-c_6^2/9$ from (\ref{3.k1k2}), and in both cases $c_6/k_1$ is real. 
With the above identifications the metric functions in (\ref{3.Eins-2}) become
\begin{align}
	(\rho^\text{E})^2 & = \frac{2a_1 c_6}{k_1}\varepsilon_\kappa\hat\kappa^2\sqrt{\frac{\hat T}{6(-\varepsilon_\kappa)\hat G}}
	&
	(f_6^E)^2 & =  -\frac{4\varepsilon_\kappa c_6^2a_1^2}{(\rho^\text{E})^2}\frac{\hat\kappa^2}{\hat T^{\eta}}
	&(f_2^E)^2 & =  \frac{4 c_2^2a_1^2\varepsilon_\kappa}{9(\rho^\text{E})^2} \hat\kappa^2\hat T^{\eta}
\end{align}
It will be convenient to define a real constant $\tilde c$ as
\begin{align}
	\tilde c&=\frac{2a_1 \varepsilon_\kappa c_6}{k_1}
\end{align}
and note that $k_1^2=-\varepsilon_\kappa$. The metric functions then take the form
\begin{align}
	(\rho^\text{E})^2 & = \tilde c\hat\kappa^2\sqrt{\frac{\hat T}{6(-\varepsilon_\kappa)\hat G}}
	&
	(f_6^E)^2 & =  \frac{\tilde c^2}{(\rho^\text{E})^2}\frac{\hat\kappa^2}{\hat T^{\eta}}
	&(f_2^E)^2 & =  \frac{\tilde c^2\hat\kappa^2}{9(\rho^\text{E})^2} \hat T^{\eta}
\end{align}
Noting the relation $\Lambda=-\varepsilon_\kappa$ which follows from the definitions, and taking $\eta=-1$, we find a match with (\ref{eq:metric-def-recall}) including the non-trivial dependence on $\Lambda=-\varepsilon_\kappa$.

\sm

Recovering the $AdS_6\times S^2$ and $S^6\times AdS_2$ solutions provides a useful consistency check. We note that recovering $\eta=-1$ is in line with the comments in footnote \ref{foot:eta} and with the earlier discussions of $AdS_6\times S^2$ and $S^6\times AdS_2$ real forms.

\subsection{$dS_{1,5}\times S^2\times\Sigma$ in Type~IIB$^\star$, Type~IIB$^\prime$}
\label{sec:dS15-S2-Sigma}

Among the new real form solutions in Table~\ref{eq:real-forms-spinor-rel} we will briefly discuss the $dS_{1,5}\times S^2\times\Sigma$ solutions, which exist in Type IIB$^\star$ and Type IIB$^\prime$ with spacetime signature $(1,9)$ and realize the real form of $F(4)$ with $\ms\mo(1,6)\oplus \ms\mo(3)$ bosonic symmetry.

\sm

For these solutions $\Sigma$ has signature $(0,2)$, which can be realized by taking the metric function $\rho$ real, so that $\Lambda_\rho=1$, with the relation $w^\star=\tilde w$ among the complex coordinates on $\Sigma$.
This fixes the conjugation relations between $\kappa_{1,2}$, $\tilde\kappa_{1,2}$ in Table~\ref{eq:real-forms-spinor-rel}. Demanding $f_6$ in (\ref{3.f6}) to be real fixes the reality condition for $c_6$, and the condition for $c_2$ follows from (\ref{3.k1k2}) with $k_1=k_2=i$. This leads to the following conditions,
\begin{align}
	\text{IIB$^\star$}:&& (\alpha^\star,\gamma^\star)&=(-\beta,+\delta) &
	(\kappa_1,\kappa_2)^\star&=(+\tilde\kappa_1,-\tilde\kappa_2)
	&c_6&\in\RR
	&
	c_2&\in\RR
	\nonumber\\
	\text{IIB$^\prime$}:&& (\alpha^\star,\beta^\star)&=(-\delta,-\gamma) &
	(\kappa_1,\kappa_2)^\star&=(-\tilde\kappa_1,+\tilde\kappa_2)&
	c_6&\in i\RR
	&
	c_2&\in i\RR
\end{align}
With these relations $f_2$ is real, as desired.
From the conjugation relations between $\kappa_{1,2}$, $\tilde\kappa_{1,2}$ we conclude from (\ref{eq:Lambda-def}), (\ref{3.Akappa}), following the arguments that led to (\ref{AdS26-eps}), that
\begin{align}
	\epsilon_\kappa&=\pm \tilde\epsilon_\kappa
	&
	\varepsilon_\kappa&=\pm 1
	&
	\xi^\star&=- \tilde\xi
\end{align}
with the upper/lower signs for Type IIB$^\star$/IIB$^\prime$.
Type IIB$^\star$ and IIB$^\prime$ both have imaginary axion, so that the combinations $\tau_\pm$ in (\ref{4.sub.1}) are imaginary and independent. With the above relations the equations for $\tau_\pm$ and $\rho$ in (\ref{4.sub.8}), (\ref{4.sub.9}) then decompose into two sets of complex conjugate equations.
The relations between $\kappa_{1,2}$, $\tilde\kappa_{1,2}$ lift via (\ref{3.Akappa}) to 
\begin{align}
	\cA_{1}^\star&= \mp\tilde\cA_{1}
	&
	\cA_{2}^\star&= \pm\tilde\cA_{2}
\end{align}
This makes $\kappa^2$ and $\cG$ in (\ref{kappa2-G-def}) imaginary and $T^2$ real for both cases. If $T^2<0$, so that $T$ is imaginary, the explicit expressions for $\tau_\pm$ in (\ref{eq:tau-pm-sum}) indeed realize both combinations as imaginary. Moreover, with this choice $(f_6^E)^2$ and $(f_2^E)^2$ in (\ref{3.Eins-2}) are real if $(\rho^E)^2$ is. The remaining regularity requirements for the metric then amount to positivity conditions.

\sm

The reality properties of the 3-form fields $H_{(3)}$ and $\tilde F_{(3)}$ are determined by (\ref{eq:g-h-sol}) in combination with the expansions (\ref{3.a.3}).
From (\ref{eq:g-h-sol}),
\begin{align}
	(h_z)^\star&=\pm h_{\tilde z} & (g_z)^\star&=\mp g_{\tilde z}
\end{align}
with the upper signs for Type IIB$^\star$ and the lower for Type IIB$^\prime$. In Type IIB$^\star$, $H_{(3)}$ is real and $\tilde F_{(3)}$ imaginary, while in IIB$^\prime$ the roles are reversed. This is the expected behavior.

\sm

The solutions have a natural analytic continuation to $S^6\times S^2\times\Sigma$ solutions via
\begin{align}
	dS_{1,5}&\rightarrow S^6
\end{align}
This continuation retains the positive curvature of $\mathcal M_6$ and does not affect any fields other than the metric.
The resulting $S^6\times S^2\times\Sigma$ solutions have to be understood in complex Type IIB$_\CC$: changing the spacetime signature changes the charge conjugation matrix and spoils consistency of the spinor reality conditions in Euclidean signature. Relatedly, the continued solutions do not realize any of the real forms of $F(4)$ listed in (\ref{listmbsa}), which do not include a form with $\ms\mo(7)\times \ms\mo(3)$ symmetry.

\subsection{$SL(2,\RR)$ symmetry}\label{sec:SL2R}

The $SL(2,\CC)$ symmetry of the complex solutions to Type~IIB$_\CC$ supergravity is reduced to a subgroup by the reality conditions imposed on the differential forms $\kappa_1, \tilde \kappa_1, \kappa_2$ and $\tilde \kappa_2$ for the real forms of the solutions. The reality conditions on $\xi$ and $\tilde \xi$ reduce to four cases,
\begin{align}
\{ \xi , \tilde \xi \} ^* & =  \{ \tilde \xi,   \xi\}    & \text{IIB}_\RR, \text{IIB}_3 
	&& \Sigma _{(0,2)}, \Sigma_{(2,0)} && M^* & = \pm M
\no \\
\{ \xi , \tilde \xi \}^* & = \{ - \tilde \xi,  - \xi \}   & \text{IIB}^\star, \text{IIB}^\prime 
	&& \Sigma _{(0,2)}, \Sigma_{(2,0)} && M^* & = \pm \sigma ^3 M \sigma ^3
\no \\
\{ \xi, \tilde \xi \}^*  & =  \{ \xi,  \tilde \xi \}   & \text{IIB}_3 ~~~ 
	&& \Sigma _{(1,1)} ~~~ && M^* & = \pm M
\no \\
 \{ \xi, \tilde \xi \}^* & =  \{ - \xi, - \tilde \xi\}  &  \text{IIB}^\star, \text{IIB}^\prime 
	&& \Sigma _{(1,1)} ~~~ && M^* &  = \pm \sigma ^3 M \sigma ^3 
\end{align} 
In the last column of the list above, we have exhibited the restriction imposed on the general form of the $M \in SL(2,\CC)$ matrices of (\ref{3.AA}). In each one of these cases, the duality group $SL(2,\CC)$ is reduced to an $SL(2,\RR)$ subgroup.

\newpage

%%%%%%%%%%%%%%%%%%%%%%%%%%%%%%%%%%%%%%%%%%%
%%%%%%%%%%%%%%%%%%%%%%%%%%%%%%%%%%%%%%%%%%%
\section{Relation with polarized IKKT matrix model}
\setcounter{equation}{0}
\label{sec:7}
%%%%%%%%%%%%%%%%%%%%%%%%%%%%%%%%%%%%%%%%%%%
%%%%%%%%%%%%%%%%%%%%%%%%%%%%%%%%%%%%%%%%%%%

In this section we discuss the relevance of our work to the IKKT model and in particular recently proposed  holographic dualities for a mass deformed (or polarized)  IKKT model. 

\subsection{The (polarized) IKKT matrix model}

We start with the Lorentzian IKKT model \cite{Ishibashi:1996xs}, whose action and supersymmetry transformations can be obtained as reduction to zero dimensions of Lorentzian ten-dimensional $\cN=1$ super Yang-Mills theory or four-dimensional $\cN=4$ super Yang-Mills theory, both with gauge group $SU(N)$. This leads to an $SO(1,9)$ global symmetry.\footnote{There is no spacetime with Lorentzian causal structure in the Lorentzian IKKT matrix model. The $SO(1,9)$ symmetry is analogous to the $SO(1,5)$ R-symmetry of the Euclidean 4d $\mathcal N=4$ SYM theory obtained by dimensional reduction of Lorentzian 10d $\mathcal N=1$ SYM, discussed e.g.\ in \cite{Pestun:2007rz}.}

\sm

The degrees of freedom of the IKKT model are Hermitian $N \times N$ matrices $X_\mu$ and $\Psi _\a$ which transform in the adjoint representation under $SU(N)$.
Here   $X_\mu$ with $  \mu=1,\cdots,10$, transforms under the defining representation of $SO(1,9)$ while $\Psi _\a$ with $ \a=1,\cdots,32 $, transforms under the Dirac  spinor representation of $SO(1,9)$, subject to the Weyl condition $\Gamma ^{11} \Psi = \Psi $. 
The action of the IKKT model is given by,
\bea\label{IKKTa}
S_\text{IKKT} = \tr \left ( - { 1 \over 4} [X_\mu, X_\nu] [X^\mu , X^\nu] + \half \Psi _\a (\cC \G^\mu) _{\a \b} [ X_\mu , \Psi _\b] \right )
\eea
where $\cC$ is the charge conjugation matrix satisfying $(\Gamma ^\mu)^t = - \cC \Gamma ^\mu \cC^{-1}$ and $\cC= \cC^*=-\cC^t$ (see appendix \ref{sec:A}).
This action is invariant under 16 linearly realized supersymmetries  parameterized by $SO(1,9)$ spinors $\ep$ with components $\ep_\alpha$ and 16 non-linearly realized  supersymmetries parameterized by spinors $v$ with components  $v_\alpha$,
\bea
\label{Susya}
\delta X^\mu & = & 
- \Psi _\a (\cC \Gamma ^\mu)_{\a \b } \ep_\b 
\no \\
\delta \Psi _\a & = & 
\half \, \Gamma ^{\mu \nu} _{\a \b} [X_\mu, X_\nu] \ep_\b + v_\alpha {\bf 1}
\eea

\sm

The Euclidean IKKT model, in which the $SO(1,9)$ symmetry becomes $SO(10)$, can be defined by analytic continuation. This model has to be understood in a complexified setting (see e.g.\ the discussion in \cite{Hartnoll:2024csr}). It may be related to configurations of $N$ D-instantons in Type~IIB string theory and corresponding supergravity solutions \cite{Gibbons:1995vg,Bergshoeff:1998ry,Ciceri:2025maa}.

\sm

In analogy with the BFSS matrix quantum mechanics model \cite{Banks:1996vh}, which arises from D0 branes in Type IIA string theory and has been proposed as non-perturbative definition of M-theory, the IKKT model has been proposed to provide a non-perturbative definition of Type~IIB string theory. However, in the absence of a natural definition of energy and a clear holographic decoupling limit, this proposal is generally regarded as more speculative.

\subsubsection{The polarized IKKT model}

The polarized IKKT model \cite{Bonelli:2002mb} (see also \cite{Martina:2025kwc}) is a deformation of the IKKT model where the action (\ref{IKKTa}) is supplemented by the following term with a single free parameter $\Omega$,
\bea
\label{massdKKT}
S_\Omega = S_\text{IKKT} 
+ \tr \left ( { \Omega ^2\over 4^3} X^A X_A 
+  {3 \Omega ^2 \over 4^3} X_a X^a 
+ i \Omega  X_8 [X_9, X_{10}] 
+ i {\Omega \over 8} \Psi _\a (\cC \, \mN) _{\a \b} \Psi _\b \right )
\eea
where $A=1,\cdots, 7$ and $a = 8,9,10$ and $\mN = - \Gamma ^8 \Gamma ^9 \Gamma ^{10} $. 
The deformation breaks the nonlinearly realized  supersymmetries, but it preserves sixteen linearly realized supersymmetries, albeit with a modified transformation law,
\bea
\delta X^\mu & = & 
- \Psi _\a (\cC \Gamma ^\mu)_{\a \b } \ep_\b 
\no \\
\delta \Psi _\a & = & 
\half \, \Gamma ^{\mu \nu} _{\a \b} [X_\mu, X_\nu] \ep_\b 
+ i {\Omega\over 8} \big ( \Gamma ^\mu \, \mN + 2 \, \mN \, \Gamma ^\mu\big ) X_\mu \ep_\b
\eea
This deformation can be defined in the Lorentzian and Euclidean IKKT models.
In the Euclidean version it breaks the $SO(10)$ symmetry to $SO(7) \times SO(3)$.

\sm
Despite the challenges in applying holography to D-instantons and to the IKKT model as a pure matrix integral, recent work has led to intriguing progress for the polarized IKKT model. In particular, \cite{Hartnoll:2024csr} and \cite{Komatsu:2024bop,Komatsu:2024ydh} matched supersymmetric localization computations in the polarized IKKT model to results obtained from proposed holographic duals. For convergence reasons, these studies used the Euclidean IKKT model, where all kinetic terms are positive definite. This choice naturally has to be reflected on the gravity side.

\sm

In \cite{Hartnoll:2024csr} it was argued that the polarized IKKT matrix integral at large $\Omega$ is dominated by a saddle that corresponds holographically to a D1-brane embedded in a Euclidean Type~IIB background, which is a finite cavity with $\ms\mo(3)\oplus\ms\mo(7)$ symmetry supported by NS-NS three-form flux. The background and brane were shown to be supersymmetric and the D1-brane was argued to arise upon polarization of D-instantons in the flux background. 

\sm

The authors of  \cite{Komatsu:2024bop,Komatsu:2024ydh}, on the other hand, engineered fully backreacted Euclidean Type~IIB configurations with $\ms\mo(3)\oplus\ms\mo(7)$ symmetry and the desired brane charges. This was accomplished by analytically continuing the general local form of supersymmetric $AdS_6\times S^2\times\Sigma$ solutions to standard Type IIB given in \cite{DHoker:2016ujz}.
This analytic continuation operates at the level of the bosonic fields, with additional phases introduced in relations implied by supersymmetry in the standard Type IIB solutions so that the results could be argued to satisfy the equations of motion. The solutions were suggested to realize a (possibly real) form of $F(4)$ but the fermionic symmetries were not discussed in detail.

\subsection{Comments on holography for polarized IKKT}

We now outline how the general form of supersymmetric solutions realizing $F(4,\CC)$ in complex Type IIB$_\CC$ and the discussion of their real forms, presented in this paper, connect to holography for the polarized IKKT model.

\subsubsection{Euclidean and Lorentzian IKKT}

The first point we (re)emphasize is that there is no real form of $F(4;\CC)$ with bosonic subalgebra $\ms \mo (7;\RR) \oplus \ms \mo (3;\RR)$. This result  was obtained  in \cite{Nahm:1977tg,Parker:1980af} and is proven in appendix~\ref{sec:F} by elementary methods (see also footnote \ref{foot:2}). Correspondingly, there are no supersymmetric solutions with $S^6\times S^2\times\Sigma$ geometry in any real form of Type IIB$_\CC$.
This does not preclude the existence of solutions with $\ms \mo (7;\RR) \oplus \ms \mo (3;\RR)$ symmetry where the bosonic fields satisfy the reality conditions of Type IIB$^\star$ but Killing spinors, if they exist, do not. This clarifies the interpretation of the solutions of \cite{Komatsu:2024bop,Komatsu:2024ydh}.
It is in line with the discussion of \cite{Hartnoll:2024csr}, which noted that the supersymmetries of the $\ms \mo (7;\RR) \oplus \ms \mo (3;\RR)$ invariant cavity (and of saddles of the Euclidean polarized IKKT model) are all complex.

\sm

Our results establish, in particular, the general local form of supersymmetric $S^6\times S^2\times\Sigma$ solutions in Type IIB$_\CC$. This is expected to capture as special cases the cavity background of \cite{Hartnoll:2024csr} and generalizations, as well as the solutions of \cite{Komatsu:2024bop,Komatsu:2024ydh} if they are indeed supersymmetric. We leave a systematic investigation of regularity conditions and global completions of the local solutions found in this paper for future work. Establishing manifestly supersymmetric backreacted global solutions with symmetries matching those of the polarized IKKT model would strengthen the connection, and having the Killing spinors provides the basis for studying supersymmetric probes. The latter was used extensively in fleshing out the holographic interpretation of the related $AdS_6\times S^2\times\Sigma$ solutions (e.g.\  \cite{Bergman:2018hin,Uhlemann:2020bek,Gutperle:2020rty}) and may be leveraged similarly in IKKT holography.

\sm

Our results also provide a useful angle on $S^6\times S^2\times\Sigma$ solutions from the perspective of real forms: Table \ref{eq:real-forms-reduced-BPS} exhibits a real form of the complex solutions in Type IIB$^\star$ with $dS_{1,5}\times S^2\times\Sigma$ geometry, which realizes the real form of $F(4;\CC)$ with  $\ms\mo(1,6;\RR)\oplus\ms\mo (3;\RR)$ symmetry. This class of solutions, discussed in section \ref{sec:dS15-S2-Sigma}, has the correct symmetries to describe the Lorentzian polarized IKKT model with $SO(1,9)$ broken to $SO(1,6)\times SO(3)$. The Lorentzian IKKT model is of interest in its own right (e.g.\ \cite{Kim:2011cr,Nishimura:2022alt,Asano:2024def}). Moreover, the $dS_{1,5}\times S^2\times\Sigma$ solutions permit an analytic continuation to $S^6\times S^2\times\Sigma$, where $dS_{1,5}\rightarrow S^6$ retains the curvature of $\cM_6$ and does not affect any supergravity fields other than the metric. This provides a simpler path to $S^6\times S^2\times\Sigma$ than starting from $AdS_6\times S^2\times\Sigma$.

\subsubsection{Type IIB$_\CC$ holography for IKKT$_\CC$}

Given that the Euclidean IKKT model has to be understood in a complexified setting, it seems natural to make this aspect explicit and take it to its logical conclusion: we can study the fully complexified IKKT model, which we denote as IKKT$_\CC$, and general holographic connections to complex Type IIB$_\CC$ supergravity.

\sm

The IKKT model can be complexified to a holomorphic theory in a similar way as Type IIB supergravity. The fermionic variable $\Psi_\alpha$ is regarded as a complex Weyl spinor, but the action of the IKKT model can be formulated without involving  the  Hermitian conjugate  $\Psi_\alpha^\dagger$ -- the dependence on $\Psi_\alpha$ is said to be holomorphic. The same interpretation can be extended to  the bosonic variable  $X^\mu$. The complexified IKKT model may thus be defined as a multiple complex contour integral,  for which the Hermitian restriction amounts to a particular choice of contours for the bosonic variables. Where supersymmetry transformations do not preserve Hermiticity, they can be interpreted as deformation of the contours along which the integrals over $X_\mu$ are performed. This facilitates supersymmetric localization computations, e.g.\ for the original IKKT model in \cite{Moore:1998et} (see also \cite{Krauth:1998xh,Green:1998yf}) and for the deformed model in \cite{Hartnoll:2024csr,Komatsu:2024bop,Komatsu:2024ydh}.
It is also in line with the view advanced in \cite{Witten:2021nzp,Kontsevich:2021dmb}, interpreting the functional integral as contour integral over complex field configurations. 

\sm

The complexified polarized IKKT model, defined this way by contour integrals, has an $F(4;\CC)$-invariant action. Recent work has focused on saddles along a particular (Euclidean) contour and their matching to saddles in Euclidean Type IIB supergravity. One may extend this to studying the fully complexified polarized IKKT integral, e.g.\ the behavior along general contours and its complex analytic structure, and aim to connect this more broadly to complexified Type IIB$_\CC$ supergravity and string theory. By providing the general local form of complex Type IIB$_\CC$ solutions with $F(4,\CC)$ symmetry and a comprehensive list of their real forms, our work can be seen as a first step in this program. It would be interesting, for example, to explore what role, if any, solutions with exotic signatures which are invariant under sixteen contour-preserving supersymmetries have to play in connection with holography for the IKKT$_\CC$ model. More ambitiously, one may hope for new insights relating to a non-perturbative definition of Type IIB string theory.

\clearpage

\appendix

%%%%%%%%%%%%%%%%%%%%%%%%%%%%%%%%%%%%%%%%%%%
%%%%%%%%%%%%%%%%%%%%%%%%%%%%%%%%%%%%%%%%%%%
\section{Clifford algebra basis}
\setcounter{equation}{0}
\label{sec:A}
%%%%%%%%%%%%%%%%%%%%%%%%%%%%%%%%%%%%%%%%%%%
%%%%%%%%%%%%%%%%%%%%%%%%%%%%%%%%%%%%%%%%%%%

The Dirac matrices  with frame indices are denoted $\Gamma^A$ where $A=0,1, \cdots, 9$. They will satisfy the Clifford algebra $\{ \Gamma ^A , \Gamma ^B \} = 2 \eta ^{AB}$ where $\eta$ is the $SO(1,9)$-invariant frame metric with signature $(-+ \cdots +)$ regardless of the spacetime signature of the real form of Type~IIB$_\CC$ or the real form of its solutions.  We shall choose a basis for  $\Gamma ^A$ that is adapted to the factorization $\mathcal M_6\times \mathcal M_2  \times \Sigma$ Ansatz, and follow the conventions that were adopted in \cite{DHoker:2016ujz} for the spacetime $\rm AdS_6\times S^2\times\Sigma$, namely,
\begin{align}
\G^m & =  \gamma^m \otimes I_2 \otimes I_2  & m&=0,1,2,3,4,5
\notag \\
\G^{i} & =  \gamma_{(1)} \otimes \gamma^{i} \otimes I_2  & i&=6,7
\notag \\
\G^a \, & =  \gamma_{(1)}  \otimes \gamma_{(2)} \otimes  \gamma ^a & a&=8,9
\end{align}
where  a convenient basis for the lower dimensional Dirac-Clifford algebras is
\begin{align}
\gamma^0 & = -i \sigma ^2 \otimes I_2 \otimes I_2
\notag \\
\gamma^1 &= \sigma ^1 \otimes I_2 \otimes I_2
\notag \\
\gamma^2 &= \sigma ^3 \otimes \sigma ^2 \otimes I_2 
\hskip 1in \gamma^6 = \sigma^1
\notag \\
\gamma^3 &= \sigma ^3 \otimes \sigma ^1\otimes I_2
\hskip 1in \gamma^7 = \sigma^2
\notag \\
\gamma^4 &= \sigma ^3 \otimes \sigma ^3 \otimes \sigma ^1
\hskip 1in \gamma^8 = \sigma^1
\notag \\
\gamma^5 &= \sigma ^3 \otimes \sigma ^3\otimes  \sigma ^2
\hskip 1in \gamma^9 = \sigma^2
\end{align}
We will also need the chirality matrices on the various components of
$\mathcal M_6\times \mathcal M_2  \times \Sigma$, and they are chosen as follows,
\begin{align}
\gamma _{(1)} & =  \sigma ^3 \otimes \sigma ^3 \otimes \sigma ^3
&
\gamma _{(2)} & =  \sigma ^3
&
\gamma _{(3)} & =  \sigma ^3
\end{align}
so that,
\begin{align}
\Gamma ^{012345} & = - \gamma _{(1)} \otimes I_2 \otimes I_2
&
\Gamma ^{67} & = i \, I_6 \otimes \gamma _{(2)} \otimes I_2
\notag \\
&&
\Gamma ^{89} & = i \, I_6 \otimes I_2 \otimes \gamma _{(3)}
\end{align}
The 10-dimensional chirality matrix in this basis is given by,
\begin{align}\label{eq:Gamma-star}
\G^{11} = \G^{0123456789}   = \gamma_{(1)} \otimes \gamma_{(2)} \otimes \gamma_{(3)}=\sigma ^3\otimes \sigma ^3\otimes \sigma ^3\otimes \sigma ^3\otimes \sigma ^3
\end{align}
The charge conjugation matrix $\cC$ is defined by,
\bea
(\Gamma ^A )^t = - \cC \Gamma ^A \cC^{-1}
\eea
It may be given by the product over all anti-symmetric $\Gamma ^A$ matrices, $\cC= -i \Gamma ^0 \Gamma ^2 \Gamma^5 \Gamma ^7 \Gamma ^9$ or in tensor product form, 
\bea
\cC = i \sigma ^2 \otimes \sigma ^1 \otimes \sigma ^2 \otimes \sigma ^1 \otimes \sigma ^2
\eea
which satisfies,
\bea
\cC^* = \cC \hskip 0.8in \cC^t = - \cC \hskip 0.8in \cC^2=- I_{32} \hskip 0.8in \{ \cC, \Gamma ^{11}\}=0
\eea
Note that the matrix obtained as the product over all symmetric $\Gamma$-matrices is proportional to $\cC \Gamma ^{11}$ and satisfies $(\Gamma ^A)^t = (\cC \Gamma^{11}) \Gamma ^A (\cC \Gamma ^{11})^{-1}$. 
The complex conjugation matrices in each component are also defined as in \cite{DHoker:2016ujz} by
\begin{align}
\label{bmatdef}
\left ( \gamma^m \right ) ^* &= +B_{(1)} \gamma ^m B_{(1)} ^{-1}
&
  (B_{(1)})^* B_{(1)} &= - I_6 
& 
 B_{(1)} &= -i \gamma^2 \gamma ^5=I_2 \otimes \sigma_1 \otimes \sigma_2
\nonumber \\
\left ( \gamma^{i} \right ) ^* &= -B_{(2)} \gamma ^{i} B_{(2)} ^{-1}
&   (B_{(2)})^* B_{(2)} &= -I_2 
&
B_{(2)} &=  \gamma^7=\sigma^2
\nonumber \\
\left ( \gamma^a \right ) ^* &= - B_{(3)} \gamma ^a B_{(3)} ^{-1}
&   (B_{(3)})^* B_{(3)} &= - I_2 
&
 B_{(3)} &=  \gamma^9 =\sigma^2
\end{align}
The 10-dimensional complex conjugation matrix $\cB$ satisfies,
\begin{align}
( \Gamma^M)^* &= \cB \Gamma^M\cB^{-1}
&
\cB ^* \cB&=I_{32} 
&
[ \cB, \Gamma^{11}  ] &= 0
\end{align}
and in this basis is given by,
\begin{align}\label{eq:cB-def}
\cB &=  -i B_{(1)}\otimes \left ( B_{(2)} \gamma _{(2)} \right ) \otimes B_{(3)}   
 =   \, I_2 \otimes \sigma ^1 \otimes \sigma ^2 \otimes \sigma ^1 \otimes \sigma ^2
\end{align}
Complex conjugation and charge conjugation matrices are related by $\cB =  \cC \Gamma ^0=  \Gamma ^0 \cC$.

\clearpage

%%%%%%%%%%%%%%%%%%%%%%%%%%%%%%%%%%%%%%%%%%%
%%%%%%%%%%%%%%%%%%%%%%%%%%%%%%%%%%%%%%%%%%%
\section{Derivation of the BPS equations for Type IIB$_\CC$}
\setcounter{equation}{0}
\label{sec:B}
%%%%%%%%%%%%%%%%%%%%%%%%%%%%%%%%%%%%%%%%%%%
%%%%%%%%%%%%%%%%%%%%%%%%%%%%%%%%%%%%%%%%%%%

In this appendix, we present a derivation of the BPS equations for the complexified version of Type IIB supergravity directly from the BPS equations of standard Type IIB supergravity. Upon complexifying the spacetime metric the signature loses its intrinsic significance and may be chosen at will. We shall adopt the $(1,9)$ signature so that we can continue to use the customary basis for the Dirac matrices of standard Type IIB given, for example,  in \cite{DHoker:2007zhm,DHoker:2016ujz}, and summarized in appendix \ref{sec:A}. 

\subsection{Standard Type IIB supergravity}

The boson fields of standard Type IIB supergravity consist of  the metric $g_{MN}$, the one-forms $P$ and $Q$ representing the axion-dilaton, a three-form field strength $G$, and a self-dual five-form $F_{(5)}$ field strength. The fields $g_{MN}, Q$ and $F_{(5)}$ are real-valued, while $P$ and $G$ are complex-valued. They satisfy the following Bianchi identities,
\begin{align}
dP-2i Q\wedge P & = 0  & d G - i Q\wedge G +  P\wedge \bar G & = 0
\no \\
d Q + i P\wedge \bar P & = 0  &  d F_{(5)} -  \tfrac{i}{8} G \wedge \bar G & = 0
\label{bianchi1}
\end{align}
where the bar stands for complex conjugation.  The fermion fields are the dilatino $\lambda$ and the gravitino $\psi_M$, both of which are complex Weyl spinors with opposite chirality $\Gamma^{11} \lambda =- \lambda$, and $\Gamma^{11} \psi_M  =\psi_M$. The supersymmetry variations of the fermions, evaluated at vanishing fermion fields and expressed in the Einstein frame metric,  are given by,
\bea
\label{susy1}
\delta \psi_M
&=& \nabla  _M  \ep -{i \over 2} Q_M \ep 
+ {i\over 480}(\G \cdot F_{(5)})  \Gamma_M  \ep
-{1\over 96} \big  ( \Gamma_M (\G \cdot G)
+ 2 (\G \cdot G) \G_M \big  ) \cB^{-1} \ep^*
\no \\
\delta\lambda
&=& 
i (\G \cdot P) \cB^{-1} \ep^*
-{i\over 24} (\G \cdot G) \ep
\eea
where ${\cal B}$ is defined in (\ref{eq:cB-def}) and $\Gamma ^{11} \ep = \ep$.  The BPS equations are obtained by setting $\delta \lambda = \delta \psi _M=0$. The remaining field equations will not be needed here. 

\sm

The equations (\ref{bianchi1}) and (\ref{susy1})  are invariant under local $U(1)$ gauge transformations under which $g_{MN}$ and $F_{(5)}$ are invariant, while the fields $P, Q, G, \lambda, \psi_M$ and the parameter $\ep$  transform as follows for an arbitrary real-valued function $\theta$,
\begin{align}
\label{su11a}
P & \to  e^{2 i \theta} \, P &
\lambda & \to e^{3 i \theta/2} \, \lambda
\no \\
Q & \to  Q + d \theta &
\psi_M & \to e^{i \theta/2} \, \psi_M
\no \\
G & \to  e^{i \theta} \, G &
\ep & \to  e^{i \theta/2} \, \ep
\end{align}
Type IIB supergravity is also invariant under a global $SU(1,1) \sim SL(2,\RR)$  which leaves $g_{MN}$ and  $F_{(5)}$ invariant, and transforms $P, Q, G, \lambda, \psi_M, \ep$ by a $U(1)$ phase given in (\ref{su11a}) whose parameter $\theta$ depends on the fields and on the $SL(2,\RR)$ transformation.

\subsection{Complexifying}

Complexifying the real fields $g_{MN}$ and $F_{(5)}$ is achieved simply by promoting each real field component to a complex-valued function. However, additional considerations enter when implementing this process on $P, Q, G, \lambda, \psi, \ep$ since they  transform non-trivially under the local $U(1)$ gauge symmetry by complex phase factors. For example, the decomposition of the complex Weyl spinor dilatino $\lambda$ into the sum of two independent Majorana-Weyl spinors is not preserved under a $U(1)$ gauge transformation. 

\sm

To decompose $P, Q, G, \lambda, \psi, \ep$ into real boson and Majorana-Weyl fermion fields, we first need to fix a gauge for the local $U(1)$ symmetry and eliminate the unphysical degree of freedom associated with the field-dependent gauge transformation $\theta$ of (\ref{su11a}). To do so, we choose to solve the Bianchi identities of (\ref{bianchi1}) in terms of the real-valued axion $\chi$ and dilaton $\phi$ fields, and $G$ in terms of real-valued 3-form fields $H_{(3)}$ and $\tilde F_{(3)}$, 
\bea
\label{B.PQG}
P  =   \half d \phi + {i \over 2} e^{ \phi} \, d \chi    
\hskip 0.6in
Q  =  - \half e^{ \phi} \, d \chi  
\hskip 0.6in
G  =   e^{- \phi/2}  H_{(3)} + i \, e^{\phi/2} \tilde F_{(3)} 
\eea
where $\tilde F_{(3)} = F_{(3)} - \chi H_{(3)}$ and the Bianchi identities imply $d H_{(3)}=d F_{(3)}=0$. The BPS equations in this gauge assume exactly the same form as (\ref{susy1}) but with $P, Q, G$ expressed in terms of $\chi, \phi, H_{(3)}$ and $F_{(3)}$ as given by (\ref{B.PQG}).

\sm

Having made the above gauge choice in (\ref{B.PQG}), we decompose the complex Weyl spinors $ \ep , \lambda,  \psi$ into Majorana-Weyl spinors $\ep_\pm, \lambda _\pm$ and $\psi_\pm$, respectively, 
\begin{align}
\label{MW1}
\ep & = \ep_+ + i \ep_- &
\cB^{-1}  \ep ^* & = \ep_+ - i \ep_- 
\no \\
\lambda & = \lambda_+ + i \lambda_- &
\cB^{-1} \lambda ^* & = \lambda_+ - i \lambda_-
\no \\
\psi & = \psi_+ + i \psi_- &
\cB^{-1} \psi ^* & = \psi_+ - i \psi_-
\end{align}
Decomposing the supersymmetry transformations of the fermions accordingly gives, 
\bea
\delta  \psi_{\pm M}
&=& \nabla  _M  \ep_\pm  
\mp {e^{\phi}  \over 4} \p _M \chi \,   \ep_\mp 
\mp {e^{- \phi/2} \over 96} \Big \{  \Gamma_M (\G \cdot H_{(3)} ) + 2 (\G \cdot H_{(3)} ) \G_M \Big \}  \ep_\pm
\no \\ &&
\mp {1 \over 480}(\G \cdot  F_{(5)})  \Gamma_M \ep_\mp 
-{e^{ \phi/2} \over 96} \Big \{  \Gamma_M (\G \cdot \tilde F_{(3)} )
+ 2  ( \G \cdot \tilde F_{(3)}  ) \G_M \Big \}  \ep_\mp
\no \\
\delta \lambda _\pm & = & 
\half (\G \cdot \p \phi) \ep_\mp 
\mp {e^{ \phi} \over 2}  (\G \cdot \p \chi) \ep _\pm 
\pm {e^{- \phi/2} \over 24 }  ( \G \cdot H_{(3)}) \ep_\mp 
+ {e^{\phi/2} \over 24}  (\G \cdot \tilde F_{(3)}) \,  \ep_\pm
\eea
Assembling the spinors into doublets of Majorana-Weyl spinors, 
\bea
\label{multiplets}
\lambda = \left ( \bma \lambda _+ \cr \lambda _- \ema \right )
\hskip 0.8in 
\psi_M = \left ( \bma \psi _{+M} \cr \psi_{-M} \ema \right )
\hskip 0.8in 
\ep = \left ( \bma \ep_+ \cr \ep_- \ema \right )
\eea
and using the real-valued matrices,
\bea
s ^1 = \left ( \bma 0 & 1 \cr 1 & 0 \ema \right )
\hskip 0.8in
s ^2 = \left ( \bma 0 & 1 \cr -1 & 0 \ema \right )
\hskip 0.8in
s ^3 = \left ( \bma 1 & 0 \cr 0 & -1 \ema \right )
\eea
we have equivalently, 
\bea
\delta  \psi_{M}
&=& \nabla  _M  \ep - { e^{ \phi}  \over 4} \p_M \chi s^2  \ep 
-{e^{- \phi/2} \over 96} \Big \{  \Gamma_M (\G \cdot H_{(3)} ) + 2 (\G \cdot H_{(3)} ) \G_M \Big \} s^3 \ep
\no \\ &&
- {1 \over 480}(\G \cdot  F_{(5)})  \Gamma_M s^2 \ep
-{e^{ \phi/2 } \over 96} \Big \{  \Gamma_M (\G \cdot \tilde F_{(3)}  ) 
+ 2 ( \G \cdot \tilde F_{(3)}  ) \G_M \Big \} s^1  \ep
\no \\
\delta \lambda  & = & 
\half (\G \cdot \p \phi) s^1\ep 
- {e^{ \phi} \over 2}  (\G \cdot \p \chi) s^3 \ep 
+ {e^{- \phi/2} \over 24 }  ( \G \cdot H_{(3)}) s^2 \ep 
+ {e^{\phi/2} \over 24} (\G \cdot \tilde F_{(3)}  )  \ep
\eea
All boson fields in this expression are real, and all fermion fields are Majorana-Weyl. Complexifying the BPS equations is now readily achieved by declaring the boson fields $g_{MN}, F_{(3)}, H_{(3)}, F_{(5)}, \chi, \phi$ to be complex-valued, and the spinor doublets $\psi_M, \lambda, \ep$ to have complex Weyl spinor entries. Note that these equations are expressed in terms of the Einstein frame metric.

\subsection{Conversion to string frame}

In 10-dimensional spacetime, the relation between the Einstein frame metric $g^E_{MN}$ and the string frame metric $g^S_{MN}$ is given in terms of the dilaton $\phi$, as follows, 
\bea
g^{E}  _{MN} = e^{- \phi/2} \, g^{S}_{MN}
\eea
Hence the effective rules for the Dirac matrices in passing from the Einstein frame to the string frame are as follows,
\bea
\Gamma^M \to e^{\phi/4} \, \Gamma^M
\hskip 1in
\Gamma _M \to  e^{- \phi/4} \, \Gamma _M
\eea 
while the transformation of the spin connection, which is contained in $\nabla _M \ep$, is given by, 
\bea
 \om_{M \, ab}^E  \Gamma ^{ab} =  \om_{M \, ab }^S \Gamma ^{ab} 
 - { 1 \over 4} \Gamma _M ( \Gamma \cdot \p \phi) 
 + { 1 \over 4} (\Gamma \cdot \p \phi ) \Gamma _M
\eea
where the right side is expressed in terms of the string frame metric. The corresponding supersymmetry transformations in string frame metric are (where the formulation is now in terms of the string frame metric throughout),
\bea
\delta \psi _M  &=& 
\nabla  _M  \ep 
- {1 \over 8} \Big ( \Gamma _M ( \Gamma \cdot \p \phi) -(\Gamma \cdot \p \phi ) \Gamma _M \Big ) \ep
- { e^{ \phi}  \over 4} \p_M \chi s^2  \ep
 \\ &&  
-{1 \over 96} \Big \{  \Gamma_M (\G \cdot H_{(3)} ) + 2 (\G \cdot H_{(3)} ) \G_M \Big \} s^3 \ep
\no \\ &&
- {e^{ \phi}  \over 480}(\G \cdot  F_{(5)})  \Gamma_M s^2 \ep
-{e^{ \phi} \over 96} \Big \{  \Gamma_M  ( \G \cdot \tilde F_{(3)}  ) 
+ 2  ( \G \cdot \tilde F_{(3)}  ) \G_M \Big \} s^1  \ep
\no \\
e^{-\phi/4} \, \delta \lambda  & = & 
\half (\G \cdot \p \phi) s^1\ep 
+ {1\over 24 }  ( \G \cdot H_{(3)}) s^2 \ep 
- {e^{ \phi} \over 2}  (\G \cdot \p \tau_1) s^3 \ep 
+ {e^{ \phi} \over 24}  (\G \cdot \tilde F_{(3)} ) \ep 
\no
\eea
Taking the following linear combination of the gravitino with the dilatino, 
\bea
\delta \tilde \psi _M = \delta \psi_M + {1 \over 4} e^{-\phi/4} s^1 \Gamma _M \delta \lambda
\eea
letting $\ep \to e^{-\phi/4} \ep$, we obtain the BPS equations in string frame given in (\ref{1.BPS.1}). 
Apart from some signs that can be reversed upon changing basis, and a factor of 4 in the coefficient of $F_{(5)}$, this result agrees with \cite{Bergshoeff:2007cg}.

\newpage

%%%%%%%%%%%%%%%%%%%%%%%%%%%%%%%%%%%%%%%%%%%
%%%%%%%%%%%%%%%%%%%%%%%%%%%%%%%%%%%%%%%%%%%
\section{Reducing the BPS equations}
\setcounter{equation}{0}
\label{sec:C}
%%%%%%%%%%%%%%%%%%%%%%%%%%%%%%%%%%%%%%%%%%%
%%%%%%%%%%%%%%%%%%%%%%%%%%%%%%%%%%%%%%%%%%%

In this appendix we provide details on the derivation of the reduced BPS equations (\ref{3.c.1}). It will be useful to have the following relations, 
\begin{align} 
\label{Gamma.rel}
\Gamma \cdot \p \phi & = \f_a \Gamma ^a & \Gamma \cdot H_{(3)} & = 3! \, h_a \, \Gamma ^{67a}
& \Gamma ^{a} & =  \gamma_{(1)} \otimes \gamma _{(2)}  \otimes \gamma ^a
\no \\
\Gamma \cdot \p \tau_1 & = \chi_a \Gamma ^a & \Gamma \cdot \tilde F_{(3)} & = 3! \, g_a \, \Gamma ^{67a}
& \Gamma ^{67} & = i I_8 \otimes \gamma _{(2)}  \otimes I_2
\end{align}
and the action of $\Gamma ^{67}$ on the basis of Killing spinors,
\bea
\Gamma ^{67} \chi ^{\eta_1, \eta_2} & =&i \chi ^{\eta_1, - \eta _2}
\no \\
\Gamma ^{67a} \chi ^{\eta_1, \eta_2} & =& i \gamma ^a \chi ^{\eta_1, - \eta _2}
\eea

\subsection{Reducing the dilatino BPS equation}

Using the relations of (\ref{Gamma.rel}) the dilatino BPS equation of (\ref{1.BPS.1}) reduces as follows,
\begin{align}
0&=  \left(\varphi_a-e^\phi\chi_a s^2 \right) \Gamma^a \ep 
- \frac{1}{2}   \left(h_a s^3  -e^\phi g_a s^1 \right) \Gamma^{a67} \ep
\end{align}
Using the $\Gamma$-matrices in appendix \ref{sec:A} and the $\tau$-matrix notation, we obtain,
\bea
0=\sum_{\eta_1,\eta_2} \chi^{\eta_1,\eta_2} \otimes \left [
\left(\varphi_a-e^\phi\chi_a s^2 \right) \gamma^a\tau^{(11)}\zeta 
- \frac{1}{2}\left(h_a s^3 + e^\phi g_a s^1 \right)i\tau^{(10)}\gamma^a\zeta\right]
\eea
Identifying the coefficients of the basis spinors $\chi^{\eta_1, \eta_2}$ and multiplying to the left by $\tau^{(10)}$  gives the $(\lambda$) equation in (\ref{3.c.1}).

\subsection{Reducing the gravitino BPS equation}

For the gravitino equation we use the relations (\ref{3.d.2}) and the following partial contractions,
\bea
H_{(3) m NP } \, \Gamma ^{NP} & = & 0
\no \\
H_{(3) i NP } \, \Gamma ^{NP} & = & 2 h_a \Gamma _i \Gamma ^{67} \Gamma ^a
\no \\
H_{(3) a NP } \, \Gamma ^{NP} & = & 2 h_a \Gamma ^{67}
\eea
Substituting also the expressions of the remaining fields in (\ref{1.BPS.1}) we obtain, 
\begin{align}
(m)&& 
0&= \hat \nabla_m \, \ep
	+\left( \frac{D_a f_6}{2f_6}
	-\frac{1}{8}e^\phi \chi_a s^2
	-\frac{1}{8}e^\phi g_a \Gamma^{67} s^1 \right) \Gamma_m \Gamma^a \ep
\no \\
(i)&& 
0&= \hat \nabla_i \, \ep
	+ \left ( \frac{D_af_2}{2f_2} 
	- \frac{1}{4}h_a \Gamma ^{67}  s^3 
	- \frac{1}{8} e^\phi \chi_a s^2
	- \frac{1}{8} e^\phi g_a \Gamma^{67} s^1  \right ) \Gamma _i \Gamma ^a \ep
\no \\
(a)&& 
0&=\nabla_a \, \ep
	- \frac{1}{4}h_a\Gamma^{67} s^3 \ep
	+\frac{1}{8}e^\phi \chi_b \Gamma^b \Gamma_a s^2 \ep
	+\frac{1}{8}e^\phi g_b \Gamma^{67} \Gamma ^b \Gamma_a s^1 \ep
\end{align}
Eliminating the covariant derivatives $\hat \nabla _m$ and $\hat \nabla _i$ with the  Killing spinor equations in (\ref{3.b.1}), the $(m)$ and $(i)$ components become, 
\begin{align}
	0&=\Gamma_m\sum_{\eta_1,\eta_2}\chi^{\eta_1,\eta_2} \otimes \left[
	\frac{k_1}{2f_6}\eta_1\zeta_{\eta_1,\eta_2}
	+\left(\frac{D_a f_6}{2f_6}-\frac{1}{8}e^\phi\chi_a s^2 \right)\gamma^a\zeta_{-\eta_1,-\eta_2}
	-\frac{i}{8}e^\phi g_a \gamma^a s^1 \zeta_{-\eta_1,\eta_2}
	\right]
\no \\
	0&=
	\Gamma_i\sum_{\eta_1,\eta_2}\chi^{\eta_1,\eta_2}\otimes\left[
	\frac{k_2}{2f_2}\eta_2\zeta_{-\eta_1,\eta_2}
	+\left(\frac{D_a f_2}{2f_2}-\frac{1}{8}e^\phi \chi_a s^2 \right)\gamma^a\zeta_{-\eta_1,-\eta_2}
	\right.
\no \\
	&\hskip 35mm
	\left.
	-\frac{i}{4}h_a\gamma^a s^3  \zeta_{-\eta_1,\eta_2}
	+\frac{i}{8}e^\phi g_a \gamma^a s^1 \zeta_{-\eta_1,\eta_2}
	\right]
\end{align}
Using the $\tau$-matrix notation, and multiplying through the final result by the matrix $\tau^{(10)}$, we readily convert this result to the $(m)$ and $(i) $ equations of (\ref{3.c.1}).  Finally, the  components of the $(a)$ equation are,
\begin{align}
	0&=\sum_{\eta_1,\eta_2} \chi^{\eta_1,\eta_2}\otimes\left[
	\left(D_a+\frac{i}{2}\hat \omega_a \sigma^3\right)\zeta_{\eta_1,\eta_2}
	- \frac{i}{4}h_a s^3 \zeta_{\eta_1,-\eta_2}
	\right.
	\nonumber\\
	&\hskip 30mm
	\left.
	+\frac{1}{8}e^\phi\chi_b\gamma^b\gamma_a s^2 \zeta_{\eta_1,\eta_2}
	+\frac{i}{8}e^\phi g_b\gamma^b\gamma_a s^1 \zeta_{\eta_1,-\eta_2}
	\right]
\end{align}
Identifying coefficients gives the $(a)$ equation of  (\ref{3.c.1}).

\newpage

%%%%%%%%%%%%%%%%%%%%%%%%%%%%%%%%%%%%%%%%%%%
%%%%%%%%%%%%%%%%%%%%%%%%%%%%%%%%%%%%%%%%%%%
\section{Conformal  gauge for a complex metric on $\Sigma_\CC$}
\setcounter{equation}{0}
\label{sec:D}
%%%%%%%%%%%%%%%%%%%%%%%%%%%%%%%%%%%%%%%%%%%
%%%%%%%%%%%%%%%%%%%%%%%%%%%%%%%%%%%%%%%%%%%

A complex metric $g_{mn}(\xi)$ on $\Sigma_\CC$ is symmetric in $m,n=1,2$ but its entries are locally holomorphic  functions on a complexified surface $\Sigma_\CC$, parametrized by two complex coordinates $\xi^1, \xi^2$. Equivalently, we may write the complex-valued metric $g_{mn}$ as follows,
\bea
g_{mn} = e_m {}^a  e_n{}^b \, \delta _{ab}
\hskip 1in a,b=1,2
\eea
where $e_m{}^a$ are holomorphic functions of $\Sigma_\CC$. The frame group is $SO(2,\CC) $. Normalizing the  constant frame metric such that $\delta _{z \tz}=\delta_{\tz z} =2$ and $\delta_{zz}=\delta_{\tz \tz}=0$, the complex-valued  metric can be decomposed as follows,
\bea
g_{mn} = 2 \, e_m {}^z  \, e_n{}^\tz  + 2 \, e_m {}^\tz  \, e_n{}^z
\eea 
where $e_m {}^z$ and $e_m{}^{\tilde z}$ are holomorphic functions on $\Sigma_\CC$. We stress that $\tilde z$ is \textit{not} the complex conjugate of $z$ and  $e_m{}^\tz$ is \textit{not} the complex conjugate of $e_m {}^z$.

\sm

The equivalent of conformal gauge for complex metrics, to which we shall refer as \textit{generalized conformal gauge}, is defined in terms of the original metric by,
\bea
g_{mn} (\xi^1,\xi^2) \, d\xi^m d \xi^n = 4\rho(w,\tilde w)^2 dw \, d\tilde w
\eea
where $w, \tilde w$ are two independent local complex coordinates on $\Sigma_\CC$ and $\rho(w,\tilde w)$ is holomorphic on $\Sigma_\CC$.  The system of local complex coordinates $w, \tilde w$ are obtained from a holomorphic diffeomorphism on $\Sigma _\CC$ and exists because all the components of the metric $g_{mn}$ are holomorphic on $\Sigma_\CC$. The arguments for their existence are similar to the ones for the existence of conformal gauge on a standard Riemann surface, except that here all dependences are holomorphic. In particular that means that in any Taylor series expansion of the diffeomorphism near the identity, the real coordinates may simply be replaced by complex coordinates. From the generalized conformal gauge of the metric, we obtain the components of the frame in this gauge, 
\begin{align}
e_m{}^z \, d\xi^m  & = \rho \, dw
\no \\
e_m{}^\tz \, d \xi^m & = \tilde \rho \, d \tw
\end{align}
Since the  frame group $SO(2,\CC)$ acts  by  $e_m {}^z \to \lambda \, e_m {}^z $ and $ e_m {}^\tz \to \lambda ^{-1} e_m {}^\tz$, where $\lambda $ is an arbitrary locally holomorphic function of $w, \tw$ that is nowhere vanishing, we can choose a gauge in which $\tilde \rho = \rho$. The corresponding differential equations  are, 
\begin{align}
D_z f & =   e_z {}^m \p_m f = \rho^{-1}  \p_w f   & D_z f & =0 \hbox{ is solved by } \quad f(\tw)
\no \\
D_\tz f & =   e_\tz {}^m \p_m f =  \rho^{-1}  \p_\tw f  & D_\tz f & =0  \hbox{ is solved by } \quad f(w)
\end{align}
respectively, where $e_a{}^m$ are the inverse frames, 
so that the covariant derivatives and the connection are given by (\ref{3.conf}).

\newpage

%%%%%%%%%%%%%%%%%%%%%%%%%%%%%%%%%%%%%%%%%%%
%%%%%%%%%%%%%%%%%%%%%%%%%%%%%%%%%%%%%%%%%%%
\section{Decoupling and solving the equations for $\rho^2$ and $\tau_\pm$}
\setcounter{equation}{0}
\label{sec:E}
%%%%%%%%%%%%%%%%%%%%%%%%%%%%%%%%%%%%%%%%%%%
%%%%%%%%%%%%%%%%%%%%%%%%%%%%%%%%%%%%%%%%%%%

In this appendix, we provide details for the calculations towards solving the differential equations for $\rho$ and $\tau_\pm$ in section \ref{sec:3-solving}. The decoupling procedure of section \ref{sec:3} and its use to obtain the general solution by quadrature  in section \ref{sec:3-integrating} are expounded in the two subsections below.

\subsection{Decoupling}

We carry out the change variables from $\tau_\pm$ to $Z_\pm$ given in the first equation of (\ref{4.Ztau1}), and repeated here for convenience, 
\bea
\label{EE.1}
Z_\pm^2 = { \tau _\pm - \tilde  \xi  \over \tau_\pm - \xi}
\hskip 1in 
X = \varepsilon_\kappa Z_+ Z_-
\hskip 0.5in
\varepsilon_\kappa=\frac{\tilde\epsilon_\kappa}{\epsilon_\kappa}
\eea
where $\epsilon_\kappa$, $\tilde\epsilon_\kappa$ are defined in (\ref{eq:Lambda-def}).
Inverting these relations gives $\tau_\pm$ as a function of $Z_\pm$ and the functions $\xi, \tilde \xi$, in  two alternative but equivalent expressions,
\bea
\label{EE.2}
\tau_\pm - \xi = {  \xi  - \tilde  \xi \over Z_\pm^2-1}
\hskip 1in
\tau_\pm - \tilde  \xi  = { Z_\pm ^2 ( \xi  - \tilde  \xi ) \over Z_\pm^2-1}
\eea
These formulas may be used to compute the following intermediate combinations,
\bea
\label{EE.3}
{ \p_w \tau_\pm \over \tau_+ - \tau_-} 
= { Z_\pm^2 (Z_\mp^2-1) \over Z_-^2 - Z_+^2} \p_w \ln ( \xi  - \tilde  \xi)
-  { Z_\mp^2 -1 \over Z_\pm^2-1} \cdot {  \p_w Z_\pm^2 \over Z_-^2 - Z_+^2}
\eea
as well as,
\bea
\label{EE.4}
\tau _\pm - \xi +{2 \over Z_+Z_-}  ( \tau _\pm - \tilde  \xi ) & = & 
{  \xi  - \tilde  \xi \over Z_\mp^2 (Z_\pm^2-1) } \left ( Z_\mp^2 + 2 Z_+Z_- \right )
\no \\
\tau _\pm - \tilde  \xi  +2 Z_+Z_-  ( \tau _\pm - \xi ) & = &
{ \xi  -  \tilde  \xi \over Z_\pm^2-1} \left ( Z_\pm^2 + 2 Z_+Z_- \right )
\eea
With the help of these intermediate equations and $X^2=Z_+^2Z_-^2$, we obtain the following rewriting of (\ref{4.sub.8}) in terms of the functions $Z_\pm$, 
\bea
\label{EE.5}
\p_w \ln \hat \rho^2 \mp { Z_\pm^2 - 2 X \over Z_-^2-Z_+^2} (Z_\mp^2-1) \p_w \ln ( \xi  - \tilde  \xi) 
\pm  { Z_\mp^2 - 2 X \over Z_-^2 - Z_+^2} \cdot { \p_w Z_\mp^2  \over Z_\mp^2} & = & 0
\no \\
\p_\tw \ln \hat \rho^2 \mp { Z_\mp^2 - 2 X \over Z_\mp^2( Z_-^2-Z_+^2)} (Z_\mp^2-1) \p_\tw \ln (\xi  - \tilde   \xi) 
\pm  { Z_\pm^2 - 2 X \over Z_-^2 - Z_+^2} \cdot { \p_\tw Z_\mp^2 \over Z_\mp^2} & = & 0
\eea

\subsubsection{Sum equations}

Taking the sums of the $\pm$ equations of (\ref{EE.5}), we obtain, 
\bea
\label{EE.6}
0 & = & 2 \p_w \ln \hat \rho^2 +  \p_w \ln { (Z_-+\varepsilon_\kappa Z_+)^3 \over (Z_- - \varepsilon_\kappa Z_+)}   - (1-2X) \p_w \ln ( \xi - \tilde  \xi) 
\no \\
0 & = & 2 \p_\tw \ln \hat \rho^2 +  \p_\tw \ln { (Z_-+\varepsilon_\kappa Z_+)^3 \over (Z_--\varepsilon_\kappa Z_+)} - 2 \p_\tw \ln X  - \left (1-{2 \over X} \right ) \p_\tw \ln ( \xi - \tilde  \xi) 
\eea
Redefining $\hat \rho^2$ as in the second equation of (\ref{4.Ztau1}),
\bea
\label{EE.7}
{ 1 \over R^2} = - {  k_1 \over c_6} \, 
{ (Z_-+\varepsilon_\kappa Z_+)^{3 \over 2} \, \hat \rho^2 \over X^\half   (Z_--\varepsilon_\kappa Z_+)^{1 \over 2} ( \xi  - \tilde  \xi)^\half  } 
\eea
the equations become,
\bea
\label{EE.8}
 \p_w \ln R^2 - \half \p_w \ln X  - X \, \p_w \ln ( \xi - \tilde  \xi) & = & 0
\no \\
\p_\tw \ln R^2 +  \half \p_w \ln X - { 1 \over X}  \p_\tw \ln ( \xi - \tilde  \xi) & = & 0
\eea
Note that both equations depend on the product $Z_+ Z_-$ but not on the ratio $Z_+/Z_-$. Combining the first two terms in each equation gives the pair (\ref{4.sum}).

\subsubsection{Difference equations}

Taking the differences of the $\pm$ equations of (\ref{EE.5}), we observe that all dependence on $\hat \rho$ cancels and we obtain, 
\bea
\label{EE.9}
0 & = & \Big (  2 X^2 + 4 X - (Z_+^2 + Z_-^2) (1+2X) \Big )  \p_w \ln ( \xi  - \tilde  \xi) 
-  \p_w \Big ( Z_+^2+Z_-^2-4X \Big ) 
\no \\
0 & = & \Big ( 2 + 4 X   - (Z_+^2 + Z_-^2 ) \big (1+ 2 X^{-1} \big ) \Big ) 
\p_\tw \ln (\xi  - \tilde   \xi) 
-  \p_\tw (Z_+^2 + Z_-^2) 
\no \\ && \qquad
+ 2 (Z_+^2 + Z_-^2 - 2X) \p_\tw \ln X   
\eea
Eliminating  $Z_+^2 + Z_-^2$ in favor of $Z_+^2 + Z_-^2=- (Y+2)X$ and regrouping contributions gives the pair of equations in (\ref{4.linear}).

\subsubsection{Remaining pair of equations}

Expressing the pair of equations in (\ref{4.sub.9}) in terms of the variables $R, Z_\pm$, we obtain,
\bea
\label{EE.10}
0 & = & -{\tilde\epsilon_\kappa \kappa_1 \over \varepsilon_\kappa R^2} \big ( Z_+^2 + Z_-^2 - 2 X \big ) X^{-\half}
- \p_w (Z_+^2 + Z_-^2) + 2 \big ( Z_+^2 + Z_-^2 \big ) \p_w \ln X 
\no \\ && \qquad 
- \big (  2X^2 -  Z_+^2 - Z_-^2 \big ) \p_w \ln ( \xi  - \tilde  \xi) 
\no \\
0 & = & {\epsilon_\kappa\tilde  \kappa_1 \over R^2} \big ( Z_+^2 + Z_-^2- 2 X \big ) X^{\half} 
- \p_\tw (Z_+^2 + Z_-^2)  + \big ( Z_+^2 + Z_-^2 -2 \big ) \p_\tw \ln ( \xi  - \tilde  \xi) 
\eea
The combination $R^{-2}(Z_+^2+Z_-^2-2X)X^{1/2}$ in the second equation is invariant under $\varepsilon_\kappa\rightarrow -\varepsilon_\kappa$, and we note the additional $\varepsilon_\kappa$ emerging naturally in the first equation.
Overall, the behavior of the equations in (\ref{EE.10}) under sign reversals of $\kappa_{1,2}$, $\tilde\kappa_{1,2}$ is consistent with (\ref{4.sub.9}).
Subtracting the equations in (\ref{EE.9}) and (\ref{EE.10}) pairwise from one another, the combination $Z_+^2+Z_-^2 + 2X $ factorizes from each equation, and gives upon simplification, 
\bea
\label{EE.11}
0 & = & - {\epsilon_\kappa\kappa _1 \over R^2X^\half } + 2 \p_w \ln X + 2 (1+X) \p_w \ln ( \xi  - \tilde  \xi) 
\no \\
0 & = & { \epsilon_\kappa\tilde \kappa _1 \over R^2} X^\half  - 2 \p_\tw \ln X + 2 \left (1 + { 1 \over X} \right ) \p_\tw \ln (\xi  - \tilde   \xi) 
\eea
Note that, just as the equations (\ref{EE.8}), namely (\ref{4.sub.9}), these equations involve only the functions $R$ and $X$. In fact, adding pairwise twice the equations of (\ref{EE.8})  produces the following pair of equations, 
\bea
0 & = & -{\epsilon_\kappa \kappa_1 \over R^2X^\half }+ 2 \p_w \ln R^2 + \p_w \ln X  + 2 \p_w \ln (\xi  - \tilde   \xi) 
\no \\
0 & = & { \epsilon_\kappa\tilde \kappa _1 \over R^2} X^\half  + 2 \p_\tw \ln R^2 - \p_\tw \ln X +2 \p_\tw \ln (\xi  - \tilde   \xi) 
\eea 
Multiplying through by $R^2 (\xi  - \tilde   \xi) X^{\pm \half}$ allows us to recast the result in the following form,
\bea
\label{EE.12}
0 & = & -\epsilon_\kappa(\kappa_2 - \kappa _1 \tilde  \xi)    +  \p_w \left ( 2 R^2 X^\half (\xi  - \tilde   \xi) \right ) 
\no \\
0 & = &  \epsilon_\kappa(\tilde \kappa _2 - \tilde \kappa _1  \xi) +  \p_\tw \left ( 2R^2 X^{-\half} ( \tilde \xi  -   \xi) \right )
\eea
or, equivalently, this gives the pair of equations of (\ref{4.dec}).  These are equations that can be integrated by quadrature, as we shall now show.

\subsection{Integrating}

To integrate the equations (\ref{4.dec}), aka (\ref{EE.12}), conveniently, we introduce the functions $\cA_1, \cA_2, \tilde \cA_1, \tilde \cA_2$ to which the one-forms $\kappa_1, \kappa_2, \tilde \kappa_1, \tilde \kappa_2$ and $\xi, \tilde \xi$ are related by (\ref{3.Akappa}). Equations (\ref{4.dec}) are then readily integrated by quadrature giving (\ref{4.dec1}) in terms of two additional as yet arbitrary functions, $\cA_0, \tilde \cA_0$ which are independent of $\tw$ and $w$, respectively. 

\sm

The functions $\cA_0, \tilde \cA_0$ are determined, up to an additive constant, by solving the pair of equations (\ref{4.sum}) next. To do so, we begin by rewriting the equations (\ref{4.sum}) as follows,
\bea
\p_w \left (  R^2 X^{- \half} (\tilde  \xi  - \xi)  \right ) - \left ( R^2 X^\half  + R^2 X^{-\half} \right ) \p_w (\tilde  \xi  -\xi) & = & 0
\no \\
\p_\tw \left (  R^2 X^{ \half} (\tilde  \xi  - \xi)  \right ) - \left ( R^2 X^{-\half}  + R^2 X^{\half} \right ) \p_\tw (\tilde  \xi  -\xi) & = & 0
\eea
Eliminating $R^2 X^{\pm \half} (\tilde  \xi  - \xi)$ in terms of $\cA_{0,1,2}$, $\tilde\cA_{0,1,2}$ using (\ref{4.dec1}), we obtain,
\bea
\label{4.AAa}
(\tilde  \xi  - \xi) { \p_w \cA_0 \over \p_w \xi } 
+ \cA_0 - \tilde \cA_0 - \cA_2 + \tilde \cA_2 + \tilde  \xi  (\cA_1   -  \tilde \cA_1)  & = & 0
\no \\
(\tilde  \xi  - \xi) { \p_\tw \tilde \cA_0 \over \p_\tw \tilde  \xi } 
+ \cA_0 - \tilde \cA_0 - \cA_2 + \tilde \cA_2 + \xi (\cA_1   - \tilde \cA_1)  & = & 0
\eea
The difference of the two equations has an overall factor of $\tilde \xi - \xi$ multiplying the relation,
\bea 
\label{EE.a.1}
{ \p_w \cA_0 \over \p_w \xi} + \cA_1 =  { \p_\tw \tilde \cA_0 \over \p_\tw \tilde  \xi } + \tilde \cA_1
\eea
Since the left side is independent of $\tw$ and the right side is independent of $w$, both sides must be constants. Since $\cA_1$ and $\tilde \cA_1$ were defined by $\kappa_1$ and $\tilde \kappa_1$ only up to additive constants, we may absorb the constant in (\ref{EE.a.1}) into the definitions of $\cA_1$ and $\tilde \cA_1$, giving,
\bea
\label{4.Adera}
\p_w \cA_0 & = & -\cA_1 \, \p_w \xi 
\no \\
\p_\tw \tilde \cA_0  & = & -\tilde \cA_1 \, \p_\tw \tilde  \xi 
\eea
Recasting the first term of the equations of (\ref{4.AAa}) using the above result, we obtain two identical equations, given by,
\bea
\label{4.AAAa}
 \tilde \cA_0  - \tilde \cA_2 + \tilde  \xi  \tilde \cA_1  = \cA_0  -  \cA_2  + \xi \cA_1
\eea
The left side being independent of $w$ and the right side being independent of $\tw$ implies that both sides must be constant. Using the fact that $\cA_0$ and $\tilde \cA_0$ were defined only up to additive constants, we readily obtain (\ref{3.AA0}).

\newpage

%%%%%%%%%%%%%%%%%%%%%%%%%%%%%%%%%%%%%%%%%%%
%%%%%%%%%%%%%%%%%%%%%%%%%%%%%%%%%%%%%%%%%%%
\section{Real forms of the Lie superalgebra $F(4)$}
\setcounter{equation}{0}
\label{sec:F}
%%%%%%%%%%%%%%%%%%%%%%%%%%%%%%%%%%%%%%%%%%%
%%%%%%%%%%%%%%%%%%%%%%%%%%%%%%%%%%%%%%%%%%%

The classical Lie superalgebra $F(4)$ has 24 bosonic and 16 fermionic generators. Its complex form $F(4;\CC)$ has maximal bosonic subalgebra $\ms \mo (3; \CC)  \oplus  \ms \mo (7; \CC)$. The different involutions of $F(4;\CC)$ give rise to different real forms of $F(4;\CC)$ whose maximal bosonic subalgebras are various real forms of  $\ms \mo (3; \CC)  \oplus   \ms \mo (7; \CC)$. However, every real form of $\ms \mo (3; \CC)  \oplus    \ms \mo (7; \CC)$ does not correspond to a real form of $F(4;\CC)$. The classification of the allowed real forms of $F(4;\CC)$ is subject to discrepancies between the results of different authors. 
Additionally, we seek a decisive result as to whether a real forms of $F(4;\CC)$ exists with the  maximal bosonic subalgebra $\ms \mo (3; \RR)  \oplus    \ms \mo (7; \RR)$. For these reasons, we present here a derivation of all possible real forms using elementary methods.

\subsection{The maximal bosonic subalgebra}

It will be convenient to formulate the structure relations for both the complexified maximal bosonic Lie algebra $\ms \mo (3; \CC) \oplus  \ms \mo (7; \CC)$ of $F(4;\CC)$ as well as for its various  real forms in terms of a  pair of metrics $(\eta, \tilde \eta)$ of arbitrary signature, parametrized as follows,
\bea
\eta _{ab} = (-)^{s_a} \delta_{ab} \hskip 1in \tilde \eta _{mn} = (-)^{\tilde s_m} \delta _{mn}
\eea 
for $a,b =1,2,3$ and $m,n= 1, \cdots, 7$ with $s_a$ and $\tilde s_m$ taking the values 0 or 1.  The Dirac matrices and corresponding spinor representation generators $\sigma _{ab} $ and $\tilde \sigma _{mn}$ satisfy, 
\begin{align}
\label{C.1}
\{ \gamma _a, \gamma_b \} & = 2 \, \eta_{ab} & 
\gamma_{ab} & = \thalf [\gamma _a, \gamma _b] & 
\sigma _{ab} &= \thalf \gamma _{ab}
\no \\
\{ \tilde \gamma _m, \tilde \gamma_n \} & = 2 \, \tilde \eta_{mn} & 
\tilde \gamma_{mn} & = \thalf [\tilde \gamma _m, \tilde \gamma _n] & 
\tilde \sigma _{mn} &= \thalf \tilde \gamma _{mn}
\end{align} 
The charge conjugation matrices $C$ and $\tilde C$ satisfy, \footnote{The signs in the left column are determined by the fact that the products $\g_1 \g_2 \g_3$ and $\tilde \g_1 \cdots \tilde \g_7$ are proportional to the identity matrices $I_2$ and $I_8$, respectively.}
\begin{align} 
\label{C.C}
\gamma _a ^t & = - C \gamma _a C^{-1} &
C^t & =  \ep_C \, C & 
\big ( C \gamma _{ab}  \big ) ^t & =  - \ep_C \, C \gamma _{ab}
\no \\
\tilde \gamma_m ^t & = - \tilde C \gamma _m C^{-1} & 
\tilde C^t & =  \ep_{\tilde C} \, \tilde C & 
\big ( \tilde C \tilde \gamma _{mn} \big )^t & =  - \ep_{\tilde C} \, \tilde C \tilde \gamma _{mn} 
\end{align}
where $\ep_C$ and $\ep_{\tilde C}$ can take the values $\pm 1$. Note that the matrices $C $ and $\tilde C$ may be chosen independently of the signatures of the metrics $\eta$, and $\tilde \eta$. Choosing a basis in which $\gamma_2$ as well as $\tilde \gamma_2, \tilde \gamma_4$ and $\tilde \gamma _6$ are anti-symmetric, we find,
\begin{align}
\label{C.C1}
C  & = \gamma _2 & \ep_C & = -1
\no \\
\tilde C & = \tilde \gamma _2 \tilde \gamma_4 \tilde \gamma _6 & \ep_{\tilde C} & =+ 1
\end{align}

\sm

The bosonic generators of $F(4;\CC)$, namely the generators of $\ms \mo (3;\CC) $ and $ \ms \mo (7;\CC)$, are denoted $\cA_{ab}$ and $\tilde \cA_{mn}$. The fermionic generators of $F(4;\CC)$ are $\cS_{\a \mu}$ where $\a=1,2$ and $\mu=1,\cdots, 7$ label the unique Dirac spinor representations of $\ms \mo (3;\CC) $ and $ \ms \mo (7;\CC)$, respectively.  The structure relations of $\ms \mo (3;\CC) \oplus \ms \mo (7;\CC)$ are $[ \cA_{ab}, \tilde \cA_{mn} ] = 0 $ and,
\bea
\label{C.2}
{} [\cA_{ab} , \cA_{cd} ] & = & \eta _{ad} \, \cA_{bc} + \eta _{bc} \, \cA_{ad} - \eta _{ac} \, \cA_{bd} - \eta _{bd} \, \cA_{ac}
\no \\
{} [\tilde \cA _{mn} , \tilde \cA _{pq} ] & = & \tilde \eta  _{mq} \, \tilde \cA _{np} + \tilde \eta _{np} \, \tilde \cA _{mq} - \tilde \eta _{mp} \, \tilde \cA _{nq} - \tilde \eta _{nq} \, \tilde \cA _{mp}
\eea
The spinor representations obey these relations with $\cA_{ab} = \sigma _{ab}$ and $\tilde \cA_{mn} = \tilde \sigma _{mn}$. Since $\cS$ is spinor it transforms under the bosonic generators as follows,
\bea
\label{C.3}
%\label{1.AMS}
{} [ \cA_{ab} , \cS_{\a \mu} ] & = & \cS_{\b \mu} \, \big ( \sigma _{ab} \big ) _{\b \a}
\no \\
{} [ \tilde \cA_{mn} , \cS_{\a \mu} ] & = & \cS_{\a \nu} \, \big ( \tilde \sigma _{mn} \big ) _{\nu \mu}
\eea
The right-action is required by the Jacobi identities of the type $[\cA, [\cA, \cS]]$ and $[\tilde \cA, [\tilde \cA, \cS]]$.

\subsection{Structure relations for $F(4;\CC)$}

The structure of the anti-commutation relation of two $\cS$ generators is dictated by its $\ms \mo (3;\CC) $ and $ \ms \mo (7;\CC)$ transformation laws,
\bea
\label{1.SS}
\{ \cS_{\a \mu} , \cS_{\b \nu} \} = s \, \tilde C_{\mu \nu} \big ( C \gamma ^{ab}  \big ) _{\a \b} \cA_{ab}
+ t \, C_{\a \b} \big ( \tilde C \tilde \g ^{mn} \big ) _{\mu \nu} \tilde \cA_{mn}
\eea
where $s$ and $t$ are complex numbers that remain to be determined, and $C$ and $\tilde C$ are the charge conjugation matrices defined in (\ref{C.C}). Both sides of (\ref{1.SS}) are invariant under swapping the pairs $(\a \mu) $ and $(\b \nu)$ thanks to the relation $\ep_C = - \ep _{\tilde C}$ obtained in (\ref{C.C1}). The Jacobi identities of the type $[\cA, \{ \cS, \cS\}]$ and $[\tilde \cA, \{ \cS, \cS\}]$  are then automatically  satisfied. 

\sm

To derive the requirements imposed by the  Jacobi identity on three $\cS$ generators,
\bea
{} [ \{ \cS_{\a \mu}, \cS_{\b \nu} \} , \cS_{\g \rho} ] +
\hbox{cycl} ( \a \mu, \b \nu, \g \rho) =0
\eea
we evaluate the anti-commutators using (\ref{1.SS}), 
\bea
s \, \tilde C_{\mu \nu} \big ( C \g^{ab} \big ) _{\a \b} [ \cA_{ab} , \cS_{\g \rho}] 
+ t \, C_{\a \b} \big ( \tilde C \tilde \gamma ^{mn} \big ) _{\mu \nu}
[ \tilde \cA _{mn} , \cS_{\g \rho} ] +
\hbox{cycl} ( \a \mu, \b \nu, \g \rho) =0
\quad
\eea
and the remaining commutators using (\ref{C.3}). Omitting the common factor $\cS_{\delta \sigma}$, we obtain, 
\bea
\label{1.tri}
s \, \big ( C \g^{ab} \big ) _{\a \b} \big ( \gamma _{ab} \big ) _{\delta \gamma} \tilde C_{\mu \nu} \, \delta _{\sigma \rho} 
+ t \, C_{\a \b} \, \delta_{\gamma \delta} \big ( \tilde C \tilde \g ^{mn} \big ) _{\mu \nu} \big ( \tilde \gamma _{mn} \big ) _{\sigma \rho}
+ \hbox{cycl} ( \a \mu, \b \nu, \g \rho) =0
\quad
\eea
To analyze these equations, we need the following Fierz identities, proven in  \cite{Freedman:2012zz}. The Dirac matrices $\g_{ab}$ for $\ms \mo (3;\CC)$ or any of its real forms satisfy the following Fierz identity for $\a, \b, \g, \delta \in \{ 1,2\}$,
\bea
\label{1.thm.1}
\big ( C \g^{ab} \big )_{\a \b} \big ( \g_{ab} \big ) _{\delta \g} & = &
2 \, C_{\a \b} \, \delta _{\g \delta} - 4 \, C_{\a \g} \, \delta _{\b \delta}
\eea
while the  matrices $\tilde \g_{mn}$ satisfy the following Fierz identity for $\mu, \nu, \rho, \sigma \in \{ 1,\cdots, 8\}$,
\bea
\label{1.lem.1}
\big ( \tilde C \tilde \gamma  ^{mn} \big ) _{\mu \nu } \big ( \tilde \gamma _{mn} \big ) _{\rho \sigma} 
+ \big ( \tilde C \tilde \gamma ^{mn} \big ) _{\mu \sigma} \big ( \tilde \gamma _{mn} \big ) _{\rho \nu} 
& = & 
6 \, \delta_{\mu \rho } \, \tilde C _{ \nu \sigma} 
- 3 \, \delta _{\nu \rho} \, \tilde C_{\mu \sigma} 
- 3 \, \delta_{\rho \sigma} \, \tilde C_{\mu \nu}
\eea
Importantly, these relations hold regardless of the signatures of the metrics $\eta$ and $\tilde \eta$. 

\sm

Substituting the Fierz identities into (\ref{1.tri}) and identifying the coefficients, subject to the relation $C_{\a \b} \, \delta _{\g \delta} + {\rm cycl} (\a, \b,\g)=0$,  we find the relation $3t+2s=0$.  We may rescale the generators $\cS$ such that (\ref{1.SS}) becomes,\bea
\label{1.SSa}
\{ \cS_{\a \mu} , \cS_{\b \nu} \} = - \half \tilde C_{\mu \nu} \big ( C \gamma ^{ab}  \big ) _{\a \b} \cA_{ab}
+ { 1 \over 3} C_{\a \b} \big ( \tilde C \tilde \gamma ^{mn} \big ) _{\mu \nu} \tilde \cA_{mn}
\eea
Equations (\ref{C.2}), (\ref{C.3}) and (\ref{1.SSa}) give all the structure relations of $F(4;\CC)$.\footnote{The expression given for (\ref{1.SSa}) agrees with the one of Rittenberg, except that our prefactor $-\half$ is incorrectly given as a factor of 2 there.}

\subsection{Parametrizing Dirac matrices for arbitrary signatures}
\label{sec:C.3}

The Dirac matrices $\gamma ^E_a$ and $\tilde \gamma ^E_m$ for the Euclidean metrics are Hermitian, so that their behavior under complex conjugation coincides with their behavior under transposition, 
\bea
(\gamma ^E_a)^* = - C \gamma _a ^E C^{-1} 
\hskip 1in
(\tilde \gamma ^E_m)^* = - \tilde C \tilde \gamma ^E_m \tilde C^{-1}  
\eea
The Dirac matrices $\gamma_a$ and $\tilde \gamma _m$ for metrics $\eta$ and $\tilde \eta$ of arbitrary signature are obtained by multiplying $\gamma_a^E$ and $\tilde \gamma _m^E$ by suitable powers of $i$, respectively,
\begin{align}
\eta_{ab} & = (-)^{s_a} \, \delta_{ab}  & \g_a & = i^{s_a} \, \g^E _a & (\gamma _a)^* &= - (-)^{s_a} C \gamma _a  C^{-1} 
\no \\
\tilde \eta _{mn} & = (-)^{\tilde s_m} \, \delta_{mn} & \tilde \g_m & = i^{\tilde s_m} \, \tilde \g^E_m & 
(\tilde \gamma_m)^* & = - (-)^{\tilde s_m} \tilde C \tilde \gamma _m \tilde C^{-1}  
\end{align}
A convenient basis for the Dirac matrices $\gamma ^E_a$ and $\tilde \gamma _m^E$ with Euclidean metric may be chosen as follows in terms of the standard Pauli matrices $\sigma _a$ for $a=1,2,3$,
\begin{align}
\g^E_1 & = \sigma _1 &
\tilde \g _1^E & = \sigma _1 \otimes I_2 \otimes I_2 & 
\tilde \g _2^E & = \sigma _2 \otimes I_2 \otimes I_2
\no \\ 
\g^E_2 & = \sigma _2 & 
\tilde \g _3^E & = \sigma _3 \otimes  \sigma_1 \otimes I_2 & 
\tilde \g _4^E & = \sigma _3 \otimes \sigma _2 \otimes I_2
\no \\ 
\g^E_3 & = \sigma _3 &
\tilde \g _5^E & = \sigma _3 \otimes  \sigma_3 \otimes \sigma _1 & 
\tilde \g _6^E & = \sigma _3 \otimes \sigma _3 \otimes \sigma_2
\no \\ C & = \sigma _2 & 
\tilde \g _7^E & = \sigma _3 \otimes  \sigma_3 \otimes \sigma _3 &
\tilde C & = \sigma _2 \otimes \sigma _1 \otimes \sigma _2
\end{align}

\subsection{Conditions for the existence of an involution}

Complex conjugation is an involution that maps the complex Lie superalgebra $F(4;\CC)$ to itself and whose square  is the identity. A real form of $F(4;\CC)$ is the quotient of $F(4;\CC)$ by a complex conjugation involution. Alternatively, a real form of $F(4;\CC)$ is characterized by the existence of generators $\cA, \tilde \cA$ and $ \cS$ that are invariant under complex conjugation, which we denote here by a star, in the following sense,
\bea
\label{1.conj}
\cA_{ab}^* = \cA_{ab} \hskip 0.8in 
\tilde \cA_{mn}^* = \tilde \cA _{mn} \hskip 0.8in 
\cS_{\a \mu}^* = S_{\b \nu} \, X_{\b \a} \, Y_{\nu \mu}
\eea
for invertible matrices $X$ and $Y$ of dimensions $2 \times 2$ and $8 \times 8$, respectively. The involution property of complex conjugation is automatic on the generators $\cA, \tilde \cA$, and on $\cS$ requires $(\cS_{\a \mu}^*)^*=\cS_{\a \mu}$, namely $X_{\g \b} X^*_{\b \a} Y_{\rho \nu} Y^* _{\nu \mu} = \delta_{\a \g} \delta _{\mu \rho}$ or, in matrix notation, 
\bea
\label{1.XY}
X X^* \otimes Y Y^* = I_2 \otimes I_8
\eea
If matrices $X$ and $Y$ compatible with the structure relations of $F(4;\CC)$ exist,  then the corresponding $\cA, \tilde \cA$ and $ \cS$ are the generators of a real form of $F(4;\CC)$. We shall now classify all possible real forms of $F(4;\CC)$ by analyzing these compatibility conditions.

\sm

The structure relations of the bosonic generators $\cA$ and $\tilde \cA$ of (\ref{C.2}) are manifestly consistent with (\ref{1.conj}), while (\ref{C.3}) impose the following conditions,  
\bea
\label{1.XGY}
\g_{ab} X = X \g_{ab}^* \hskip 1in \tilde \g_{mn} Y = Y \tilde \g _{mn}^*
\eea
Compatibility of these equations with their complex conjugates requires $[ \gamma_{ab}, XX^*]=0$ and $[\tilde \g _{mn}, YY^*]=0$. 
Since $\half \g_{ab}$ and $\half \tilde \g_{mn}$ generate irreducible representations of $\ms \mo(3)$ and $\ms \mo(7)$, respectively, $XX^*$ and $YY^*$ must be proportional to the identities, 
\bea
XX^* = x_0 I_2 \hskip 1in YY^* = y_0 I_8
\eea
where $x_0, y_0 \in \RR$. Finally, consistency of (\ref{1.conj}) with (\ref{1.SSa}) requires, 
\bea
\tilde C^* _{\mu \nu} \big ( C \gamma ^{ab} \big )^* _{\a \b} & = & 
\tilde C_{\rho \sigma} \big ( C \gamma ^{ab} \big ) _{\g \delta} X_{\g \a} X_{\delta \b} Y_{\rho \mu} Y_{\sigma \nu}
\no \\
C^*_{\a \b} \big ( \tilde C \tilde \g ^{mn} \big ) ^*_{\mu \nu} & = &
C_{\g \delta } \big ( \tilde C \tilde \g ^{mn} \big ) _{\rho \sigma} X_{\g \a} X_{\delta \b} Y_{\rho \mu} Y_{\sigma \nu}
\eea 
Factorizing the equations into independent $\ms \mo(3)$ and $\ms \mo(7)$ parts, and expressing the result in matrix notation, we obtain, 
\begin{align}
\label{C.8}
\tilde C^* & = \lambda \, Y^t \tilde C Y & (C \gamma^{ab})^* & = \lambda ^{-1} X^t C \gamma ^{ab} X
\no \\
C^* & = \mu \, X^t C X  & ( \tilde C \tilde \g ^{mn})^* & = \mu^{-1} Y^t \tilde C \tilde \g ^{mn} Y
\end{align}
The equations on the left  imply the ones on the right thanks to (\ref{1.XGY}) provided $\lambda \mu =1$.

\subsection{Solving the conditions for an involution}

In the basis of Dirac matrices adopted in section \ref{C.3}, the relations (\ref{1.XGY}) become, 
\bea
\gamma ^E_{ab} \, X & = & (-)^{s_a + s_b +a+b} X \, \gamma ^E_{ab}
\no \\
\tilde \gamma ^E_{mn} \, X & = & (-)^{\tilde s_m + \tilde s_n + m +n} X \, \tilde \gamma _{mn}^E
\eea
These relations are invariant under an overall sign flip of either metrics $\eta$ or $\tilde \eta$, so that we shall list only the cases with mostly positive signature, and we find, 
\begin{align}
& (+++) & \ms \mo (3;\RR) ~ && X &= x \, \sigma ^2 && XX^*=- x^2 I_2
\no \\
& (++-) & \ms \mo (2,1;\RR) && X &= x \, \sigma ^3 && XX^*=+ x^2 I_2
\end{align}
and 
\begin{align}
& (+++++++) & \ms \mo (7;\RR) ~ && 		Y & = y \, \sigma ^2 \otimes \sigma ^1 \otimes \sigma ^2 & YY^*&=+ y^2 I_8
\no \\
& (++++++-) & \ms \mo (6,1;\RR) && 	Y & = y \, \sigma ^3 \otimes \sigma ^1 \otimes \sigma ^2 & YY^*&=- y^2 I_8
\no \\
& (+++++--) & \ms \mo (5,2;\RR) &&		Y & = y \, \sigma ^1 \otimes \sigma ^1 \otimes \sigma ^2 & YY^*&=- y^2 I_8
\no \\
& (++++---) & \ms \mo (4,3;\RR) &&		Y & = y \, \sigma ^2 \otimes \sigma ^1 \otimes \sigma ^2 & YY^*&=+ y^2 I_8
\end{align}
where $ x,y \in \CC$. In all the above cases, the equations on the left of (\ref{C.8}) are consistent provided $x^2 \mu= y^2 \lambda$, so that the condition $\lambda \mu=1$ requires $x^2y^2=1$.  Imposing the condition (\ref{1.XY}) gives all possible real forms of $F(4;\CC)$, which we characterize here by their maximal bosonic subgroups,
\bea
\ms \mo (3;\RR) ~ & \oplus & \ms \mo (1,6;\RR)
\no \\
\ms \mo (3;\RR) ~ &  \oplus & \ms \mo (2,5;\RR)
\no \\
\ms \mo (1,2;\RR) &  \oplus & ~ \ms \mo (7;\RR)
\no \\
 \ms \mo (1,2;\RR) &  \oplus & \ms \mo (3,4;\RR)
\eea
Any other real form of the maximal bosonic subalgebra $\ms \mo (3;\CC) \oplus \ms \mo (7;\CC)$ does not correspond to a real form of $F(4;\CC)$. In particular, there exists no real form with maximal bosonic subalgebra $\ms \mo (3;\RR) \oplus \ms \mo (7;\RR)$.

\clearpage

\bibliographystyle{utphys}
\bibliography{s6s2sigma}

\providecommand{\href}[2]{#2}\begingroup\raggedright\begin{thebibliography}{10}

\bibitem{Pestun:2007rz}
V.~Pestun, ``{Localization of gauge theory on a four-sphere and supersymmetric
  Wilson loops},'' \href{https://dx.doi.org/10.1007/s00220-012-1485-0}{{\em
  Commun. Math. Phys.} {\bfseries 313} (2012) 71--129},
  \href{https://arxiv.org/abs/0712.2824}{{\ttfamily arXiv:0712.2824 [hep-th]}}.

\bibitem{Freedman:2013oja}
D.~Z. Freedman and S.~S. Pufu, ``{The holography of $F$-maximization},''
  \href{https://dx.doi.org/10.1007/JHEP03(2014)135}{{\em JHEP} {\bfseries 03}
  (2014) 135}, \href{https://arxiv.org/abs/1302.7310}{{\ttfamily
  arXiv:1302.7310 [hep-th]}}.

\bibitem{Bobev:2013cja}
N.~Bobev, H.~Elvang, D.~Z. Freedman, and S.~S. Pufu, ``{Holography for $N =
  2^*$ on $S^4$},'' \href{https://dx.doi.org/10.1007/JHEP07(2014)001}{{\em
  JHEP} {\bfseries 07} (2014) 001},
  \href{https://arxiv.org/abs/1311.1508}{{\ttfamily arXiv:1311.1508 [hep-th]}}.

\bibitem{Pestun:2016zxk}
V.~Pestun {\em et~al.}, ``{Localization techniques in quantum field
  theories},'' \href{https://dx.doi.org/10.1088/1751-8121/aa63c1}{{\em J. Phys.
  A} {\bfseries 50} no.~44, (2017) 440301},
  \href{https://arxiv.org/abs/1608.02952}{{\ttfamily arXiv:1608.02952
  [hep-th]}}.

\bibitem{Gibbons:1978ac}
G.~W. Gibbons, S.~W. Hawking, and M.~J. Perry, ``{Path Integrals and the
  Indefiniteness of the Gravitational Action},''
  \href{https://dx.doi.org/10.1016/0550-3213(78)90161-X}{{\em Nucl. Phys. B}
  {\bfseries 138} (1978) 141--150}.

\bibitem{Halliwell:1989dy}
J.~J. Halliwell and J.~B. Hartle, ``{Integration Contours for the No Boundary
  Wave Function of the Universe},''
  \href{https://dx.doi.org/10.1103/PhysRevD.41.1815}{{\em Phys. Rev. D}
  {\bfseries 41} (1990) 1815}.

\bibitem{Louko:1995jw}
J.~Louko and R.~D. Sorkin, ``{Complex actions in two-dimensional topology
  change},'' \href{https://dx.doi.org/10.1088/0264-9381/14/1/018}{{\em Class.
  Quant. Grav.} {\bfseries 14} (1997) 179--204},
  \href{https://arxiv.org/abs/gr-qc/9511023}{{\ttfamily arXiv:gr-qc/9511023}}.

\bibitem{Witten:2021nzp}
E.~Witten, ``{A Note On Complex Spacetime Metrics},''
  \href{https://arxiv.org/abs/2111.06514}{{\ttfamily arXiv:2111.06514
  [hep-th]}}.

\bibitem{Hull:1998vg}
C.~M. Hull, ``{Timelike T duality, de Sitter space, large N gauge theories and
  topological field theory},''
  \href{https://dx.doi.org/10.1088/1126-6708/1998/07/021}{{\em JHEP} {\bfseries
  07} (1998) 021}, \href{https://arxiv.org/abs/hep-th/9806146}{{\ttfamily
  arXiv:hep-th/9806146}}.

\bibitem{Hull:1998ym}
C.~M. Hull, ``{Duality and the signature of space-time},''
  \href{https://dx.doi.org/10.1088/1126-6708/1998/11/017}{{\em JHEP} {\bfseries
  11} (1998) 017}, \href{https://arxiv.org/abs/hep-th/9807127}{{\ttfamily
  arXiv:hep-th/9807127}}.

\bibitem{Dijkgraaf:2016lym}
R.~Dijkgraaf, B.~Heidenreich, P.~Jefferson, and C.~Vafa, ``{Negative Branes,
  Supergroups and the Signature of Spacetime},''
  \href{https://dx.doi.org/10.1007/JHEP02(2018)050}{{\em JHEP} {\bfseries 02}
  (2018) 050}, \href{https://arxiv.org/abs/1603.05665}{{\ttfamily
  arXiv:1603.05665 [hep-th]}}.

\bibitem{Ishibashi:1996xs}
N.~Ishibashi, H.~Kawai, Y.~Kitazawa, and A.~Tsuchiya, ``{A Large N reduced
  model as superstring},''
  \href{https://dx.doi.org/10.1016/S0550-3213(97)00290-3}{{\em Nucl. Phys. B}
  {\bfseries 498} (1997) 467--491},
  \href{https://arxiv.org/abs/hep-th/9612115}{{\ttfamily
  arXiv:hep-th/9612115}}.

\bibitem{Bonelli:2002mb}
G.~Bonelli, ``{Matrix strings in pp wave backgrounds from deformed
  superYang-Mills theory},''
  \href{https://dx.doi.org/10.1088/1126-6708/2002/08/022}{{\em JHEP} {\bfseries
  08} (2002) 022}, \href{https://arxiv.org/abs/hep-th/0205213}{{\ttfamily
  arXiv:hep-th/0205213}}.

\bibitem{Hartnoll:2024csr}
S.~A. Hartnoll and J.~Liu, ``{The Polarised IKKT Matrix Model},''
  \href{https://arxiv.org/abs/2409.18706}{{\ttfamily arXiv:2409.18706
  [hep-th]}}.

\bibitem{Hartnoll:2025ecj}
S.~A. Hartnoll and J.~Liu, ``{Statistical Physics of the Polarised IKKT Matrix
  Model},'' \href{https://arxiv.org/abs/2504.06481}{{\ttfamily arXiv:2504.06481
  [hep-th]}}.

\bibitem{Komatsu:2024bop}
S.~Komatsu, A.~Martina, J.~a. Penedones, A.~Vuignier, and X.~Zhao, ``{Einstein
  gravity from a matrix integral -- Part I},''
  \href{https://arxiv.org/abs/2410.18173}{{\ttfamily arXiv:2410.18173
  [hep-th]}}.

\bibitem{Komatsu:2024ydh}
S.~Komatsu, A.~Martina, J.~Penedones, A.~Vuignier, and X.~Zhao, ``{Einstein
  gravity from a matrix integral -- Part II},''
  \href{https://arxiv.org/abs/2411.18678}{{\ttfamily arXiv:2411.18678
  [hep-th]}}.

\bibitem{DHoker:2016ujz}
E.~D'Hoker, M.~Gutperle, A.~Karch, and C.~F. Uhlemann, ``{Warped $AdS_6\times
  S^2$ in Type IIB supergravity I: Local solutions},''
  \href{https://dx.doi.org/10.1007/JHEP08(2016)046}{{\em JHEP} {\bfseries 08}
  (2016) 046}, \href{https://arxiv.org/abs/1606.01254}{{\ttfamily
  arXiv:1606.01254 [hep-th]}}.

\bibitem{Corbino:2017tfl}
D.~Corbino, E.~D'Hoker, and C.~F. Uhlemann, ``{$AdS_{2} \times S^{6}$ versus
  $AdS_{6} \times{} S^{2}$ in Type IIB supergravity},''
  \href{https://dx.doi.org/10.1007/JHEP03(2018)120}{{\em JHEP} {\bfseries 03}
  (2018) 120}, \href{https://arxiv.org/abs/1712.04463}{{\ttfamily
  arXiv:1712.04463 [hep-th]}}.

\bibitem{Bergshoeff:2007cg}
E.~A. Bergshoeff, J.~Hartong, A.~Ploegh, J.~Rosseel, and D.~Van~den Bleeken,
  ``{Pseudo-supersymmetry and a tale of alternate realities},''
  \href{https://dx.doi.org/10.1088/1126-6708/2007/07/067}{{\em JHEP} {\bfseries
  07} (2007) 067}, \href{https://arxiv.org/abs/0704.3559}{{\ttfamily
  arXiv:0704.3559 [hep-th]}}.

\bibitem{Bobev:2018ugk}
N.~Bobev, P.~Bomans, and F.~F. Gautason, ``{Spherical Branes},''
  \href{https://dx.doi.org/10.1007/JHEP08(2018)029}{{\em JHEP} {\bfseries 08}
  (2018) 029}, \href{https://arxiv.org/abs/1805.05338}{{\ttfamily
  arXiv:1805.05338 [hep-th]}}.

\bibitem{Bobev:2019bvq}
N.~Bobev, P.~Bomans, F.~F. Gautason, J.~A. Minahan, and A.~Nedelin,
  ``{Supersymmetric Yang-Mills, Spherical Branes, and Precision Holography},''
  \href{https://dx.doi.org/10.1007/JHEP03(2020)047}{{\em JHEP} {\bfseries 03}
  (2020) 047}, \href{https://arxiv.org/abs/1910.08555}{{\ttfamily
  arXiv:1910.08555 [hep-th]}}.

\bibitem{DHoker:2007zhm}
E.~D'Hoker, J.~Estes, and M.~Gutperle, ``{Exact half-BPS Type IIB interface
  solutions. I. Local solution and supersymmetric Janus},''
  \href{https://dx.doi.org/10.1088/1126-6708/2007/06/021}{{\em JHEP} {\bfseries
  06} (2007) 021}, \href{https://arxiv.org/abs/0705.0022}{{\ttfamily
  arXiv:0705.0022 [hep-th]}}.

\bibitem{DHoker:2007mci}
E.~D'Hoker, J.~Estes, and M.~Gutperle, ``{Gravity duals of half-BPS Wilson
  loops},'' \href{https://dx.doi.org/10.1088/1126-6708/2007/06/063}{{\em JHEP}
  {\bfseries 06} (2007) 063}, \href{https://arxiv.org/abs/0705.1004}{{\ttfamily
  arXiv:0705.1004 [hep-th]}}.

\bibitem{Corbino:2020lzq}
D.~Corbino, ``{Warped AdS$_{2}$ and SU(1, 1|4) symmetry in Type IIB},''
  \href{https://dx.doi.org/10.1007/JHEP03(2021)060}{{\em JHEP} {\bfseries 03}
  (2021) 060}, \href{https://arxiv.org/abs/2004.12613}{{\ttfamily
  arXiv:2004.12613 [hep-th]}}.

\bibitem{DHoker:2008lup}
E.~D'Hoker, J.~Estes, M.~Gutperle, and D.~Krym, ``{Exact Half-BPS Flux
  Solutions in M-theory. I: Local Solutions},''
  \href{https://dx.doi.org/10.1088/1126-6708/2008/08/028}{{\em JHEP} {\bfseries
  08} (2008) 028}, \href{https://arxiv.org/abs/0806.0605}{{\ttfamily
  arXiv:0806.0605 [hep-th]}}.

\bibitem{Nahm:1977tg}
W.~Nahm, ``{Supersymmetries and Their Representations},''
  \href{https://dx.doi.org/10.1201/9781482268737-2}{{\em Nucl. Phys. B}
  {\bfseries 135} (1978) 149}.

\bibitem{Parker:1980af}
M.~Parker, ``{CLASSIFICATION OF REAL SIMPLE LIE SUPERALGEBRAS OF CLASSICAL
  TYPE},'' \href{https://dx.doi.org/10.1063/1.524487}{{\em J. Math. Phys.}
  {\bfseries 21} (1980) 689--697}.

\bibitem{Kac}
V.~Kac, ``{Lie Superalgebras},''
  \href{https://dx.doi.org/10.1016/0001-8708(77)90017-2}{{\em Advances in
  Mathematics} {\bfseries 26} (1977) 8--96}.

\bibitem{Sorba}
L.~Frappat, A.~Sciarrino, and P.~Sorba, {\em {Dictionnary on Lie Algebras and
  Superalgebras}}.

\bibitem{Gomis:2006cu}
J.~Gomis and C.~Romelsberger, ``{Bubbling Defect CFT's},''
  \href{https://dx.doi.org/10.1088/1126-6708/2006/08/050}{{\em JHEP} {\bfseries
  08} (2006) 050}, \href{https://arxiv.org/abs/hep-th/0604155}{{\ttfamily
  arXiv:hep-th/0604155}}.

\bibitem{DGU-to-appear}
E.~D'Hoker, M.~Gutperle, and C.~F. Uhlemann {\em to appear} .

\bibitem{2016JGP...104..163P}
V.~{Pessers} and J.~{Van der Veken}, ``{On holomorphic Riemannian geometry and
  submanifolds of Wick-related spaces},''
  \href{https://dx.doi.org/10.1016/j.geomphys.2016.02.009}{{\em Journal of
  Geometry and Physics} {\bfseries 104} (June, 2016) 163--174},
  \href{https://arxiv.org/abs/1503.07354}{{\ttfamily arXiv:1503.07354
  [math.DG]}}.

\bibitem{2020arXiv200200810B}
F.~{Bonsante} and C.~{El Emam}, ``{On immersions of surfaces into SL(2,C) and
  geometric consequences},''
  \href{https://dx.doi.org/10.48550/arXiv.2002.00810}{{\em arXiv e-prints}
  (Feb., 2020) arXiv:2002.00810},
  \href{https://arxiv.org/abs/2002.00810}{{\ttfamily arXiv:2002.00810
  [math.DG]}}.

\bibitem{DHoker:2017mds}
E.~D'Hoker, M.~Gutperle, and C.~F. Uhlemann, ``{Warped $AdS_6\times S^2$ in
  Type IIB supergravity II: Global solutions and five-brane webs},''
  \href{https://dx.doi.org/10.1007/JHEP05(2017)131}{{\em JHEP} {\bfseries 05}
  (2017) 131}, \href{https://arxiv.org/abs/1703.08186}{{\ttfamily
  arXiv:1703.08186 [hep-th]}}.

\bibitem{DHoker:2017zwj}
E.~D'Hoker, M.~Gutperle, and C.~F. Uhlemann, ``{Warped $AdS_6\times S^2$ in
  Type IIB supergravity III: Global solutions with seven-branes},''
  \href{https://dx.doi.org/10.1007/JHEP11(2017)200}{{\em JHEP} {\bfseries 11}
  (2017) 200}, \href{https://arxiv.org/abs/1706.00433}{{\ttfamily
  arXiv:1706.00433 [hep-th]}}.

\bibitem{Corbino:2018fwb}
D.~Corbino, E.~D'Hoker, J.~Kaidi, and C.~F. Uhlemann, ``{Global half-BPS
  $AdS_2\times S^6$ solutions in Type IIB},''
  \href{https://dx.doi.org/10.1007/JHEP03(2019)039}{{\em JHEP} {\bfseries 03}
  (2019) 039}, \href{https://arxiv.org/abs/1812.10206}{{\ttfamily
  arXiv:1812.10206 [hep-th]}}.

\bibitem{Gibbons:1995vg}
G.~W. Gibbons, M.~B. Green, and M.~J. Perry, ``{Instantons and seven-branes in
  type IIB superstring theory},''
  \href{https://dx.doi.org/10.1016/0370-2693(95)01565-5}{{\em Phys. Lett. B}
  {\bfseries 370} (1996) 37--44},
  \href{https://arxiv.org/abs/hep-th/9511080}{{\ttfamily
  arXiv:hep-th/9511080}}.

\bibitem{Bergshoeff:1998ry}
E.~Bergshoeff and K.~Behrndt, ``{D - instantons and asymptotic geometries},''
  \href{https://dx.doi.org/10.1088/0264-9381/15/7/002}{{\em Class. Quant.
  Grav.} {\bfseries 15} (1998) 1801--1813},
  \href{https://arxiv.org/abs/hep-th/9803090}{{\ttfamily
  arXiv:hep-th/9803090}}.

\bibitem{Ciceri:2025maa}
F.~Ciceri and H.~Samtleben, ``{Holography for the
  Ishibashi-Kawai-Kitazawa-Tsuchiya Matrix Model},''
  \href{https://dx.doi.org/10.1103/fb8g-b8fd}{{\em Phys. Rev. Lett.} {\bfseries
  135} no.~6, (2025) 061601},
  \href{https://arxiv.org/abs/2503.08771}{{\ttfamily arXiv:2503.08771
  [hep-th]}}.

\bibitem{Banks:1996vh}
T.~Banks, W.~Fischler, S.~H. Shenker, and L.~Susskind, ``{M theory as a matrix
  model: A conjecture},''
  \href{https://dx.doi.org/10.1201/9781482268737-37}{{\em Phys. Rev. D}
  {\bfseries 55} (1997) 5112--5128},
  \href{https://arxiv.org/abs/hep-th/9610043}{{\ttfamily
  arXiv:hep-th/9610043}}.

\bibitem{Martina:2025kwc}
A.~Martina, ``{Massive deformations of supersymmetric Yang-Mills matrix
  models},'' \href{https://arxiv.org/abs/2507.17813}{{\ttfamily
  arXiv:2507.17813 [hep-th]}}.

\bibitem{Bergman:2018hin}
O.~Bergman, D.~Rodr\'\i{}guez-G\'omez, and C.~F. Uhlemann, ``{Testing
  AdS$_{6}$/CFT$_{5}$ in Type IIB with stringy operators},''
  \href{https://dx.doi.org/10.1007/JHEP08(2018)127}{{\em JHEP} {\bfseries 08}
  (2018) 127}, \href{https://arxiv.org/abs/1806.07898}{{\ttfamily
  arXiv:1806.07898 [hep-th]}}.

\bibitem{Uhlemann:2020bek}
C.~F. Uhlemann, ``{Wilson loops in 5d long quiver gauge theories},''
  \href{https://dx.doi.org/10.1007/JHEP09(2020)145}{{\em JHEP} {\bfseries 09}
  (2020) 145}, \href{https://arxiv.org/abs/2006.01142}{{\ttfamily
  arXiv:2006.01142 [hep-th]}}.

\bibitem{Gutperle:2020rty}
M.~Gutperle and C.~F. Uhlemann, ``{Surface defects in holographic 5d SCFTs},''
  \href{https://dx.doi.org/10.1007/JHEP04(2021)134}{{\em JHEP} {\bfseries 04}
  (2021) 134}, \href{https://arxiv.org/abs/2012.14547}{{\ttfamily
  arXiv:2012.14547 [hep-th]}}.

\bibitem{Kim:2011cr}
S.-W. Kim, J.~Nishimura, and A.~Tsuchiya, ``{Expanding (3+1)-dimensional
  universe from a Lorentzian matrix model for superstring theory in
  (9+1)-dimensions},''
  \href{https://dx.doi.org/10.1103/PhysRevLett.108.011601}{{\em Phys. Rev.
  Lett.} {\bfseries 108} (2012) 011601},
  \href{https://arxiv.org/abs/1108.1540}{{\ttfamily arXiv:1108.1540 [hep-th]}}.

\bibitem{Nishimura:2022alt}
J.~Nishimura, ``{Signature change of the emergent space-time in the IKKT matrix
  model},'' \href{https://dx.doi.org/10.22323/1.406.0255}{{\em PoS} {\bfseries
  CORFU2021} (2022) 255}, \href{https://arxiv.org/abs/2205.04726}{{\ttfamily
  arXiv:2205.04726 [hep-th]}}.

\bibitem{Asano:2024def}
Y.~Asano, J.~Nishimura, W.~Piensuk, and N.~Yamamori, ``{Defining the Type IIB
  Matrix Model without Breaking Lorentz Symmetry},''
  \href{https://dx.doi.org/10.1103/PhysRevLett.134.041603}{{\em Phys. Rev.
  Lett.} {\bfseries 134} no.~4, (2025) 041603},
  \href{https://arxiv.org/abs/2404.14045}{{\ttfamily arXiv:2404.14045
  [hep-th]}}.

\bibitem{Moore:1998et}
G.~W. Moore, N.~Nekrasov, and S.~Shatashvili, ``{D particle bound states and
  generalized instantons},''
  \href{https://dx.doi.org/10.1007/s002200050016}{{\em Commun. Math. Phys.}
  {\bfseries 209} (2000) 77--95},
  \href{https://arxiv.org/abs/hep-th/9803265}{{\ttfamily
  arXiv:hep-th/9803265}}.

\bibitem{Krauth:1998xh}
W.~Krauth, H.~Nicolai, and M.~Staudacher, ``{Monte Carlo approach to M
  theory},'' \href{https://dx.doi.org/10.1016/S0370-2693(98)00557-7}{{\em Phys.
  Lett. B} {\bfseries 431} (1998) 31--41},
  \href{https://arxiv.org/abs/hep-th/9803117}{{\ttfamily
  arXiv:hep-th/9803117}}.

\bibitem{Green:1998yf}
M.~B. Green and M.~Gutperle, ``{D instanton partition functions},''
  \href{https://dx.doi.org/10.1103/PhysRevD.58.046007}{{\em Phys. Rev. D}
  {\bfseries 58} (1998) 046007},
  \href{https://arxiv.org/abs/hep-th/9804123}{{\ttfamily
  arXiv:hep-th/9804123}}.

\bibitem{Kontsevich:2021dmb}
M.~Kontsevich and G.~Segal, ``{Wick Rotation and the Positivity of Energy in
  Quantum Field Theory},'' \href{https://dx.doi.org/10.1093/qmath/haab027}{{\em
  Quart. J. Math. Oxford Ser.} {\bfseries 72} no.~1-2, (2021) 673--699},
  \href{https://arxiv.org/abs/2105.10161}{{\ttfamily arXiv:2105.10161
  [hep-th]}}.

\bibitem{Freedman:2012zz}
D.~Z. Freedman and A.~Van~Proeyen,
  \href{https://dx.doi.org/10.1017/CBO9781139026833}{{\em {Supergravity}}}.
\newblock Cambridge Univ. Press, Cambridge, UK, 5, 2012.

\end{thebibliography}\endgroup
\end{document}